\include{BoxedEPS}

\documentclass[11pt]{amsart}

\usepackage{diagbox}
\usepackage{multirow}
\usepackage{mathrsfs}
\usepackage{amssymb}
\usepackage{amsfonts}
\usepackage{amsbsy}
\usepackage{latexsym}
\usepackage{amssymb,latexsym,amsmath,amsthm}
\usepackage{xcolor}
\usepackage[ruled,linesnumbered]{algorithm2e}
\SetKwRepeat{Do}{do}{while}%
\usepackage{graphicx}
\usepackage{float} 
\usepackage{subfigure} %
\usepackage{epstopdf}
\usepackage{booktabs}% make tabular

\evensidemargin=\oddsidemargin \textwidth=160mm

\newtheorem{thm}{Theorem}[section]
\newtheorem{cor}[thm]{Corollary}

\newtheorem{prop}[thm]{Proposition}

\newtheorem{assump}[thm]{Assumption}

\theoremstyle{definition}

\theoremstyle{remark}
\newtheorem{rem}{Remark}[section]

\numberwithin{equation}{section}

\begin{document}
\bibliographystyle{plain}

\title [Random sampling and reconstruction]
{Random sampling and reconstruction of concentrated signals in a reproducing kernel space}

%%%%%%%%%%%%%%%%%%%%%%%%%%%%%%%%%%%%%%%%%%%%%%%%%%%%%%%%%%%%%%%%%%%%%%%%
\author{Yaxu Li}
\address{Y. Li, Department of Mathematics, Sun
Yat-sen University, Guangzhou 510275, China.} \email{liyx97@mail2.sysu.edu.cn}

\author{Qiyu Sun}
\address{Q. Sun, Department of Mathematics, University of Central Florida,
Orlando, FL 32816, USA}
\email{qiyu.sun@ucf.edu }

\author{Jun Xian}
\address{J. Xian, Department of Mathematics, Sun
Yat-sen University, Guangzhou 510275, China.} \email{xianjun@mail.sysu.edu.cn}

%%%%%%%%%%%%%%%%%%%%%%%%%%%%%%%%%%%%%%%%%%%%%%%%%%%%%%%%%%%%%%%%%%%%%%%%

\date{\today }

\thanks{The project is partially supported  by  the National Science Foundation (DMS-1816313) and Guangdong Provincial Government of China through the Computational Science Innovative Research Team Program and Guangdong Province Key Laboratory of Computational Science at the Sun Yat-sen University.
 }

\subjclass[2010] {94A12, 42C40, 65T60}  %{\color{red} ???, 42A70, 45Q05, 44A60, 94A20}

\keywords{ Random sampling and reconstruction; Reproducing kernel space;  Corkscrew domain}

\maketitle

%%%%%%%%%%%%%%%%%%%%%%%%%%%%%%%%%%%%%%%%%%%%%%%%%%%%%%%%%%%%%%%%%%%%%%%%
\begin{abstract}
%Sampling signals of interest on a sampling set  in a stable way and reconstructing the original signal  exactly or approximately from its sampling data  are fundamental problems in sampling theory.
In this paper, we consider (random) sampling  of signals concentrated on a bounded Corkscrew domain  $\Omega$  of a metric measure space,
and reconstructing  concentrated signals  approximately from their (un)corrupted sampling data taken on a sampling set  contained in $\Omega$.
We 
establish a weighted stability
of bi-Lipschitz type for a (random) sampling scheme on the set of  concentrated signals  in a reproducing kernel space.
The weighted stability of bi-Lipschitz type
 provides a weak robustness to the sampling scheme, however due to the nonconvexity of the set of concentrated signals,
it
   does not imply the unique signal reconstruction.
From
 (un)corrupted samples taken on a finite sampling set  contained in $\Omega$, we
  propose an algorithm to find   approximations to  signals  concentrated on
  a bounded Corkscrew domain  $\Omega$.
Random sampling is a sampling scheme where sampling positions
are randomly taken according to a probability distribution.
%, which has received a lot of attention recently
% in the communities of signal processing,  compressive sensing, learning theory and sampling theory.
Next we show that, with high probability,  signals  concentrated on
  a bounded Corkscrew domain  $\Omega$ can be reconstructed approximately from their
 uncorrupted (or randomly corrupted) samples taken at
 i.i.d. random
positions drawn on  $\Omega$, provided that the sampling size is at least  of the order $\mu(\Omega) \ln (\mu(\Omega))$, where $\mu(\Omega)$
is the measure of the concentrated domain $\Omega$.
Finally, we demonstrate the
performance of proposed approximations to the original concentrated
signal when the sampling procedure is taken either with small density or
randomly with large size.
\end{abstract}

\maketitle

\section{Introduction}
Sampling signals of interest %on a sampling set $\Gamma$
 in a stable way and reconstructing the original signals  exactly or approximately from
their  (un)corrupted sampling data  % taken on $\Gamma$
are  fundamental problems in sampling theory. 
A common assumption is that signals of interest have some additional properties, such as residing in a linear space, or having sparse representation in a dictionary, or having finite rate of innovation \cite{aldroubi2001SIAM, Candes2008, chengacha2019, Foucartbook, nashedsun2010, smalezhou2004, sun2006SIAMJMA,  unser2000, vetterli2002}. In this paper, we consider (random) sampling and reconstruction of  signals in a reproducing kernel space concentrated on a  bounded Corkscrew domain.

Let $(X, \rho, \mu)$ be a metric measure space and $L^p:=L^p(X,\rho,\mu), 1\le p\le \infty$, be the
linear space of all $p$-integrable functions on the  metric measure space $(X, \rho, \mu)$ with the standard $p$-norm  denoted by $\|\cdot\|_p$.
In this paper, we use the range space
\begin{equation}\label{Vrks}
  V_p:=\{Tf :\  f\in L^p \}=\{f\in L^p : \ Tf=f \}, \ 1\le p \le \infty,
\end{equation}
 of an idempotent integral operator
\begin{equation}\label{integraloperator.def}
Tf(x)=\int_X K(x,y)f(y)d\mu(y), \ f\in L^p,
\end{equation}
as  the reproducing kernel space for  %concentrated {\color{blue}delete ``concentrated''?, since till now people do not know what the concentrated signal is}
 signals to reside in,
where  the integral kernel $K$ having certain off-diagonal decay and H\"older continuity, see Assumption \ref{kernel.assumption}.
The  above range space  $V_p, 1\le p\le \infty$, was introduced by Nashed and Sun  in the Euclidean setting
\cite{nashedsun2010}, and it has rich geometric structure and lots of flexibility to
approximate real data set in signal processing and learning theory.
 Our illustrative  examples are
 spaces of  $p$-integrable \mbox{(non-)uniform} splines \cite{aldroubi1998, sun2008ACM, unser2007},
 shift-invariant spaces with their generators having certain regularity and decay  at infinity
 \cite{aldroubi2001SIAM, aldroubi2005sun,  unser2000}, and spaces of signals with finite rate of innovation  \cite{sun2016galerkin, dragotti2007,  sun2006SIAMJMA, vetterli2002}.
   Sampling and reconstruction of signals in the range spaces of  integral operators in the Euclidean space has been well studied,
see \cite{sun2016galerkin, jiang2020, nashedsun2010} and references therein.
  %  and Remark \eqref{rkssampling.rem} of this paper
For signals in %the reproducing kernel space
 $V_p, 1\le p\le \infty$, as shown in  Proposition
\ref{rkssampling.prop}, they
can be recovered exactly
via  an exponentially convergent algorithm
from their samples taken on a sampling set  with small density.

For some engineering  applications, signals of interest are concentrated on a bounded domain
$\Omega$
and only finitely many sampling data taken inside the domain $\Omega$ are available \cite{adcock2014, bass2013random, bass2010random, sun2016galerkin, hogan2012, jaming2014}.
    This motivates us to consider sampling and reconstruction of signals in the space $V_p, 1\le p\le \infty$, concentrated on a  bounded domain $\Omega$,
\begin{equation}\label{VOmega}
  V_{p, \Omega,\varepsilon}:=\big\{f \in V_p : \  \|f\|_{p,\Omega^c} \le \varepsilon\|f\|_p\big\},
\end{equation}
where $\varepsilon \in (0,1)$, $\Omega^c \subseteq X$ is the complement of the domain $\Omega$, and     $\|f\|_{p,\Omega^c}$ is the standard $p$-norm on  the complement $\Omega^c$.
The set $V_{p, \Omega, \varepsilon}$ of $\varepsilon$-concentrated signals  has been introduced  for time-frequency analysis \cite{grochenig2001fundations, velasco2017}, phase retrieval \cite{alaifrai2018stable, alaifrai2019gabor, grohs19stable}, and (random) sampling of bandlimited and wavelet signals \cite{bass2013random, bass2010random, sun2016galerkin, xian2019random,  yang2013}.
   As signals in $V_{p, \Omega,\varepsilon}$ are essentially supported on  the domain $\Omega$,
it is more natural to consider a sampling procedure taken on a finite sampling  set $\Gamma_\Omega$ contained inside the domain $\Omega$ only.
  %region where $f$ is small do not contribute anything significant to the reconstruction process.
         In Section \ref{deterministicsampling.section},  we  show that the sampling procedure
    $f\mapsto (f(\gamma))_{\gamma\in \Gamma_\Omega}$ for $\varepsilon$-concentrated signals   in $V_{p, \Omega, \varepsilon}$
   has weighted stability of bi-Lipschitz type  when
the  Hausdorff distance
   \begin{equation*}\label{deltaGamma.def}
 d_H(\Gamma_\Omega,\Omega):=\sup_{x\in \Omega} \rho(x, \Gamma_\Omega)\end{equation*}
    between the sampling set $\Gamma_\Omega\subset \Omega$ and the bounded Corkscrew domain $\Omega$ is small, see
  Theorem \ref{stability.thm} and Corollary \ref{stability.cor}.
 For signals in a linear space, stability of a sampling scheme %is an important concept for
   guarantees  robustness and uniqueness of reconstructing signals
   from their (noisy) samples, see \cite{aldroubi2001SIAM, sun2016galerkin, nashedsun2010, sunxian2014JFAA}.
However,
the weighted stability in Theorem \ref{stability.thm} does not imply the
  unique reconstruction even  it provides a weak robustness  for the sampling scheme on $V_{p, \Omega, \varepsilon}$, see \cite{bass2013random} and Remark \ref{stability.rem}. Therefore we should consider reconstructing  $\varepsilon$-concentrated signals in $V_{p, \Omega, \varepsilon}$ approximately, instead of exactly,
from their samples inside the domain.
A challenge to  derive such good approximations
to  $\varepsilon$-concentrated signals in $ V_{p,  \Omega, \varepsilon}$
% from their samples
%taken on a sampling set $\Gamma_\Omega$ contained in $\Omega$
is that
 the set $V_{p, \Omega, \varepsilon}$ is a nonconvex subset of the reproducing kernel space $V_p$
(and hence it is not a linear space),
 which prevents the direct application of
     reconstruction algorithms
 used for  signals in a linear space
 \cite{aldroubi1998, aldroubi2001SIAM, aldroubiCA2004,sun2016galerkin, feichtinger92, %grochenigfeichtinger1992,
 nashedsun2010, sun2006SIAMJMA, unser2007, xianli2007ACHA}.
%$\varepsilon$-concentrated signals
 % approximately from their samples insider the domain
   In Theorems \ref{approximation.thm} and \ref{deterministicnoise.thm},
  we propose an algorithm to construct  suboptimal approximations  % in $V_{p, \Omega, 9C_0\varepsilon}$
   to  $\varepsilon$-concentrated signals in $V_{p, \Omega, \varepsilon}$
  from  their (un)corrupted samples taken on a finite sampling set $\Gamma_\Omega\subset \Omega$.
  %, where $C_0$ is an absolute constant.

\medskip

Random sampling is a sampling scheme where
 sampling  positions % $\gamma\in \Gamma$
  are randomly taken according to a probability distribution
 \cite{Beutler1966, Marsy71, Vitter1985}.
  %Due to the advantage on frequency anti-aliasing and low sampling rate,
 It has received a lot of attention in the communities of  %in imaging,  speech and radar
 signal processing, compressive sensing, learning theory and sampling theory, %in last decade
see \cite{%er1985, Vitter1984,
  bass2005random,  Candes2008, chan2014, cucker2002,  %kunisrauhut2008random,
    Marvasti86, Particke2017,     % smale2009,  % candes2006a, candes2006b,
  rauhut2007random, smalezhou2005, smalezhou2004, Zarmehi2017}  and references therein.
Random sampling of concentrated signals was first discussed by Bass and Gr\"ochenig, and they proved the following result
for bandlimited signals concentrated on the cube $C_R:=[-R/2, R/2]^d$, see  \cite[Theorem 3.1]{bass2010random}.

\begin{thm}
Let  $R\ge 2$,    $\varepsilon\in (0, 1)$ and $\mu\in (0, 1-\varepsilon)$.
If  sampling  positions  $\gamma\in \Gamma_R$
are i.i.d. random variables that are uniformly distributed over the cube $C_R$, then there exist
 absolute  positive constants $A$ and $B$ such that the sampling inequalities
\begin{equation}\label{samplinginequality.def0}
 \frac{N}{R^d} (1-\varepsilon-\mu) \|f\|_2^2 \le \sum_{\gamma\in \Gamma_R} |f(\gamma)|^2\le \frac{N}{R^d} (1+\mu) \|f\|_2^2, \
 f\in  {\mathcal B}_{\varepsilon, R}\end{equation}
hold  with probability  at  least $1- A \exp(-B \mu^2 N/R^d)$, where $N=\# \Gamma_R$ is the size of the sampling set $\Gamma_R$,
${\mathcal B}$ is the space of signals bandlimited to $[-1/2, 1/2]^d$, and ${\mathcal B}_{\varepsilon, R}=\{ f\in {\mathcal B},  \ \|f\|_{2, {\mathbb R}^d\backslash C_R} \le \sqrt{\varepsilon} \|f\|_2\}$
is the set of  bandlimited signals  concentrated on $C_R$. %[-R/2, R/2]^d$.
\end{thm}

The sampling inequality of the form \eqref{samplinginequality.def0} has been %recently
 extended to signals in a shift-invariant space, with finite rate of innovation and in a reproducing kernel space on the Euclidean space
   ${\mathbb R}^d$, % concentrated to $[-R/2, R/2]^d$,
 see \cite{bass2013random, xian2019random,   jiang2020, lu2020, patel2019, yang2013}. In this paper,
we introduce a completely different approach to obtain a weighted sampling inequality, see Theorem \ref{randomstability.thm} and Remarks %\ref{samplinginequality.remark}.
\ref{randomsamplinginequality.rem1} and \ref{randomsamplinginequality.rem2}. 
The sampling inequality of the form \eqref{samplinginequality.def0} provides an estimate to the signal energy with high probability,
however it does not yield a stable reconstruction of $\varepsilon$-concentrated signals in a reproducing kernel space.
    To the best
of our knowledge, there is no algorithm available to perform the  reconstruction of $\varepsilon$-concentrated signals in a reproducing kernel space approximately from their random samples in the considered domain.
In Theorems \ref{maintheorem.tm} and  \ref{random.thm}, we show that the algorithm proposed in  Theorem \ref{approximation.thm}
provides  good approximations to  the $\varepsilon$-concentrated signal
from its uncorrupted
(or randomly corrupted)
 random samples, with high probability, when the sampling size is large enough.

The main contributions of this paper are as follows:
 (i)\ We consider sampling  and reconstruction of signals concentrated on a bounded Corkscrew domain $\Omega$ of a metric measure space, instead of signals concentrated on a cube $[-R/2, R/2]^d$ of the $d$-dimensional Euclidean space in the literature \cite{bass2013random, bass2010random, %bass2005random,
  xian2019random, jiang2020,   lu2020, yang2013}.
{(ii)}\ For a (random) sampling scheme for signals concentrated on a bounded Corkscrew domain, we
%observe that the difference of two $\varepsilon$-concentrated signals is not necessarily a $\varepsilon$-concentrated signals
%and hence we
establish a weighted stability
of bi-Lipschitz type
 instead of the sampling inequality of the form \eqref{samplinginequality.def0},
 which
  provides  weak robustness of  the sampling scheme.
 {(iii)} \  The set of $\varepsilon$-concentrated signals is nonconvex and the (random) sampling operator is not one-to-one in general.
  We propose an algorithm to construct suboptimal approximations
to the original $\varepsilon$-concentrated signals from their (random) samples on the considered domain.
(iv)\ We show that, with high probability,  signals  concentrated on
  a bounded Corkscrew domain  $\Omega$ can be reconstructed approximately from their
 samples taken at
 i.i.d. random
positions drawn on  $\Omega$, provided that the sampling size is at least  of the order $\mu(\Omega) \ln (\mu(\Omega))$, where $\mu(\Omega)$
is the measure of the concentrated domain $\Omega$.
{(v)}\ We show that with high probability,  an original  $\varepsilon$-concentrated signal  can be reconstructed approximately
  from its random samples corrupted by i.i.d. random noises,
  when the random sampling size is large enough.

The paper is organized as follows.
In Section \ref{preliminary.section},  we present some preliminaries on Corkscrew domains $\Omega$ of  a metric measure space on which  sampling is taken,
and   reproducing kernel spaces  $V_p,  1\le p\le \infty$, in which $\varepsilon$-concentrated  signals on a Corkscrew domain $\Omega$ reside.
In Section \ref{deterministicsampling.section},  we  consider the sampling procedure  $f\mapsto (f(\gamma))_{\gamma\in \Gamma_\Omega}$
  taken on a finite sampling set $\Gamma_\Omega$ contained in a Corkscrew domain $\Omega$ for $\varepsilon$-concentrated signals  $f$ in the reproducing kernel space $V_p$. We establish    the stability of bi-Lipschitz type for the above sampling procedure in Theorem \ref{stability.thm},
  and
we construct suboptimal  approximations to the original  $\varepsilon$-concentrated signals
 from  their (un)corrupted samples, see Theorems \ref{approximation.thm} and \ref{deterministicnoise.thm}.
In Section \ref{randomsampling.section},  we consider random sampling of $\varepsilon$-concentrated signals
in the reproducing kernel space $V_p$ with large sampling size, and we show that, with high probability, any $\varepsilon$-concentrated signal
can be reconstructed approximately from its (un)corrupted samples
taken randomly on the Corkscrew domain $\Omega$, see Theorems \ref{maintheorem.tm} and \ref{random.thm}.
  In Section \ref{simulations.section}, we demonstrate the performance of the proposed approximations
    to the original
  $\varepsilon$-concentrated signals when the sampling procedure is taken either with sufficient density or randomly with large size.
In Section \ref{proofs.section}, we include the proofs of all theorems and propositions.
% include the proofs of the main theorems.

\section{Preliminaries on Corkscrew domains and  reproducing kernel spaces}
\label{preliminary.section}

In this section, we present some preliminaries on  Corkscrew domains  $\Omega$ of  a metric measure space  $(X, \rho, \mu)$
%for finite (random) sampling,
and the
 range space $V_p$ of an idempotent integral operator  $T$  for $\varepsilon$-concentrated signals on $\Omega$ to reside in.
 Our illustrative model of Corkscrew domains is  Lipschitz domains in ${\mathbb R}^d$, such as
 rectangular regions $[-R/2, R/2]^d$ with side length $R\ge 1$ or  balls $B(0, R)$ with center at the origin and radius $R\ge 1$.
 For a bounded Corkscrew domain $\Omega$ with
${\rm diam}(\partial \Omega)\ge 1$,
 we observe that for any $\delta\in (0, 1)$ there exist
a finite set $\Omega_\delta$  and a disjoint partition $I_\gamma, \gamma\in \Omega_\delta$, of the domain $\Omega$
 with the property that
\begin{equation}\label{unitpartition.cor.eq1}
 B(\gamma, c \delta)\subset I_{\gamma}\subset B(\gamma, \delta)\  \ {\rm for \ all} \  \gamma\in \Omega_\delta,
\end{equation}
where $c\in (0,1)$ is an absolute constant, see Proposition \ref{unitpartition.corprop}. For the range space $V_p$  with  the integral kernel
$K$ having certain off-diagonal decay and H\"older continuity,
we show that it is
 a reproducing kernel space
 and signals in $V_p$ can be reconstructed
 from their samples  by an exponentially convergent iterative
algorithm,
 see Propositions \ref{RksProperty.pr} and
\ref{rkssampling.prop}.

\subsection{Corkscrew domains in  a metric measure space}
\label{corkcrew.subsection}

A {\em metric} $\rho$ on a set $X$  is a function $\rho: X\times X\longmapsto [0,
\infty)$ such that (i)
$\rho(x, y)= 0$ if and only if $x=y$; (ii)
$\rho(x,y)=\rho(y,x)$ for all $x, y\in X$; and (iii)
$\rho(x,y)\le \rho(x, z)+\rho(z, y)$  for all $x, y, z\in X$.
A {\em metric measure space}
$(X, \rho, \mu)$
is a metric space $(X, \rho)$ with a non-negative Borel measure $\mu$ compatible with the topology
generated by open balls
$\{y\in X, \  \rho(x,y)< r\}$ %{\color{blue}``open balls
%$\{y\in X, \  \rho(x,y)< r\}$'' to ``closed balls
%$\{y\in X, \  \rho(x,y)\le r\}$?}
with center $x\in X$ and radius $r>0$.
For a  metric measure space $(X, \rho, \mu)$, we denote the diameter of a set $Y\subset X$
by ${\rm diam}(Y)$, and define the closed  ball
with center $x\in X$ and radius $r\ge 0$ by
$$B(x,r):=\big\{y\in X, \  \rho(x,y)\le r \big\}.$$
  In this paper, we  always assume  the following: % for the metric measure space $(X, \rho, \mu)$.

 \begin{assump}\label{metricspace.assump}
 The metric measure space $(X, \rho, \mu)$ has dimension $d>0$
 in the sense that
\begin{equation}\label{regular.def}
  D_1r^d\le \mu\big(B(x, r)\big) \le D_2r^d \quad  {\rm for\  all} \ x\in X \ {\rm and} \ 0\le r\le {\rm diam}(X),
\end{equation}
where $D_1$ and $D_2$ are  positive constants. %${\rm diam}(X)$ is  the diameter of the metric space $(X, \rho)$,
\end{assump}

We call a Borel measure $\mu$ satisfying \eqref{regular.def}  to be {\em Ahlfors $d$-regular},
and denote the maximal lower bound and minimal
  upper bound in \eqref{regular.def} by $D_1(\mu)$ and $D_2(\mu)$ respectively \cite{denghan, tyson2001metric}.
Our models of  metric measure spaces are the Euclidean space $\mathbb R^d$, the  sphere $S^d\subset \mathbb R^{d+1}$ and
 the  torus ${\mathbb T}^d$. % and  a $d$-dimensional compact manifold.
%$d$-dimensional complete manifold with bounded curvature. {\bf Ask Junho Lee for complete manifold with Gaussian curvature?}

\medskip
For the metric measure space $(X, \rho, \mu)$, we can find a finite overlapping cover by
 balls $B(x_i, \delta), x_i\in X_\delta$, for all $\delta>0$ such that $B(x_i, \delta/2), x_i\in X_\delta$, are mutually disjoint.
In particular, given a dense subset $Y\subset X$, let $X_\delta$ be a maximal subset of $Y$ such that
$B(x_i, \delta/2), x_i\in X_\delta$, are mutually disjoint, i.e.,
\begin{equation}\label{disjoint.def}
B(x_i, \delta/2)\cap B(x_j, \delta/2)=\emptyset \ \ {\rm for \ all\ distinct} \ x_i, x_j\in X_\delta,
\end{equation}
and
\begin{equation}\label{maximal.eq}
B(y, \delta/2)\cap \big(\cup_{x_i\in X_\delta} B(x_i, \delta/2)\big)\ne \emptyset \ \ {\rm for \ all} \ y\in Y.\end{equation}
Then one may verify that the above family of closed balls $\{B(x_i, \delta), x_i\in X_\delta\}$   covers the whole space $X$ with finite overlapping,
\begin{equation}\label{covering.eq1}
1\le \sum_{x_i\in X_\delta} \chi_{B(x_i, \delta)}(x)\le  \frac{3^d  D_2(\mu)}{D_1(\mu)}, \ x\in X,
\end{equation}
where the first inequality holds as $X_\delta$ is closed
and
$$\rho(x, X_\delta)=\inf_{y\in X_\delta} \rho(x, y)\le \delta, \ x\in X$$
 by \eqref{maximal.eq},
and
the  second one follows since
\begin{eqnarray*}\sum_{x_i\in X_\delta} \chi_{B(x_i, \delta)}(x)
 &\hskip-0.08in  \le & \hskip-0.08in  \sum_{x_i\in  B(x, \delta)}\frac{\mu(B(x_i, \delta/2))}{D_1(\mu) (\delta/2)^d}
= \frac{\mu\big(\cup_{x_i\in B(x, \delta)} B(x_i, \delta/2)\big)}{D_1(\mu) (\delta/2)^d}\\
& \hskip-0.08in \le &  \hskip-0.08in   \frac{\mu(B(x, 3\delta/2))}{D_1(\mu) (\delta/2)^d} \le 3^d \frac{D_2(\mu)}{D_1(\mu)}
\end{eqnarray*}
by \eqref{regular.def} and \eqref{disjoint.def}.

\smallskip

We say that a domain $D$ of the metric measure space $(X, \rho, \mu)$ is a  {\em Corkscrew  domain} if
any ball $B(x, r)$ with center  at the boundary $x\in \partial D$ and radius $0<r\le \ {\rm diam}\ \partial D$ contains
one ball inside the domain with a fraction of radius,
\begin{equation} \label{corkscrew.condition}
B(y, cr)\subset D \cap B(x, r) \ {\rm for  \ some} \ y\in  D,\end{equation}
where $c\in (0, 1)$ is an absolute constant.
 Our illustrative model is Lipschitz domains in ${\mathbb R}^d$, such as
 rectangular regions $[-R/2, R/2]^d$ with side length $R\ge 1$ or  balls $B(0, R)$ with center  at the origin and radius $R\ge 1$. In this paper, we consider %(random) sampling and reconstruction of signals concentrated   on a
  bounded Corkscrew  domain  $\Omega$ satisfying the following:

 \begin{assump}\label{domain.assump}
 The bounded  domain  $\Omega$ and its complement $\Omega^c$ satisfy
 the Corkscrew condition \eqref{corkscrew.condition} and
 \begin{equation*}\label {corkscrew.condition2}
{\rm diam}(\partial \Omega)\ge 1.
\end{equation*}
\end{assump}

For a bounded domain $\Omega$ satisfying  the above assumption, we  find a  nice  covering in the following proposition, see Section \ref{Omegacovering.pr.pfsection} for the proof, which plays a crucial role in our consideration of  sampling and reconstruction of
signals concentrated on the domain $\Omega$.

\begin{prop}\label{Omegacovering.pr}
Let $(X, \rho, \mu)$ be a  $d$-dimensional metric measure space and   $\Omega$  be  the  bounded Corkscrew  domain satisfying Assumption
\ref{domain.assump}.
Then
for any $\delta\in (0, 1)$ there exists  a discrete set $\Omega_\delta\subset \Omega$ such that
\begin{subequations} \label{Omegacovering.pr.eq}
\begin{equation} \label{Omegacovering.pr.eq1}
B(x_i, c\delta)\subset \Omega \ \ {\rm for \ all} \ x_i\in \Omega_\delta,
\end{equation}
\begin{equation}\label{Omegacovering.pr.eq2}
B(x_i, \delta/2) \cap B(x_j, \delta/2)=\emptyset \ \ {\rm for \ all \ distinct }\ x_i, x_j\in \Omega_\delta,
\end{equation}
and
\begin{equation}\label{Omegacovering.pr.eq3}
1\le \sum_{x_i\in \Omega_\delta}\chi_{B(x_i, 5\delta)}(x) \le
\frac{ 11^d D_2(\mu)}{ D_1(\mu)}\ \
  {\rm for \ all} \ x\in \Omega,
\end{equation}
\end{subequations}
where
 $D_1(\mu)$ and $D_2(\mu)$ are  the maximal lower bound and minimal
  upper bound in \eqref{regular.def} respectively, and $c$ is the ratio in the Corkscrew condition \eqref{corkscrew.condition} for the domain $\Omega$.
\end{prop}

Given a discrete set $\Gamma_\Omega\subset \Omega$, we say that
$I_\gamma, \gamma\in \Gamma_\Omega$, is a {\em Voronoi partition} of the domain $\Omega$ if
\begin{subequations}\label{voronoipartition.def}
\begin{equation}\label{weakbupu.defX}  \cup_{\gamma\in \Gamma_\Omega} I_\gamma=\Omega, \  \
I_\gamma \cap I_{\gamma^{\prime}}= \emptyset \ \  {\rm for \ all \ distinct} \
\gamma, \gamma^{\prime} \in \Gamma_\Omega,
\end{equation}
and
\begin{equation}\label{Igamma.assumptionX1}
I_\gamma\subset \big\{x\in \Omega,\  \rho(x, \gamma)=   \rho(x, \Gamma_\Omega)\big\} \ \ {\rm for \ all}\  \gamma\in \Gamma_\Omega.
\end{equation}
\end{subequations}
By  Proposition \ref{Omegacovering.pr}, we have the following unit partition of the Corkscrew domain $\Omega$.

\begin{prop}\label{unitpartition.corprop}
Let $(X, \rho, \mu)$ be a   metric measure space and
$\Omega$  be  a  bounded  Corkscrew domain satisfying Assumption
\ref{domain.assump}.
Then for any $\delta\in (0, 1)$ there exists
a discrete set $\Omega_\delta\subset \Omega$ such that the corresponding
Voronoi partition $I_\gamma, \gamma\in \Omega_\delta$, of the domain $\Omega$ satisfies
\eqref{unitpartition.cor.eq1}, where $c$ is the ratio in the Corkscrew condition \eqref{corkscrew.condition} for the domain $\Omega$.
\end{prop}

\subsection{Sampling and reconstruction of signals in a reproducing kernel space}
\label{rks.subsection}

For  a kernel function $K$ on $X\times X$, we define
its {\em Schur norm}  $\|K\|_{\mathcal S}$  and {\em modulus of continuity} $\omega_\delta(K)$
 by
$$\|K\|_{\mathcal S}
=\max\Big(\sup_{x\in X} \| K(x, \cdot)\|_1,\
\sup_{y\in X} \| K(\cdot, y)\|_1\Big)$$
and
\begin{equation*} \label{modulus.def} \omega_\delta (K)(x,y)=\sup\limits_{\rho(x^\prime, x)\le \delta,\ \rho(y^\prime, y)\le \delta }|K(x^\prime, y^\prime)-K(x, y)|,  \ x, y\in X,
\end{equation*}
respectively. % {\color{blue}``$x, y\in X$ or ``$x,\ y\in X$, I mean the use slash after comma or not}
To consider sampling and reconstruction  of $\varepsilon$-concentrated signals in $V_{p, \Omega, \varepsilon}$, we always assume the following:

\begin{assump}\label{kernel.assumption}  The integral kernel $K$
of the idempotent operator $T$ in \eqref{integraloperator.def}
 has certain off-diagonal decay and H\"older continuity,
\begin{equation}\label{KStheta.def}
\|K\|_{{\mathcal S}, \theta}:=\|K\|_{\mathcal S}+ \sup_{0<\delta\le 1} \delta^{-\theta} \|\omega_\delta(K)\|_{\mathcal S} <\infty
\end{equation}
for some  $0<\theta\le 1$.
\end{assump}
 One may verify that Assumption \ref{kernel.assumption} is met for  kernels $K$
being H\"older continuous,
\begin{equation*}
|K(x, y)-K(x', y')|\le C (\rho(x, x')+\rho(y, y'))^\theta (1+\rho(x, y)+\rho(x', y'))^{-\alpha}
\end{equation*}
for all $x, x', y, y'\in X$,
and having polynomial decay,
\begin{equation*}
|K(x, y)|\le  C (1+\rho(x, y))^{-\alpha}
\end{equation*}
for all $x, y\in X$,
 where  $\alpha>d$ and $C$ is a positive constant.
It is well known that the integral operator with its kernel having finite Schur norm
 is bounded operator on $L^p$. Hence  the range space $V_p$ in \eqref{Vrks} is a closed subspace of $L^p$.
 In the following proposition, we show that it is a reproducing kernel space of $L^p$,
 which is established in \cite{nashedsun2010} for  the  Euclidean space setting, see  Section \ref{RksProperty.pr.pfsection}
 for a sketch of the proof.
  %\cite{sun2016galerkin, jiang2017sampling, xian2014average}.

\begin{prop} \label{RksProperty.pr}
Let $(X, \rho, \mu)$ be a metric measure space, $T$ be an idempotent operator whose kernel $K$ satisfies  Assumption \ref{kernel.assumption},  and $V_p, 1\le p\le \infty$, be the range space of the operator $T$ defined by \eqref{Vrks}.
 Then $V_p$ is a reproducing kernel space of $L^p$, and
 %. Moreover, for any $f \in V_{\Omega,\varepsilon}$ and $\delta_0 >0$,
 \begin{equation}\label{fx.bound}
   \|f\|_q \le  (D_1(\mu))^{-1/p+1/q} \| K\|_{{\mathcal S}, \theta}^{1-p/q} \|f\|_{p}, \  p\le q\le \infty,
 \end{equation}
 where
 $D_1(\mu)$ is  the maximal lower bound  in \eqref{regular.def}
 and  $\|K\|_{{\mathcal S}, \theta}$ is given in \eqref{KStheta.def}.
\end{prop}

To consider sampling and reconstruction of signals $f\in V_p$ concentrated
 on a bounded domain $\Omega$, we
recall the iterative algorithm
  \begin{equation}\label{frameX.algorithm}
f_0=  S_\Gamma f
\ \ {\rm and} \ \  %=\sum_{\gamma \in \Gamma_\Omega}|I_\gamma|f(\gamma)  K(\cdot,\gamma)\\
f_n=f_0 + f_{n-1} - S_{\Gamma}f_{n-1},\   n\ge 1,
%\end{array}\right.
\end{equation}
to reconstruct signals  $f\in V_p$  from their samples
 $f(\gamma), \gamma\in \Gamma$,
 taken on the sampling set $\Gamma\cap \Omega$  in  the domain  $\Omega$
 and the sampling set $\Gamma\cap {\Omega^c}$ outside the domain $\Omega$, where   $\{I_\gamma, \gamma\in \Gamma\cap \Omega\}$ and
$\{I_\gamma, \gamma\in \Gamma\cap \Omega^c\}$ are  Voronoi partitions of the domain $\Omega$
and its complement $\Omega^c$ respectively, and the preconstruction  operator $S_\Gamma$  on $L^p$ is defined by
 \begin{equation}\label{S.Gamma.U}
   S_{\Gamma}g(x)=\sum_{\gamma \in \Gamma} \mu(I_\gamma) (Tg)(\gamma) K(x,\gamma),\ g \in  L^p.
   \end{equation}
The above iterative algorithm \eqref{frameX.algorithm} has been widely used in
reconstructing  signals  in various linear spaces, see for instance
 \cite{aldroubi1998, aldroubi2001SIAM, aldroubiCA2004,sun2016galerkin, feichtinger92, %grochenigfeichtinger1992,
 nashedsun2010, sun2006SIAMJMA}.
In the following proposition, we show that the above algorithm converges exponentially,
see Section \ref{rkssampling.prop.pfsection} for the proof.

\begin{prop}\label{rkssampling.prop}
Let $(X, \rho, \mu)$ be a metric measure space,  $T$ be an idempotent operator whose kernel $K$ satisfies  Assumption \ref{kernel.assumption},  $V_p, 1\le p\le \infty$, be the range space \eqref{Vrks} of the operator $T$, and
$\Omega$ be a bounded domain.
 %{\color{blue} If $\Omega$ is only a bounded domain, can we guarantee that there exists a Voronoi partition of $\Omega$?}
 If   $\Gamma$
 is a sampling set satisfying
\begin{equation*}\label{delta.Gamma.def}
\delta(\Gamma):=\max\Big(\sup_{x\in \Omega } \rho(x, \Gamma\cap \Omega), \sup_{y\in \Omega^c } \rho(y, \Gamma\cap \Omega^c)\Big) < \|K\|_{{\mathcal S}, \theta}^{-2/\theta},
\end{equation*}
then for any $f\in V_p$, the sequence  $f_n, n\ge 0$, in the iterative algorithm  \eqref{frameX.algorithm}
converges to $f$  exponentially,
\begin{equation*}
\|f_n-f\|_p\le  \frac{ 1+ \|K\|_{{\mathcal S}, \theta}^2 (\delta(\Gamma))^\theta}  {1- \|K\|_{{\mathcal S}, \theta}^2 (\delta(\Gamma))^\theta}
\big(\|K\|_{{\mathcal S}, \theta}^2 (\delta(\Gamma))^\theta\big)^{n+1} \|f\|_p.
\end{equation*}
 \end{prop}

\section{Sampling and reconstruction of concentrated signals}
\label{deterministicsampling.section}

Stability of a sampling scheme is an important concept for the robustness and uniqueness for sampling and reconstruction of signals in a linear space, see \cite{aldroubi2001SIAM,  aldroubi2005sun, %aldroubiCA2004,
sun2016galerkin, nashedsun2010, sun2006SIAMJMA, sunxian2014JFAA,  unser2000}.   In this section, we first consider  weighted stability of bi-Lipschitz type for the sampling procedure
   on a sampling set $\Gamma_\Omega\subset\Omega$ for $\varepsilon$-concentrated signals on the domain $\Omega$.

\begin{thm}\label{stability.thm}
Let $1\le p\le \infty$, $\varepsilon\in (0, 1)$, $(X, \rho, \mu)$ be a metric measure space, $T$ be an idempotent integral operator whose kernel $K$ satisfies Assumption \ref{kernel.assumption}, $V_p$ be the range space of the operator $T$ defined by \eqref{Vrks},  $\Omega$ be a bounded domain,
and   $V_{p, \Omega, \varepsilon}$  be the set of $\varepsilon$-concentrated signals given in \eqref{VOmega}.
If
$\Gamma_\Omega\subset \Omega$ is a discrete sampling set
satisfying
\begin{equation}\label{approximation.prop.eq1}
d_H(\Gamma_\Omega, \Omega)< \Big(\frac{1-\varepsilon}{\|K\|_{{\mathcal S}, \theta}}\Big)^{1/\theta},
\end{equation}
then  for all $f, g\in V_{p, \Omega, \varepsilon}$,
 \begin{eqnarray}\label{discrete.pr.pfinite}
& \hskip-0.08in & \hskip-0.08in \big( 1-\varepsilon- \|K\|_{{\mathcal S}, \theta} (d_H(\Gamma_\Omega, \Omega))^\theta\big)\|f-g\|_p
-2\varepsilon \min(\|f\|_p, \|g\|_p) \nonumber\\
\qquad & \hskip-0.08in \le & \hskip-0.08in  \big\|\big(f(\gamma)-g(\gamma)\big)_{\gamma\in \Gamma_\Omega}\big\|_{p, \mu(\Gamma_\Omega)}
%\Big(\sum_{\gamma\in \Gamma_\Omega}|f(\gamma)-g(\gamma)|^p  |I_\gamma|\Big)^{1/p}
\le \big(1+ \|K\|_{{\mathcal S}, \theta} (d_H(\Gamma_\Omega, \Omega))^\theta\big) \|f-g\|_p,
\end{eqnarray}
where  $I_\gamma, \gamma\in \Gamma_\Omega$, is a Voronoi partition of the  domain $\Omega$, and
\begin{equation} \label{stability.thm.pf.eq1.addyaxu}
 \big\|\big(h(\gamma)\big)_{\gamma\in \Gamma_\Omega}\big\|_{p, \mu(\Gamma_\Omega)}=\left\{\begin{array}{ll}
\big(\sum_{\gamma\in \Gamma_\Omega}|h(\gamma)|^p  \mu(I_\gamma)\big)^{1/p}  & {\rm if} \ 1\le p<\infty\\
 \sup_{\gamma\in \Gamma_\Omega} |h(\gamma)| & {\rm if} \ p=\infty.\end{array}\right.
 \end{equation}
\end{thm}

By  \eqref{regular.def} and \eqref{Igamma.assumptionX1}, we have
\begin{equation*}\label{muIgamma.estimate}
\mu(I_\gamma)\le D_2(\mu) (d_H(\Gamma_\Omega, \Omega))^d, \ \gamma\in \Gamma_\Omega,
\end{equation*}
where $D_2(\mu)$ is the minimal upper bound in \eqref{regular.def}.
Therefore we have the  following unweighted inequalities for the sampling scheme $f\longmapsto (f(\gamma))_{\gamma\in\Gamma_\Omega}$, cf. \eqref{samplinginequality.def0}.

\begin{cor} \label{stability.cor} Let the metric measure space $(X, \rho, \mu)$ and the set $V_{p, \Omega, \varepsilon}$ of $\varepsilon$-concentrated signals
 be as in Theorem \ref{stability.thm}.
If $\Gamma_\Omega\subset \Omega$ is a discrete sampling set of the domain $\Omega$
satisfying \eqref{approximation.prop.eq1},
then
 \begin{equation*}\label{discrete.cor.pfinite}
\Big(\sum_{\gamma\in \Gamma_\Omega}|f(\gamma)|^p \Big)^{1/p}
\ge \frac{1-\varepsilon- \|K\|_{{\mathcal S}, \theta} (d_H(\Gamma_\Omega, \Omega))^\theta}
{(D_2(\mu))^{1/p} (d_H(\Gamma_\Omega, \Omega))^{d/p} }  \|f\|_p
\end{equation*}
hold for all $f\in V_{p, \Omega, \varepsilon}, 1\le p <\infty$.
\end{cor}

Given  a sampling set $\Gamma_\Omega$ and  noiseless samples $h(\gamma), \gamma\in \Gamma_\Omega$, of  $h\in V_{p}$,  we define
\begin{equation}\label{f0.def0}
h_I=\sum_{\gamma\in \Gamma_\Omega} h(\gamma) \chi_{I_\gamma},\  h\in V_p, \end{equation}
where $\chi_E$ is the indicator function on a set $E$.
One may verify easily that
\begin{equation}\label{f0.interpolation}
h_I(\gamma)=h(\gamma), \ \gamma\in \Gamma_\Omega,
\end{equation}
and
\begin{equation} \label{stability.thm.pf.eq1}
\|h_I\|_{p, \Omega}=
\big\|(h(\gamma))_{\gamma\in \Gamma_\Omega}\big\|_{p, \mu(\Gamma_\Omega)}.
%\left\{\begin{array}{ll}
% \big(\sum_{\gamma\in \Gamma_\Omega} |h(\gamma)|^p \mu(I_\gamma)\big)^{1/p}  & {\rm if} \ 1\le p<\infty,\\
%\sup_{\gamma\in \Gamma_\Omega}|f(\gamma)| & {\rm if} \ p=\infty. \end{array}\right.
\end{equation}
By %\eqref{approximation.prop.eq1} and
\eqref{stability.thm.pf.eq1} with $h$ replaced by $f-g$,  the proof of Theorem \ref{stability.thm} reduces to showing
\begin{equation} \label{stability.thm.pf.eq2}
\|h_I-h\|_{p, \Omega}\le \|K\|_{{\mathcal S}, \theta} (d_H(\Gamma_\Omega, \Omega))^\theta \|h\|_p, \ \ h\in V_p,
\end{equation}
see Section \ref{stability.thm.pfsection} for the detailed argument.

\medskip

Given  noiseless samples $f(\gamma), \gamma\in \Gamma_\Omega$, of an  $\varepsilon$-concentrated signal $f\in V_{p, \Omega, \varepsilon}\subset V_p$,
the signal $f_I$ in \eqref{f0.def0} coincides with the original signal $f$ on the sampling set $\Gamma_\Omega$
by \eqref{f0.interpolation}. However
the signal $f_I$ does not reside in the reproducing kernel space $V_p$ and it does not provide an
 approximation
to the original  signal $f$, except that  the Hausdorff
distance  $d_H(\Gamma_\Omega, \Omega)$ between $\Gamma_\Omega$ and $\Omega$ is small,  since
in that case
\begin{equation*}
\|f_I-f\|_p\le \|f_I-f\|_{p, \Omega}+\|f\|_{p, \Omega^c}\le  \big(\|K\|_{{\mathcal S}, \theta} (d_H(\Gamma_\Omega, \Omega))^\theta+\varepsilon\big)  \|f\|_p
\end{equation*}
by   \eqref{VOmega} and \eqref{stability.thm.pf.eq2}.

Based on the iterative algorithm \eqref{frameX.algorithm}, for any $f\in V_p$ we define
\begin{subequations}\label{framealgorithm}
\begin{equation}\label{frame.algorithm.eq0}
g_0=\sum_{\gamma \in \Gamma_\Omega}\mu(I_\gamma) f(\gamma)  K(\cdot,\gamma)\in V_p,\end{equation}
and  $g_n\in V_p, n\ge 1$, inductively by
\begin{equation}\label{frame.algorithm.eq1}
g_n=g_0 + g_{n-1} - S_{\Gamma}g_{n-1},\   n\ge 1,
\end{equation}
\end{subequations}
where the preconstruction  operator $S_\Gamma$  on $L^p$ is given in
\eqref{S.Gamma.U} with
$\Gamma=\Gamma_\Omega\cup \Gamma_{\Omega^c}$,
$\Gamma_{\Omega^c}$
is a discrete sampling set of the complement $\Omega^c$
of the domain $\Omega$ %{\color{blue} remind you to notice `` the complement $\Omega^c$''}
satisfying
\begin{equation}\label{approximation.thm.eq4}
d_H(\Gamma_{\Omega^c}, \Omega^c)\le \min\big( \varepsilon^{1/\theta}\|K\|_{{\mathcal S}, \theta}^{-1/\theta}, (2\|K\|_{{\mathcal S}, \theta}^2)^{-1/\theta}\big),
\end{equation}
%{\color{blue} I think (3.12) is $d_H(\Gamma_{\Omega^c}, \Omega^c)\le \min\big( \varepsilon^{1/\theta}\|K\|_{{\mathcal S}, \theta}^{-1/\theta},(2\|K\|_{{\mathcal S}, \theta}^2)^{-1/\theta}\big)$}
and  $\{I_\gamma, \gamma\in \Gamma_\Omega\}$ and $\{I_\gamma, \gamma\in \Gamma_{\Omega^c}\}$ are  Voronoi partitions of the domain $\Omega$ and its complement $\Omega^c$ respectively.
In the following theorem, we show that $g_n, n\ge 1$,
reconstructed from  samples $f(\gamma), \gamma\in \Gamma_\Omega$, inside the domain $\Omega$   provide good approximations
to the original $\varepsilon$-concentrated signal $f\in V_{p, \Omega, \varepsilon}$, see Section \ref{approximation.thm.pfsection}
for the detailed argument.

\begin{thm}\label{approximation.thm}
Let $1\le p\le \infty$, $\varepsilon\in (0, 1)$, $(X, \rho, \mu)$ be a metric measure space, $T$ be an idempotent operator whose kernel $K$ satisfies Assumption \ref{kernel.assumption},  $V_p$ be the range space of the operator $T$ defined by \eqref{Vrks},
$\Omega$ be a bounded  Corkscrew domain  satisfying
Assumption \ref{domain.assump},
%with its complement $\Omega^c$ satisfying the Corkscrew condition \eqref{corkscrew.condition},
$\Gamma_\Omega$ be a discrete sampling set of the domain $\Omega$,
and  $V_{p, \Omega, \varepsilon}$ be the set of $\varepsilon$-concentrated signals given in \eqref{VOmega}.
If the Hausdorff distance between the sampling set $\Gamma_\Omega$ and  the domain $\Omega$ satisfies
\begin{equation}\label{approximation.thm.eq1}
d_H(\Gamma_\Omega, \Omega)<\|K\|_{{\mathcal S}, \theta}^{-2/\theta},
\end{equation}
then for any $\varepsilon$-concentrated signal $f\in V_{p, \Omega, \varepsilon}$ and
\begin{equation}  \label{approximation.thm.eq2}
n+1\ge \frac{\ln (1/\varepsilon) -\ln \|K\|_{{\mathcal S}, \theta} } {\theta \ln (1/d_{H}(\Gamma_\Omega, \Omega))-2 \ln \|K\|_{{\mathcal S}, \theta}},
\end{equation}
 the reconstructed  signals  $g_n, n\ge 1$, in \eqref{framealgorithm}
 are
$(9C_0\varepsilon)$-concentrated signals in $V_p$,
 \begin{equation}\label{approximation.thm.eq3a}
g_n\in V_{p, \Omega,  9C_0\varepsilon},
\end{equation}
and they  provide good approximations to the original  signal  $f$,
\begin{equation}\label{approximation.thm.eq3}
\|g_n-f\|_p  \le  4C_0 \varepsilon \|f\|_p, %  \|K\|_{{\mathcal S}, \theta}}  {1- \|K\|_{{\mathcal S}, \theta}^2 (d_{H}(\Gamma_\Omega, \Omega))^\theta}  \varepsilon \|f\|_p,
\end{equation}
where
\begin{equation}\label{C0.def0}
C_0=\frac{\|K\|_{{\mathcal S}, \theta}}{1- \|K\|_{{\mathcal S}, \theta}^2 (d_{H}(\Gamma_\Omega, \Omega))^\theta}.\end{equation}
 \end{thm}

 The  reconstructed  signals  $g_n, n\ge 1$, in \eqref{framealgorithm}
 do not interpolate the sampling data  $f(\gamma), \gamma\in \Gamma_\Omega$, however they have small sampling difference to the original signal $f$ on the sampling set $\Gamma_\Omega$, since
 it follows from  \eqref{stability.thm.pf.eq2} and \eqref{approximation.thm.eq3}
 that
  \begin{eqnarray}\label{approximation.prop.eq4a}
\big\|\big(g_n(\gamma)-f(\gamma)\big)_{\gamma\in \Gamma_\Omega}\big\|_{p, \mu(\Gamma_\Omega)}
% \Big(\sum_{\gamma\in \Gamma_\Omega} | g_n(\gamma)-f(\gamma)|^p |I_\gamma|\Big)^{1/p}
& \hskip-0.08in \le & \hskip-0.08in    \big(1+ \|K\|_{{\mathcal S}, \theta} (d_H(\Gamma_\Omega, \Omega))^\theta\big) \|g_n-f\|_p\nonumber\\
 \qquad \quad& \hskip-0.08in \le & \hskip-0.08in
  4  C_0\big(1+ \|K\|_{{\mathcal S}, \theta} (d_H(\Gamma_\Omega, \Omega))^\theta\big)
    %{1- \|K\|_{{\mathcal S}, \theta}^2 (d_{H}(\Gamma_\Omega, \Omega))^\theta}
    \varepsilon \|f\|_p  %, 1\le p\le \infty.
\end{eqnarray}
for all $f\in V_{p, \Omega, \varepsilon}, 1\le  p\le \infty$. %, see Figure \ref{reconstructionerrordeterminisiticsampling}.
% and

\begin{rem}\label{stability.rem}
Take the hat function $h(x)=\max(1-|x|, 0)$, the  concentration domain $\Omega_R=[-R, R]$ for some integer $R\ge 2$, and signals
$f_\pm(x)= h(x)\pm \delta h(x-R-1), \delta \in (0, 1)$, in the shift-invariant space
 $$V_p(h)=\Big\{\sum_{k\in \mathbb Z} c(k) h(x-k), \ (c(k))_{k\in {\mathbb Z}}\in \ell^p\Big\}, \ 1\le p\le \infty,$$
 generated by the integer shifts of the hat function $h$. One may verify that the shift-invariant space $V_p(h)$
 is the range space of some idempotent integral operator with kernel satisfying
 Assumption \ref{kernel.assumption},
 and
 $f_\pm$ are $\varepsilon$-concentrated signals onto $\Omega_R$ with $\varepsilon= \delta$ for $p=\infty$ and
 $\varepsilon=\delta (1+\delta^p)^{-1/p}$ for $1\le p<\infty$, since
 %$\|f_+\|_p=\|f_-\|_p$,
$ \|f_\pm\|_{p, {\mathbb R}\backslash \Omega_R}=\delta \|h\|_p$ and
$\|f_\pm\|_p= \delta \|h\|_p/\varepsilon$.
As signals $f_\pm$ coincide on the domain $\Omega_R$, the signals
$g_{n, \pm}$ constructed in \eqref{framealgorithm} from their samples inside the domain $\Omega_R$ are the same, which implies that
$$\max(\|g_{n, +}-f_+\|_p, \|g_{n, -}-f_-\|_p)\ge \frac{1}{2} \|f_+-f_-\|_p=\delta \|h\|_{p}
=\varepsilon \|f_\pm\|_p.
$$
Therefore the error estimate in \eqref{approximation.thm.eq3}
 is suboptimal in the sense that the constant
  $4 C_0$ cannot be replaced by
  a positive constant strictly less than one in general. 
\end{rem}

Reconstructing a signal from noisy data and estimating the reconstruction error are important problems in sampling theory \cite{adcock2014, aldroubi2001SIAM, aldroubiIeee2008, aldroubiCA2004, nashedsun2010, sun2006SIAMJMA, unser2000}.
%where $I_\gamma, \gamma\in \Gamma_\Omega$ is a Voronoi partition of the domain $\Omega$.
In this paper, we propose the following algorithm  $\tilde g_n, n\ge 0$,  for signal reconstruction   when  samples  $f(\gamma), \gamma\in \Gamma_\Omega$,
of  $f\in V_p$ are corrupted by some  deterministic noise  $\pmb\xi=(\xi(\gamma))_{\gamma\in \Gamma_\Omega}$:
\begin{subequations}\label{noiseframealgorithm}
\begin{equation}\label{noiseframe.algorithm.eq1}
\tilde g_n=\tilde g_0 + \tilde g_{n-1} - S_{\Gamma}\tilde g_{n-1},\   n\ge 1,
\end{equation}
where
\begin{equation}\label{noiseframe.algorithm.eq0}
\tilde g_0=\sum_{\gamma \in \Gamma_\Omega}\mu(I_\gamma)(f(\gamma)+\xi(\gamma))  K(\cdot,\gamma)\in V_p,\end{equation}
\end{subequations}
and the preconstruction  operator $S_\Gamma$  on $L^p$ is given in
\eqref{S.Gamma.U} with
$\Gamma=\Gamma_\Omega\cup \Gamma_{\Omega^c}$,
and
$\Gamma_{\Omega^c}$
is a discrete sampling set of the complement $\Omega^c$
satisfying
\eqref{approximation.thm.eq4}.
In the following theorem, we show that the reconstructed signals $\tilde g_n$ with large $n$ provide good approximations
 to the original $\varepsilon$-concentrated signal $f$,  see Section \ref{deterministicnoise.thm.pfsection} for the proof.

\begin{thm}\label{deterministicnoise.thm}
Let $1\le p\le \infty$, $\varepsilon\in (0, 1)$, $(X, \rho, \mu)$ be a metric measure space, $T$ be an idempotent operator whose kernel $K$ satisfies Assumption \ref{kernel.assumption},  $V_p$ be the range space of the operator $T$ defined by \eqref{Vrks},
 $\Omega$ be a bounded  Corkscrew domain  satisfying
Assumption \ref{domain.assump}, $\Gamma_\Omega$ be a discrete sampling set of the domain $\Omega$ satisfying \eqref{approximation.thm.eq1},
$V_{p, \Omega, \varepsilon}$ be the set of $\varepsilon$-concentrated signals given in  \eqref{VOmega}, and $\pmb\xi=(\xi(\gamma))_{\gamma\in \Gamma_\Omega}$
be a deterministic noise vector with $\|\pmb \xi\|_{p, \mu(\Gamma_\Omega)}<\infty$.
Then for any $\varepsilon$-concentrated signal $f\in V_{p, \Omega, \varepsilon}$,
    signals  $\tilde g_n$ in \eqref{noiseframealgorithm}  with $n$ satisfying \eqref{approximation.thm.eq2}
  provide  approximations to the original  signal  $f$,
\begin{equation}\label{deterministicnoise.thm.eq3}
\|\tilde g_n-f\|_p  \le  4 C_0\varepsilon \|f\|_p +
 C_0\|\pmb\xi\|_{p, \mu(\Gamma_\Omega)},
\end{equation}
 where $C_0$ is given in \eqref{C0.def0}
and the norm $\|\cdot\|_{p, \mu(\Gamma_\Omega)}$ is defined by
\eqref{stability.thm.pf.eq1.addyaxu}. %{stability.thm.pf.eq1}.
\end{thm}

\section{Random sampling and reconstruction of concentrated signals}
\label{randomsampling.section}

In this section, we consider sampling $\varepsilon$-concentrated signals  in $V_{p, \Omega, \varepsilon}$
at i.i.d. random positions drawn on  $\Omega$, and reconstructing the original $\varepsilon$-concentrated signals  in $V_{p, \Omega, \varepsilon}$ from their samples taken on these random positions.
 We establish  a weighted stability inequality of bi-Lipschitz type for the random sampling  procedure in
Theorem \ref{randomstability.thm}.
In  Theorem \ref{maintheorem.tm} and Corollary \ref{maintheorem.cor}, we show that, with high probability,  signals  concentrated on
  a bounded Corkscrew domain  $\Omega$ can be reconstructed approximately from their
 samples taken at
 i.i.d. random
positions drawn on  $\Omega$, provided that the sampling size is at least  of the order $\mu(\Omega) \ln (\mu(\Omega))$.
Finally in Theorem \ref{random.thm}
 we prove that with high probability,  an original  $\varepsilon$-concentrated signal  can be reconstructed approximately
  from its random samples corrupted by i.i.d. random noises,
  when the random sampling size is large enough.

\begin{thm}\label{randomstability.thm}
Let  $(X, \rho, \mu)$ be a metric measure space,   $V_p, 1\le p\le \infty$, be the range space of an idempotent integral operator $T$ whose kernel $K$ satisfies Assumption \ref{kernel.assumption},
 $\Omega$ be a bounded  Corkscrew domain  satisfying
Assumption \ref{domain.assump},
and let  $V_{p, \Omega,\varepsilon}, \varepsilon\in (0, 1)$, be the set of $\varepsilon$-concentrated  signals given in \eqref{VOmega}.
If  $\{\gamma, \gamma \in \Gamma_\Omega\}$ are i.i.d. random positions drawn on $\Omega$ with respect to the probability measure $(\mu(\Omega))^{-1} d\mu$, then for any  $\tilde \varepsilon\in (0, 1-\varepsilon)$,
%and   $A\le D_2(\mu) (\tilde \varepsilon/\|K\|_{{\mathcal S}, \theta})^{d/\theta}$,
 the following weighted stability inequalities of bi-Lipschitz type
 \begin{eqnarray}\label{randomdiscrete.cor.pfinite.addyaxu}
 & \hskip-0.08in & \hskip-0.08in \big( 1-\varepsilon-\tilde \varepsilon\big)\|f-g\|_p
-2\varepsilon \min(\|f\|_p, \|g\|_p) \nonumber\\
\qquad & \hskip-0.08in \le & \hskip-0.08in
\big\|\big(f(\gamma)-g(\gamma)\big)_{\gamma\in \Gamma_\Omega}\big\|_{p, \mu(\Gamma_\Omega)}
\le \big(1+ \tilde \varepsilon\big) \|f-g\|_p, \ f, g\in V_{p,\Omega,  \varepsilon}
\end{eqnarray}
%for  $f, g\in V_{\infty, \varepsilon, \Omega}$,
hold  with  probability  at  least
$$1-
\frac{ 10^d   {\mu(\Omega)} }{c^d D_1(\mu) (\tilde\varepsilon/\|K\|_{{\mathcal S}, \theta})^{d/\theta}}
 \Big(1-\frac{c^d D_1(\mu)   (\tilde \varepsilon/\|K\|_{{\mathcal S}, \theta})^{d/\theta} } { 10^d {\mu(\Omega)}}\Big)^N,
 $$
where %$I_\gamma, \gamma\in \Gamma_\Omega$, is a Voronoi partition of the  domain $\Omega$ and
 $N$ is the size  of the sampling set $\Gamma_\Omega$.
\end{thm}

By Theorem \ref{stability.thm}, the proof of Theorem \ref{randomstability.thm} reduces to
  the following crucial  estimate on the probability on
the Hausdorff distance  $d_H(\Gamma_\Omega, \Omega)>\delta_1$
where $0<\delta_1<1$, see Section \ref{A.Gamma.Omega.Gap.proofsection} for the proof.

\begin{prop}\label{A.Gamma.Omega.Gap}
Let $(X, \rho, \mu)$ be a  $d$-dimensional metric measure space and
$\Omega$ be a bounded  Corkscrew domain  satisfying
Assumption \ref{domain.assump}. Suppose that $\{\gamma, \gamma \in \Gamma_\Omega\}$ are i.i.d. random positions drawn on $\Omega$ with respect to the probability measure $(\mu(\Omega))^{-1} d\mu$.
%{\color{blue}``${\mu(\Omega)}^{-1} d\mu$'' or ``${|\mu(\Omega)}|^{-1} d\mu$''?}
Then for $0<\delta_1\le 1$,
%        Then $A_{\Gamma_{\Omega}}(\delta_1)\ge 1$ holds with probability  with  least $1-\tau$.
        \begin{equation} \label{A.Gamma.Omega.Gap.eq00}
        \mathbb{P}\big \{ d_H(\Gamma_\Omega, \Omega)> \delta_1
        %\inf_{x\in \Omega} \sum_{\gamma\in \Gamma_\Omega} \chi_{B(x, \delta_1)}(\gamma)< 1
        \big\}
 \le  \frac{ 10^d {\mu(\Omega)}}{c^d D_1(\mu) \delta_1^d}
 \Big(1-\frac{ c^d D_1(\mu)   \delta_1^d}{10^d{\mu(\Omega)}}\Big)^N.
 \end{equation}
\end{prop}

\begin{rem}\label{randomsamplinginequality.rem1}
Applying  Corollary \ref{stability.cor} and  Proposition \ref{A.Gamma.Omega.Gap} with
$\delta_1$ replaced by $(2\|K\|_{{\mathcal S}, \theta})^{-1/\theta}$, we obtain that the following sampling inequalities
%{\color{blue}In (4.3), there is only one inequality}
 \begin{equation}\label{randomdiscrete.cor.pfinite}
\Big(\sum_{\gamma\in \Gamma_\Omega}|f(\gamma)|^p \Big)^{1/p}
\ge (1/2-\varepsilon)(D_2(\mu))^{-1/p} (2\|K\|_{{\mathcal S}, \theta})^{d/(p\theta)}  \|f\|_p, \ f\in V_{p, \Omega, \varepsilon}
\end{equation}
hold  %{\color{blue}``hold'' or ``holds''?}
with probability  at least
$$1-\frac{ 10^d (2\|K\|_{{\mathcal S}, \theta})^{d/\theta} {\mu(\Omega)}}{c^d D_1(\mu) }
 \Big(1-\frac{ c^d D_1(\mu)   }{ 10^{d}(2\|K\|_{{\mathcal S}, \theta})^{d/\theta} {\mu(\Omega)}}\Big)^N,$$ where $\varepsilon\in (0, 1/2)$ and $1\le p<\infty$.
We remark that the above sampling inequalities \eqref{randomdiscrete.cor.pfinite}
for random sampling of $\varepsilon$-concentrated signals in $V_{p, \Omega, \varepsilon}$ can be considered as a weak version of the corresponding sampling inequalities  for bandlimited/wavelet signals concentrated on  $[-R/2, R/2]^d$ in \cite{bass2013random, bass2010random, %bass2005random,
 xian2019random, jiang2020, lu2020, patel2019, yang2013}. % {\color{red}  Yaxu: please add necessary reference here}. {\color{blue}Dr. Sun, \cite{ bass2013random, bass2010random, xian2019random, lu2020} These three papers are typical papers due to they present the original proof techniques. The methods used in other papers, such as, \cite{jiang2020, lu2020, patel2019, yang2013}, are all from them.}
\end{rem}

\begin{rem}\label{randomsamplinginequality.rem2}  Let $\tau\in (0, 1/2], 1\le p<\infty$, and
\begin{equation}\label{N.def0-1}
N\ge  N_0( {\mu(\Omega)}, \tau):= \frac{ 5^{d} 2^{d+1+d/\theta}\|K\|_{{\mathcal S}, \theta}^{d/\theta}{\mu(\Omega)}} { c^d D_1(\mu)}\ln
\Big(\frac{ 10^d (2 \|K\|_{{\mathcal S}, \theta})^{d/\theta} {\mu(\Omega)}}{c^d D_1(\mu)\tau}\Big).
%\frac{ 20^d  \|K\|_{{\mathcal S}, \theta}^{d/\theta} {\mu(\Omega)}}{c^d D_1(\mu)
%\tau \varepsilon^{d/\theta}}.
 \end{equation}
Applying %\eqref {muIgamma.estimate}, %Theorem \ref{stability.thm}
Corollary \ref{stability.cor}
 and  Proposition \ref{A.Gamma.Omega.Gap}
 %{\color{blue}(3.4) is not necessary, just Applying Corollary \ref{stability.cor}
 %and  Proposition \ref{A.Gamma.Omega.Gap} }
  with
$$\delta_1=\Big(\frac{10^d}{c^d D_1(\mu)} \frac{{\mu(\Omega)}}{N} \ln \Big(\frac{ N}{ \tau}\Big)\Big)^{1/d},$$
we conclude that the following sampling inequalities  %{\color{blue}In (4.5), there is only one inequality}
 \begin{equation}\label{randomdiscrete.cor.pfinite2}
\sum_{\gamma\in \Gamma_\Omega}|f(\gamma)|^p
\ge \frac{ (1/2-\varepsilon)^p c^d D_1(\mu)}{ 10^d D_2(\mu)} \Big(  \ln \frac{ N}{\tau}\Big)^{-1}
\frac{N}{{\mu(\Omega)}}
  \|f\|_p^p, \ f\in V_{p, \Omega, \varepsilon}
\end{equation}
hold with probability  at least  $1-\tau$.
We remark that the sampling inequalities \eqref{randomdiscrete.cor.pfinite2}
for random sampling of $\varepsilon$-concentrated signals in $V_{p, \Omega, \varepsilon}$ can be considered as a weak version
 of the corresponding sampling inequalities  for bandlimited/wavelet signals concentrated on  $[-R/2, R/2]^d$ in \cite{ bass2013random, bass2010random, xian2019random, lu2020}, where the lower bound in \eqref{randomdiscrete.cor.pfinite2}
 is replaced by a  multiple of  $N\|f\|_p^p/{\mu(\Omega)}$. % \cite{bass2013random,  ???} {\color{red} Yaxu: please add necessary reference here}, {\color{blue} see above in the same page in blue color}
 \end{rem}

To the best
of our knowledge, there is no algorithm available to find good approximations to   $\varepsilon$-concentrated signals
 from their random samples inside the domain $\Omega$.
 %{\color{red} Yaxu: please add necessary reference here}  {\color{blue} \cite{jiang2020} consider the reconstruction problem. Might have some problem about their paper.}
  By  Theorem \ref{approximation.thm} and Proposition \ref{A.Gamma.Omega.Gap} with   $\delta_1$ replaced by $(2 \|K\|_{{\mathcal S}, \theta}^2)^{-1/\theta}$,  such approximations are constructed explicitly.

\begin{thm} \label{maintheorem.tm}
Let the metric measure space
$(X, \rho, \mu)$, the domain $\Omega$,  the
 set $V_{p, \Omega,\varepsilon}$ of $\varepsilon$-concentrated signals, and the sequence $g_n\in V_p, n\ge 0$,
 be as in Theorem \ref{approximation.thm}.
%and  let   $I_\gamma, \gamma\in \Gamma_{\Omega^c}$ be a  Voronoi partition of the complement $\Omega^c$.
  Suppose that   $\{\gamma, \gamma \in \Gamma_\Omega\}$ are i.i.d. random  positions drawn on $\Omega$ with respect to probability measure $(\mu(\Omega))^{-1} d\mu$,  and denote the size of $\Gamma_\Omega$ by $N$. %set $N=\#\Gamma_\Omega$.
Then  for
\begin{equation}\label{maintheorem.tm.eq1+}
  n+1\ge
\frac{\ln (1/\varepsilon) -\ln \|K\|_{{\mathcal S}, \theta} } { \ln 2},
\end{equation}
the following reconstruction error estimates
%between the reconstruction signal $g_n$ in  \eqref{frame.algorithm.eq0} and \eqref{frame.algorithm.eq1} and the original $\varepsilon$-concentrated signal $f\in V_{p, \Omega, \varepsilon}$,
\begin{equation}\label{maintheorem.tm.eq2.addyaxu}
\|g_n-f\|_p  \le 8\|K\|_{\mathcal S,\theta}\varepsilon\|f\|_p,  \ f\in V_{p, \Omega, \varepsilon},
\end{equation}
%and
%\begin{equation}\label{hatf.def2}
%\Big\|\sum_{\gamma \in \Gamma_\Omega}|I_\gamma|f(\gamma)  K(\cdot,\gamma)-f\Big\|_\infty  \le   2 (D_1(\mu))^{-1/p} \|K\|_{{\mathcal S}, \theta}^2 \varepsilon \|f\|_p
%\end{equation}
hold with probability  at  least
\begin{equation}  \label{maintheorem.tm.eq3}
1- \tau({\mu(\Omega)}, N):=
1-\frac{ 10^d (2 \|K\|_{{\mathcal S}, \theta}^{2})^{d/\theta} {\mu(\Omega)}}{c^d D_1(\mu)}
 \Big(1-\frac{ c^d D_1(\mu) }{ 10^{d}(2\|K\|_{{\mathcal S}, \theta}^2)^{d/\theta}{\mu(\Omega)}}\Big)^N.\end{equation}
 \end{thm}

For any $0<\tau<1$, one may verify that
$$\tau( {\mu(\Omega)}, N)\le \tau$$
when
\begin{equation}\label{N.def00}
N\ge  N_1( {\mu(\Omega)}, \tau):= \frac{ 10^d (2 \|K\|_{{\mathcal S}, \theta}^{2})^{d/\theta} {\mu(\Omega)}}{c^d D_1(\mu)}\ln
\frac{ 10^d (2 \|K\|_{{\mathcal S}, \theta}^{2})^{d/\theta} {\mu(\Omega)}}{c^d D_1(\mu)\tau}.
 \end{equation} 
 Therefore by  Proposition \ref{RksProperty.pr} and Theorem \ref{maintheorem.tm}, we have the following corollary.

 \begin{cor}\label{maintheorem.cor}
 Let
$\varepsilon, \tau \in (0, 1)$, and let
 the metric measure space
$(X, \rho, \mu)$, the domain    $\Omega$, the
 set $V_{p, \Omega,\varepsilon}$ of $\varepsilon$-concentrated signals,  the random sampling set $\Gamma_\Omega$, and the reconstructed signals $g_n, n\ge 0$, be as in Theorem \ref{maintheorem.tm}. If
the size $N$  of the random sampling set $\Gamma_\Omega$ satisfies
\eqref{N.def00},
then for any
integer $n$ satisfying \eqref{maintheorem.tm.eq1+} and $p\le q\le \infty$,
%the following reconstruction error estimate between the reconstruction signal $g_n$ and the original $\epsilon$-concentrated signal $f$,
\begin{equation}\label{maintheorem.tm.eq2}
\|g_n-f\|_q  \le 8 (D_1(\mu))^{-1/p+1/q} \| K\|_{{\mathcal S}, \theta}^{2-p/q}\varepsilon\|f\|_p, \ f\in V_{p, \Omega, \varepsilon},
\end{equation}
hold  with probability  at  least $1-\tau$.
 \end{cor}

Next,  we consider signal reconstruction  when  random samples  $f(\gamma), \gamma\in \Gamma_\Omega$, of a signal $f\in V_p$
are corrupted by some bounded   noise $\pmb\xi=(\xi(\gamma))_{\gamma \in \Gamma_\Omega}$,
\begin{equation}\label{noisydata.eq000.addyaxu}
 \tilde{f}_\gamma=f(\gamma)+\xi(\gamma),\ {\gamma \in \Gamma_\Omega}.
\end{equation}
Following the argument used in the proofs of Theorem \ref{deterministicnoise.thm} and Corollary
\ref{maintheorem.cor}, we have the following result when random samples are corrupted by  bounded deterministic noises.

\begin{cor}\label{noisymaintheorem.cor}
Let $\varepsilon, \tau \in (0, 1)$,  and let the metric measure space
$(X, \rho, \mu)$, the domain    $\Omega$, the set
  $V_{p, \Omega, \varepsilon}$ of $\varepsilon$-concentrated signals and the random sampling set $\Gamma_\Omega$ be as in Theorem \ref{maintheorem.tm},
   $\pmb\xi=(\xi(\gamma))_{\gamma \in \Gamma_\Omega}$
  be bounded noise vector with bound $\|\pmb \xi\|_\infty=\sup_{\gamma\in \Gamma_\Omega}|\xi(\gamma)|$,  and   the reconstructed signals $\tilde g_n, n\ge 0$, be as in Theorem \ref{deterministicnoise.thm}. If
the size $N$  of the random sampling set $\Gamma_\Omega$ satisfies
\eqref{N.def00},
then for any
integer $n$ satisfying \eqref{maintheorem.tm.eq1+},
%the following reconstruction error estimate between the reconstruction signal $g_n$ and the original $\epsilon$-concentrated signal $f$,
\begin{equation}\label{noisymaintheorem.tm.eq1}
\|\tilde g_n-f\|_{\infty}  \le 8 (D_1(\mu))^{-1/p} \| K\|_{{\mathcal S}, \theta}^{2}\varepsilon\|f\|_p+
2 \|K\|_{{\mathcal S}, \theta}
 \|\pmb\xi\|_{\infty}, \  f\in V_{p, \Omega, \varepsilon}
\end{equation}
%and
%\begin{equation}\label{hatf.def2}
%\Big\|\sum_{\gamma \in \Gamma_\Omega}|I_\gamma|f(\gamma)  K(\cdot,\gamma)-f\Big\|_\infty  \le   2 (D_1(\mu))^{-1/p} \|K\|_{{\mathcal S}, \theta}^2 \varepsilon \|f\|_p
%\end{equation}
hold  with probability  at  least $1-\tau$.
\end{cor}

\begin{rem} \label{randomnoise.rem0} Let $0\ne  h_0\in V_{p, \Omega, \varepsilon_0}$ satisfy
\begin{equation}\label{optimal.eqyaxuadd}
32 (D_1(\mu))^{-1/p} \| K\|_{{\mathcal S}, \theta}^{2}\|h_0\|_p \varepsilon_0\le \|h_0\|_\infty.
\end{equation}
Such a signal exists for sufficiently small $\varepsilon_0$ when $V_p$ is the shift-invariant space generated by the integer shifts of the hat function and $\Omega=[-R/2, R/2]$, $R\ge 2$, see Remark \ref{stability.rem}.
Take $x\in \Omega$ with $|h_0(x)|\ge \|h_0\|_\infty/2$ and let
 $\tilde g_n, n\ge 0$, in Theorem \ref{deterministicnoise.thm}
 reconstructed from noisy sampling data
 \eqref{noisydata.eq000.addyaxu} with $f=0$ and  $\xi(\gamma)=h_0(\gamma), \gamma\in \Gamma_\Omega$. Then we obtain from Corollary  \ref{maintheorem.cor}
that
\begin{equation}\label{optimal.eq11yaxuadd}
|\tilde g_n(x)-h_0(x)|\le 8 (D_1(\mu))^{-1/p} \| K\|_{{\mathcal S}, \theta}^{2}\varepsilon_0\|h_0\|_p
\end{equation}
hold with probability  at  least $1-\tau$ for large $n$.
Therefore
for large $n$,
\begin{equation*} \label{optimal.eq11a}
|\tilde g_n(x)-f(x)|=|\tilde g_n(x)|\ge  \|\pmb \xi\|_\infty/4
\end{equation*}
hold with probability  at  least $1-\tau$, since
$\|\pmb \xi\|_\infty\le \|h_0\|_\infty$ by the definition of the noise vector $\pmb \xi$, and
$$|\tilde g_n(x)|\ge |h_0(x)|-8 (D_1(\mu))^{-1/p} \| K\|_{{\mathcal S}, \theta}^{2}\varepsilon_0\|h_0\|_p\ge \|h_0\|_\infty/4$$
 by \eqref{optimal.eqyaxuadd} and \eqref{optimal.eq11yaxuadd}.
This demonstrates that the error estimate in \eqref{noisymaintheorem.tm.eq1} is suboptimal in the sense that the second part of the bound estimate $2 \|K\|_{{\mathcal S}, \theta}  \|\pmb\xi\|_{\infty}$ cannot be replaced by $A\|\pmb\xi\|_{\infty}$ for some small constant $A$. 
\end{rem}

By Remark \ref{randomnoise.rem0}, the term $2 \|K\|_{{\mathcal S}, \theta}
 \|\pmb\xi\|_{\infty}$  related to the noise vector ${\pmb \xi}$ % $\|\pmb \xi\|_\infty$
can not be ignored in the error estimate  \eqref{noisymaintheorem.tm.eq1} of Corollary \ref{noisymaintheorem.cor}, no matter how large  the sampling size $N$ is. In the following theorem, we show that scenario will be {\em completely different} if
the noise vector $\pmb \xi$ has its components being i.i.d. random variables,
see \cite{ aldroubi2018dynamical, aldroubiIeee2008, chen2015randomnoise,  % chen2014randomnoise,
 duffy2019capacity} and references therein for reconstruction of signals in various linear spaces from their samples corrupted by random noises. 
 
\begin{thm}\label{random.thm}
Let the metric measure space
$(X, \rho, \mu)$, the domain $\Omega$,  the
 set $V_{p, \Omega,\varepsilon}$ of $\varepsilon$-concentrated signals,
 the random sampling set $\Gamma_\Omega$, and the  sequence $\tilde g_n, n\ge 0$
 be as in Theorem \ref{maintheorem.tm}.
%and  let   $I_\gamma, \gamma\in \Gamma_{\Omega^c}$ be a  Voronoi partition of the complement $\Omega^c$.
  Suppose that   $\tau\in (0, 1/2)$ and $\xi(\gamma), {\gamma \in \Gamma_\Omega}$,  are  i.i.d. random variables with  mean zero and  variance $\sigma^2$,
\begin{equation}\label{random.thm.eq1}
 {\mathbb E}(\xi(\gamma))=0, \ \ {\rm Var}(\xi(\gamma))=\sigma^2, \ \gamma\in \Gamma_\Omega.
\end{equation}
Let $f\in V_{p,  \Omega, \varepsilon}$ and set  \begin{equation}\label{random.thm.pfeq0}
\tilde \delta_1=\min \left( (2 \|K\|_{{\mathcal S}, \theta}^2)^{-1/\theta}, %\Big(\frac{ \varepsilon}{\|K\|_{{\mathcal S}, \theta}}\Big)^{1/\theta},
\Big(
\frac{ \tau \varepsilon^2   \sigma^{-2} \|f\|_p^2} {  D_2(\mu) (D_1(\mu))^{2/p-1}}
          \Big)^{1/d}\right ).
\end{equation}
If the size  $N$ of the random sampling set $\Gamma_\Omega$ satisfies
\begin{equation} \label{random.thm.pfeq1}
N\ge  \frac{ 10^d {\mu(\Omega)}} { c^d D_1(\mu) \tilde \delta_1^{d}} \ln \frac{ 10^d  {\mu(\Omega)}}{c^d D_1(\mu)
 \tau \tilde \delta_1^{d}  },
\end{equation}
 then for any  integer  $n$ with
  \begin{equation}\label{noiseinteger} n+1\ge \frac{\ln (1/\varepsilon) -\ln \|K\|_{{\mathcal S}, \theta} } {\ln 2 },\end{equation}
the approximation error estimates
\begin{equation}
\|g_n-f\|_{\infty}\le 10 (D_1(\mu))^{-1/p} \| K\|_{{\mathcal S}, \theta}^{2}\varepsilon\|f\|_p \end{equation}
hold  with probability  at  least $1-2\tau$,
where
 $D_1(\mu)$ and $D_2(\mu)$ are  the maximal lower bound and minimal
  upper bound in \eqref{regular.def} respectively,  and  $c$ is the ratio in the Corkscrew condition \eqref{corkscrew.condition}
 for the domain $\Omega$.
\end{thm}

\section{Numerical demonstrations}
\label{simulations.section}

In this section, we demonstrate effectiveness of  the
%{\color{blue}shall we drop ``the'', since ``algorithms'' is plural form}
 algorithms \eqref{framealgorithm} and
 \eqref{noiseframealgorithm} to  approximate  concentrated signals
  in  the reproducing kernel space
\begin{equation*}
  V_2(\Phi)=\Big\{ \sum_{i\in {\mathbb Z}} c_i \phi_i: \  \sum_{i\in \mathbb Z} |c(i)|^2<\infty \Big \}
\end{equation*}
generated by  (non)uniform shifts of the Gaussian function $\exp (-x^2)$,
 where $\Phi:=\{\phi_i(x)=\exp(-(x-i-\theta_i)^2),  \ i\in \mathbb Z\}$ and
$\theta_i\in [-1/10, 1/10], i\in {\mathbb Z}$, are randomly selected \cite{atreas2012, sun2016galerkin, Kumar2020, sun2008ACM}. %{\color{red} references here}
Our numerical simulations indicate
that
 the correlation matrix
$  A_\Phi:=(\langle \phi_i,\phi_j\rangle)_{i,j\in {\mathbb Z}}$ has bounded inverse on $\ell^2$, and hence
the inverse
$A^{-1}_\Phi=(b_{ij})_{i,j\in \mathbb Z}$  has polynomial off-diagonal decay of any order by  Wiener's lemma for infinite matrices
\cite{ grochenig2006,  jaffard90, shinsun2019, sun2011,  sun2007}.
 Therefore
the linear space $V_2(\Phi)$ is
 the range space of an idempotent integral operator   with
 integral kernel   function
\begin{equation*}
  K_\Phi(x,y) = \sum_{i,j\in \mathbb Z}b_{ji} \phi_i(x)\phi_j(y)
\end{equation*}
satisfying  Assumption \ref{kernel.assumption} with $\theta=1$.

\begin{figure}[t] %tbp]
%\centering
%\subfigure[pic1.]
\includegraphics[width=58mm, height=42mm]{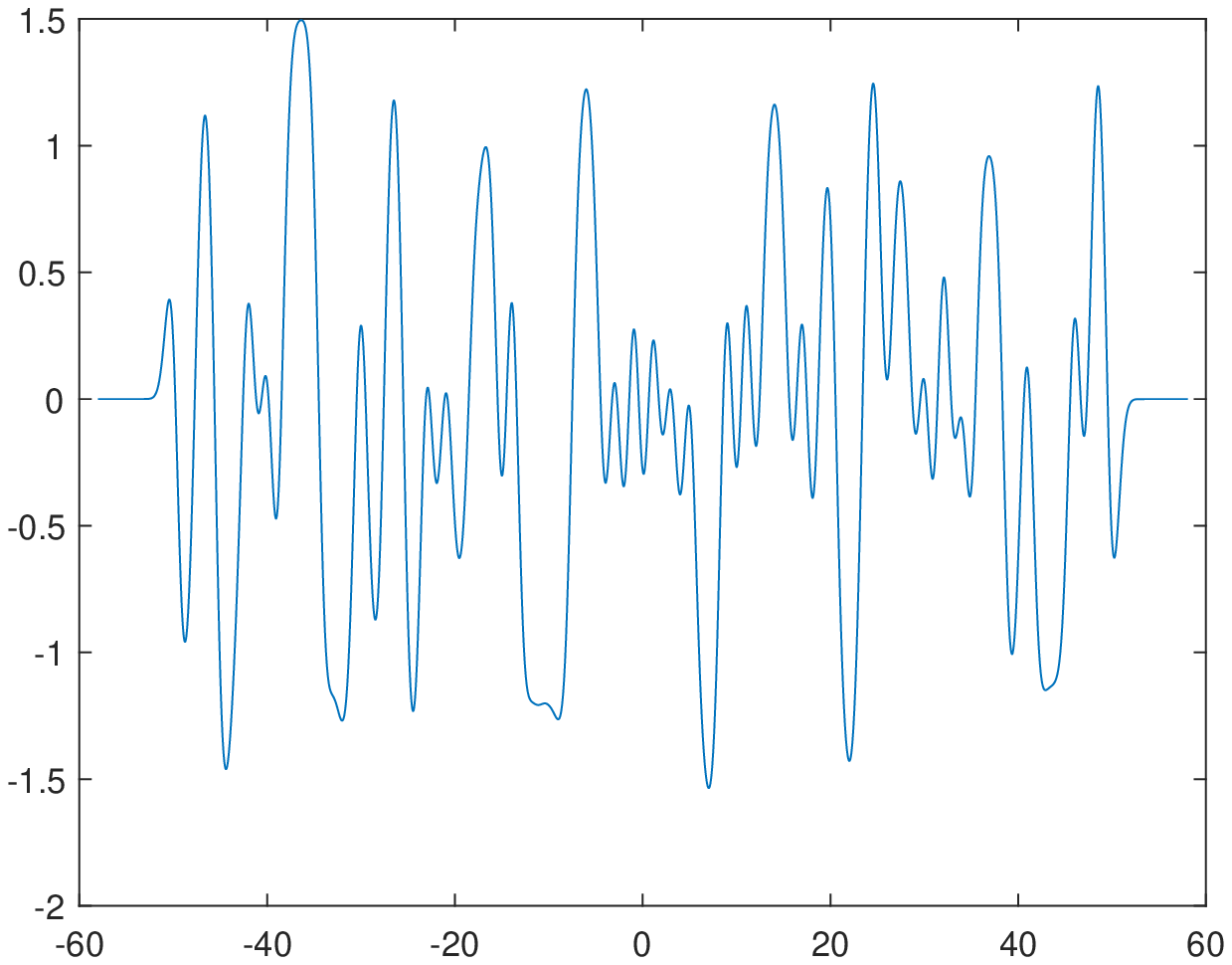}
 \hskip5mm
\includegraphics[width=58mm, height=42mm]{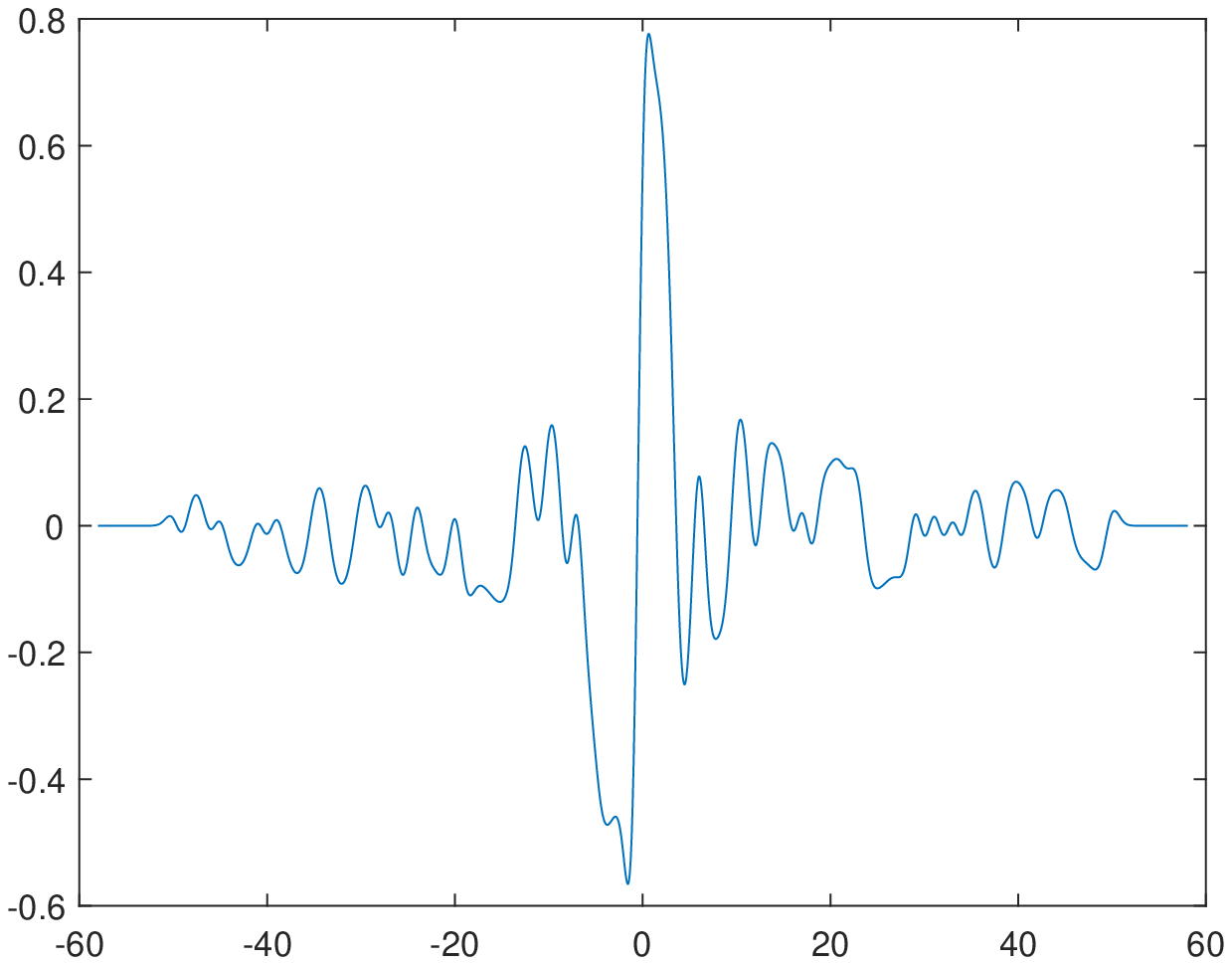}
\caption{\small Plotted on the left is
  a concentrated signal $f_{L, \alpha}$ in \eqref{simulationsignal.def} with $L=50$ and  $\alpha=0$,
  while on the right is another concentrated  signal $f_{L, \alpha}$ in \eqref{simulationsignal.def} with $L=50$ and $\alpha=0.8$,
   where the concentration ratio
   $\|f_{L, \alpha}\|_{2, \Omega_L^c}/\|f_{L, \alpha}\|_2=0.6433/7.2628=0.0886$  for  the signal in the left figure and
 $%\|f_{L, \alpha}\|_{2, \Omega_L^c}/|f_{L, \alpha}\|_2=
 0.0244/1.6869=0.0145$  for  the signal in the right figure. }
\label{originalsignal}
\end{figure}

In our simulations,
we consider the following family of  signals
 \begin{equation}\label{simulationsignal.def}
 f_{L, \alpha}=\sum_{i=-L}^L  r_i (1+|i|)^{-\alpha} \phi_i\in V_2(\Phi)
 \end{equation}
 concentrated on the interval $\Omega_L=[-L, L]$,
 where  $L\ge 1, \alpha\ge 0$, and  %random variables {\color{blue}delete ``random variables''?}
  random variables $r_i, -L\le i\le L$, are  independently selected in  $ [-1, 1]\backslash (-1/2, 1/2)$ with uniform distribution,
see Figure \ref{originalsignal} for two examples of concentrated signals $f_{L, \alpha}$ with $L=50$ and $\alpha=0, 0.8$ respectively.

%\fi

\begin{figure}[t] %[htbp]
\centering
\includegraphics[width=58mm, height=42mm]{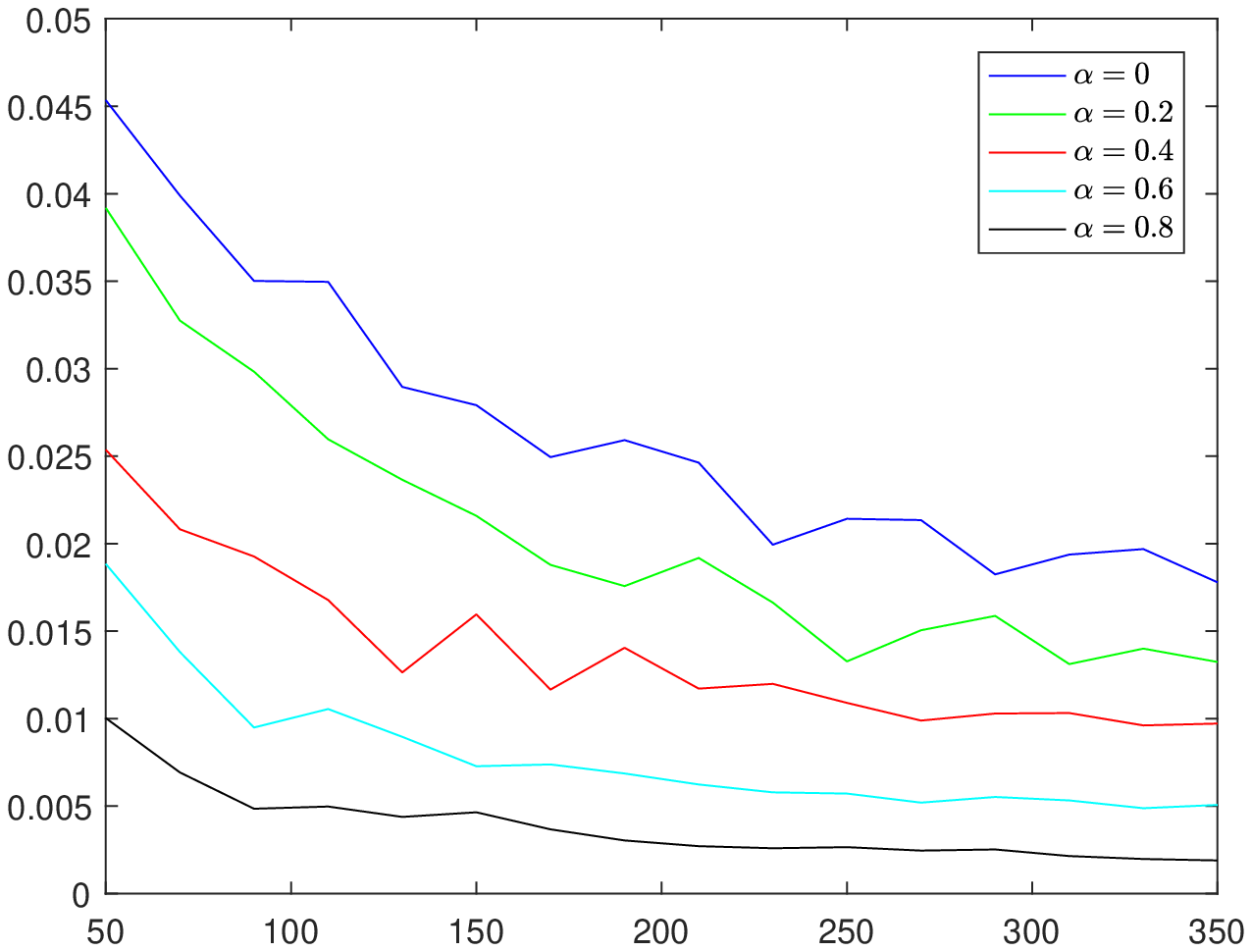}
 \hskip5mm
%\subfigure{\includegraphics[width=9cm,height=5.5cm]{G_Min_c_epsilon_67890.eps}}
%\subfigure{\includegraphics[width=9cm,height=5.5cm]{G_Ave_c_epsilon_67890.eps}}
\includegraphics[width=58mm, height=42mm]{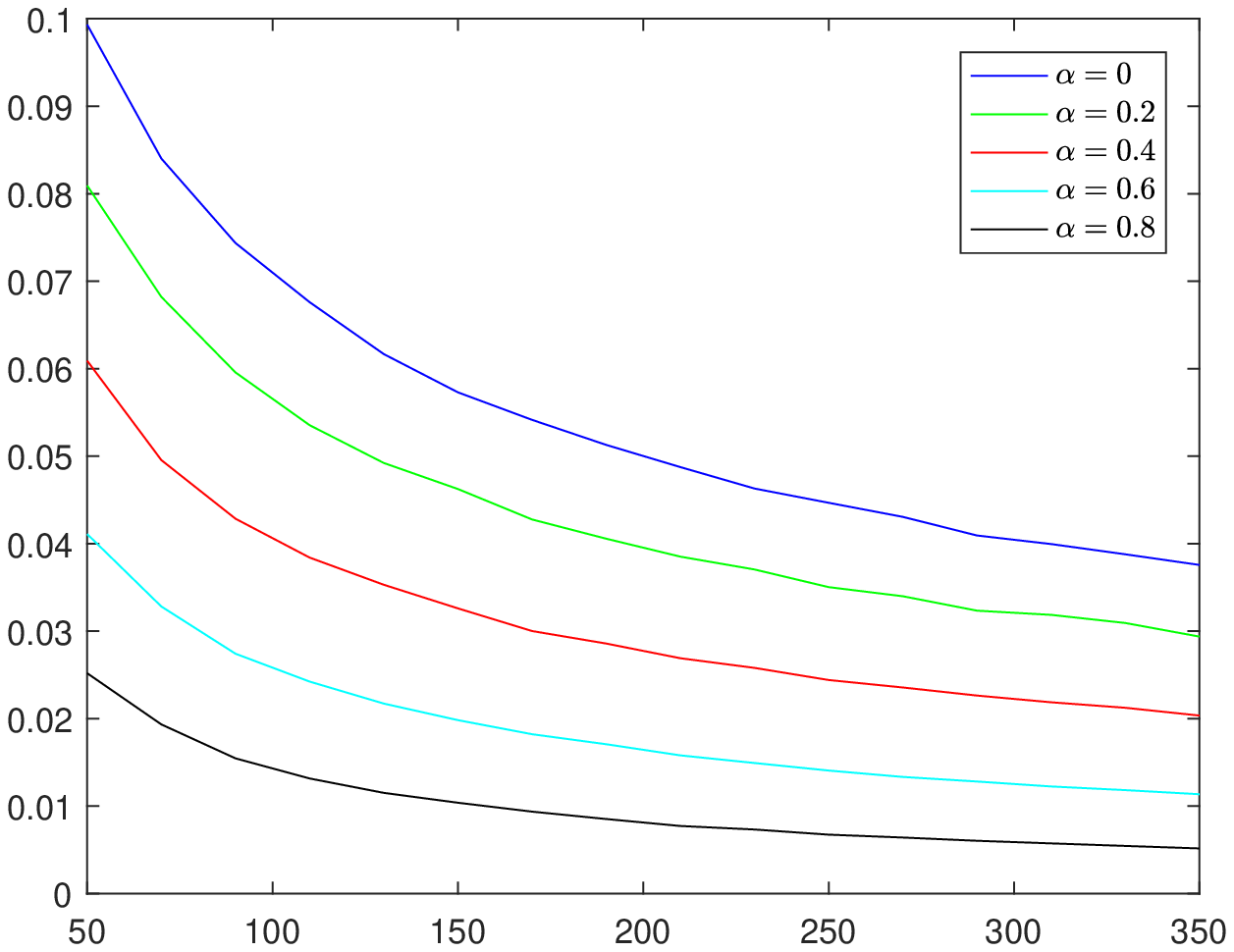}
\\
\vskip0.1in
\includegraphics[width=58mm, height=42mm]{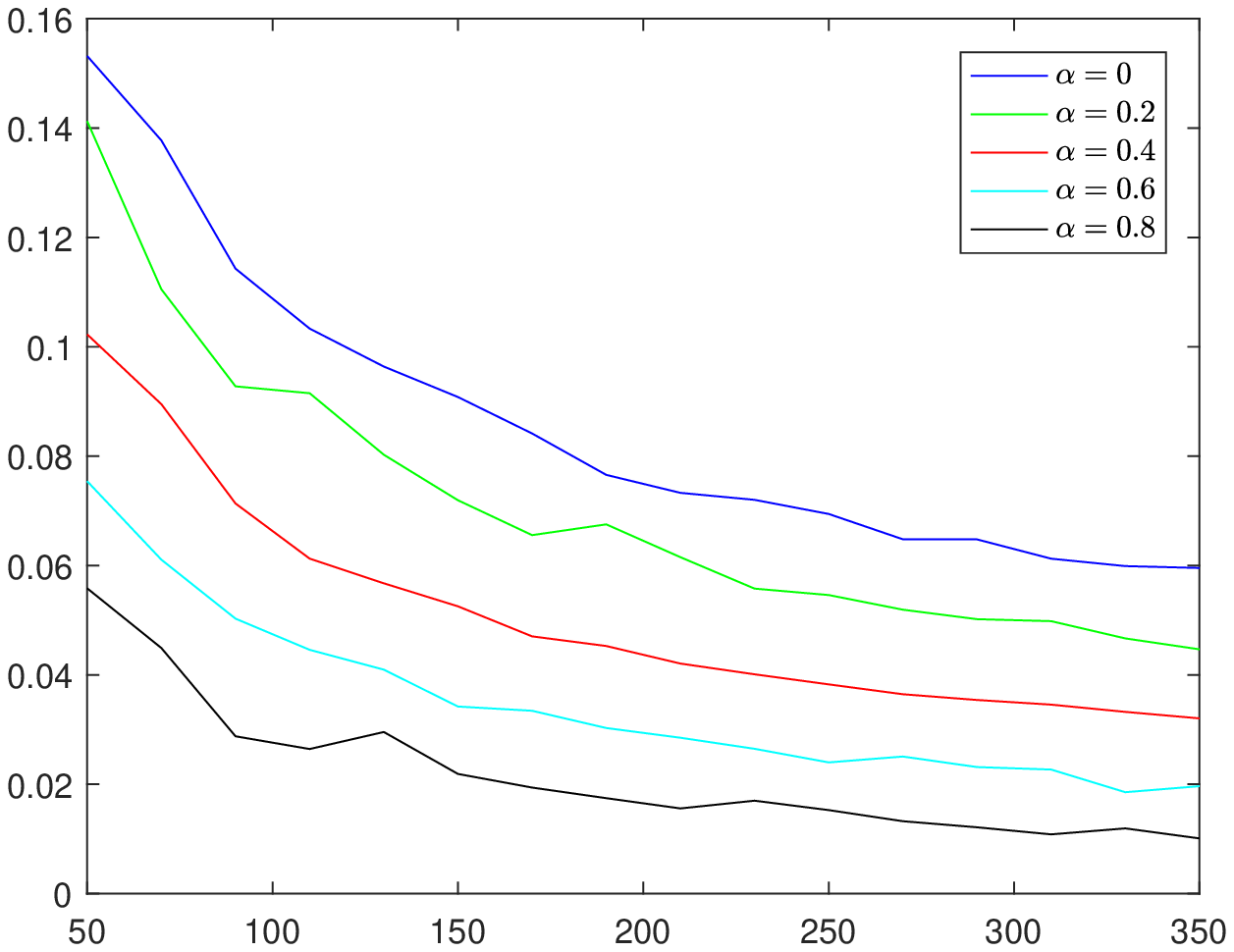}
 \hskip5mm
 \includegraphics[width=58mm, height=42mm]{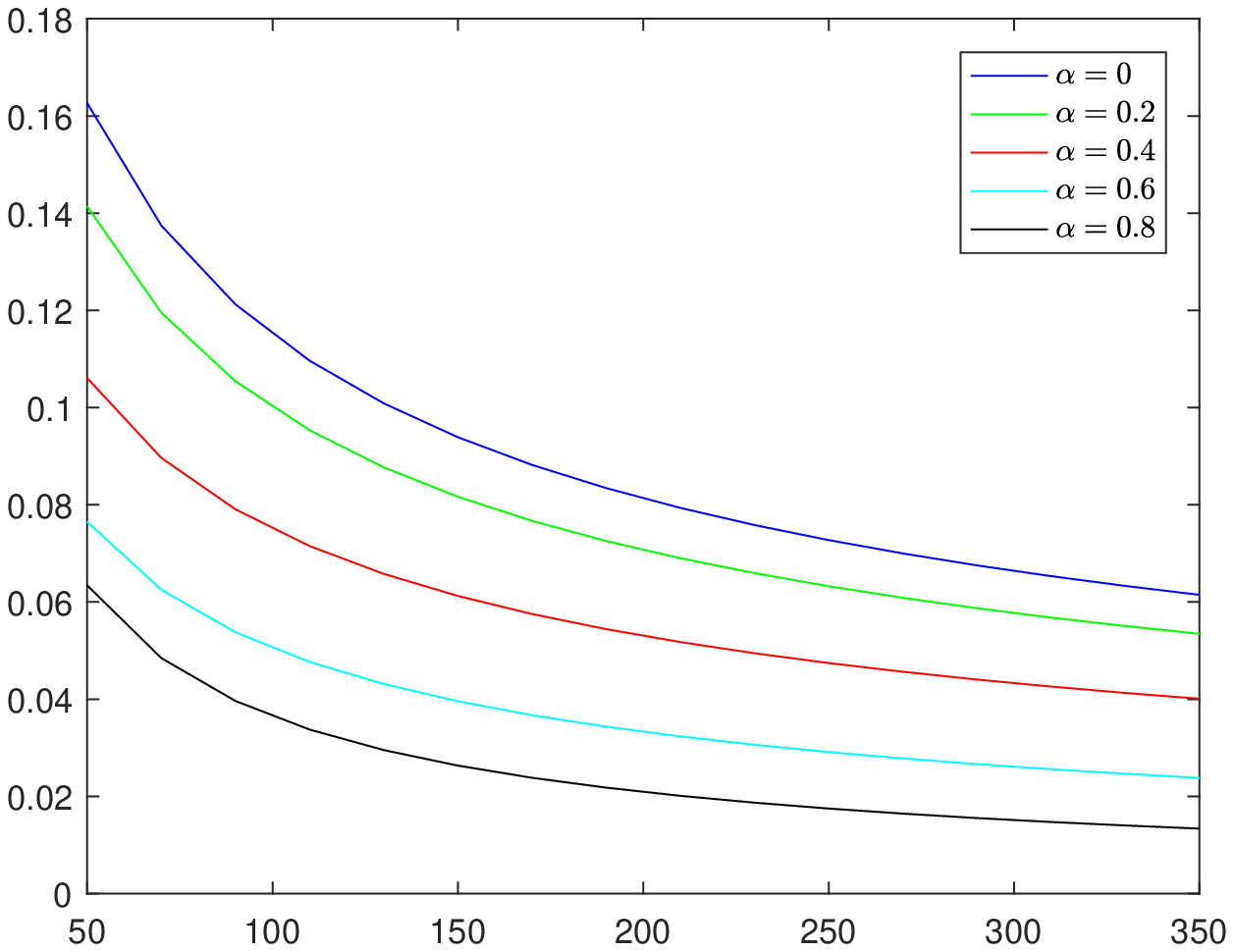}
\caption{  %M$\varepsilon_{L, \alpha}$ and
 Plotted on the top left/top right/bottom left are  the  minmal/average/maximal concentration ratio $%L^{\max(\alpha, 1/2)}
 \|f_{L, \alpha}\|_{2, \Omega_L^c}/{\|f_{L, \alpha}\|_{2}}$  over 1000 trials for $\alpha=0, 0.2, 0.4, 0.6, 0.8$ respectively. On
 the bottom right is the  concentration ratio $\varepsilon_{L, \alpha}$ selected in \eqref {calpha.selection} for the family of concentrated
 signals $f_{L, \alpha}, 50\le L\le 350$, which is approximately the maximal concentration ratio plotted on the bottom left figure. It is observed that the
 average concentration ratio $ \|f_{L, \alpha}\|_{2, \Omega_L^c}/{\|f_{L, \alpha}\|_{2}}$ is almost proportional to the selected concentration ratio $\varepsilon_{L, \alpha}$ for $\alpha=0, 0.2, 0.4, 0.6, 0.8$.
}
\label{varepsilon.fig.addyaxu}
\end{figure}

    Due to the Riesz basis property for the generator   $\Phi$ and randomness of $r_i, -L\le i\le L$, we have
\begin{equation*}\label{varepsilon.alphaeq00}
%\varepsilon_{L, \alpha}:=
\frac{\|f_{L, \alpha}\|_{2, \Omega_L^c}}
{\|f_{L, \alpha}\|_{2}}
\lesssim  \frac{L^{-\alpha}} {\big(\sum_{|i|\le L} (1+|i|)^{-2\alpha}\big)^{1/2}}
\lesssim \left\{\begin{array}{ll}  L^{-1/2}  & {\rm if} \ \alpha<1/2\\
(L\ln L)^{-1/2} & {\rm if} \ \alpha=1/2\\
L^{-\alpha} & {\rm if} \ \alpha>1/2,
 \end{array}\right.
\end{equation*}
and
\begin{eqnarray*} \label{varepsilon.alphaeq01}
\frac{\sqrt{\mathbb{E}\|f_{L, \alpha}\|_{2, \Omega_L^c}^2}}
{\sqrt{\mathbb{E}\|f_{L, \alpha}\|_{2}^2}} & \hskip-0.08in = & \hskip-0.08in \frac{\big(\sum_{|i|\le L} (1+|i|)^{-2\alpha} \int_{{\mathbb R}\backslash [-L, L]} |\phi_i(x)|^2 dx\big)^{1/2}}
{\big(\sum_{|i|\le L} (1+|i|)^{-2\alpha} \int_{{\mathbb R}} |\phi_i(x)|^2 dx\big)^{1/2}}\nonumber\\
& \hskip-0.08in \approx & \hskip-0.08in \left\{\begin{array}{ll}  L^{-1/2}  & {\rm if} \ \alpha<1/2\\
(L\ln L)^{-1/2} & {\rm if} \ \alpha=1/2\\
L^{-\alpha} & {\rm if} \ \alpha>1/2.
 \end{array}\right.
\end{eqnarray*}
Here for two positive items $A$ and $B$,  $A\lesssim  B$ means that $A/B$ is bounded by an absolute constant, and $A\approx B$ if
both $A/B$ and $B/A$ are bounded by an absolute constant.
Therefore signals $f_{L, \alpha}$ in \eqref{simulationsignal.def}
are  concentrated on $\Omega_L$ with concentration ratio %$\varepsilon_{L, \alpha}$
being about  a multiple of $L^{-\max(\alpha, 1/2)}$ for $\alpha\ne 1/2$.
% see  Table \ref{varepsilon.table}  and Figure \ref{varepsilon.fig}
 The above estimate  on the concentration ratio $%L^{\max(\alpha, 1/2)}
 \|f_{L, \alpha}\|_{2, \Omega_L^c}/{\|f_{L, \alpha}\|_{2}}$  is confirmed by our numerical simulations, see Figure
 \ref{varepsilon.fig.addyaxu}. % for   the  minmal/average/maximal concentration ratio  over 1000 trials.
So in our simulations,  we  consider that the family of concentrated  signals $f_{L, \alpha}$ in \eqref{simulationsignal.def}
%concentrated onto the domain $\Omega_L$
have concentration ratio
 \begin{equation}\label{calpha.selection}
 \varepsilon_{L, \alpha}= C_\alpha L^{-\max(\alpha, 1/2)},\end{equation}
 where $C_\alpha= 1.15, 1, 0.75, 0.80, 1.45$ for $\alpha=0, 0.2, 0.4, 0.6, 0.8$ respectively.

In the first part of our numerical simulations, we
 consider
  the sampling set
$\Gamma_L=\{\gamma_k, 1\le k\le N\}$ with
$\gamma_{i+1}-\gamma_i, 0\le i\le N-1$, being independently selected on
 %in  {\color{blue} ``independent selected in'' to ``independently selected on''}
 $[1/4, 3/4]$ with uniform distribution, where
$\gamma_0=-L, \gamma_{N+1}=L$ and $N$ is chosen so that $\gamma_{N+1}-\gamma_N\in [0, 1/4]$.
The size of  the sampling set $\Gamma_L$ is between $8L/3$ and $8L$,
%{\color{blue}replace ``is between $8L/3+1$ and $8L+1$'' to ``is between $\lfloor 8L/3 \rfloor+1$ and $8L-2$''?}
while most of them have their sizes around $4L$.
To construct the preconstruction operator   in \eqref{S.Gamma.U} associated with the above sampling set $\Gamma_{L}$, we take
uniformly  sampling set on the complement $\Omega_L^c={\mathbb R}\backslash [-L, L]$ with  gap
$\delta_{L, \alpha}:=C_\alpha L^{-\max(\alpha, 1/2)}/2$, where  $C_\alpha$ is given in \eqref{calpha.selection}.
%$\varepsilon_{L, \alpha}/2$.
 Under the above setting,
 the preconstruction operator in \eqref{S.Gamma.U} becomes
\begin{eqnarray*}
  S_{L, \alpha} f(x) & \hskip-0.08in = & \hskip-0.08in  \sum_{k=1}^N |I_{\gamma_k}| f(\gamma_k)K_\Phi(x,\gamma_k)  +
\delta_{L, \alpha} \sum_{m=0}^{\infty} f(\gamma_m^+)  K_\Phi(x, \gamma_m^+)\\
& \hskip-0.08in & \hskip-0.08in  +
\delta_{L, \alpha}\sum_{m=0}^{\infty} f(\gamma_m^-)  K_\Phi(x, \gamma_m^-), \ f\in V_2(\Phi),
\end{eqnarray*}
where $\gamma_m^\pm= \pm (L+(m+1/2)\delta_{L, \alpha})$, $m\ge 0$, $|I_{\gamma_1}|=L+\frac{\gamma_2+\gamma_1}{2}$,
$|I_{\gamma_N}|=L-\frac{\gamma_{N}+\gamma_{N-1}}{2}$,
and $|I_{\gamma_k}|= \frac{\gamma_{k+1}-\gamma_{k-1}}{2}, 2\le k\le N-1$.
Let $g_{n, L, \alpha}, n \ge 0$, be  the $n$-th term  in the iterative algorithm \eqref{framealgorithm} with the original concentrated signal being $f_{L, \alpha}$ and the above sampling set $\Gamma_L$
 with the Hausdorff distance
$d_H(\Gamma_L, [-L, L])\le 3/8$.
Shown in Table \ref{tab:samplingone} is
the average  of the relative approximation error  (RAE)
\begin{equation}\label{ELalpha.def00}
 E_{L, \alpha}(n)=\|g_{n, L, \alpha}-f_{L, \alpha}\|_2/\|f_{L, \alpha}\|_{2}\end{equation}
over 500 trials for $n=0, 3$.
%%%%%%%%%%%%%%%%%%%%%%%%%%%%%%%%%%%%%%%%%%%%%%%

\begin{figure}[t] %[htbp]
\centering
%\subfigure[pic1.]
%\includegraphics[width=38mm, height=33mm]{Paperf0L=50alpha=0.jpg}
\includegraphics[width=58mm, height=42mm]{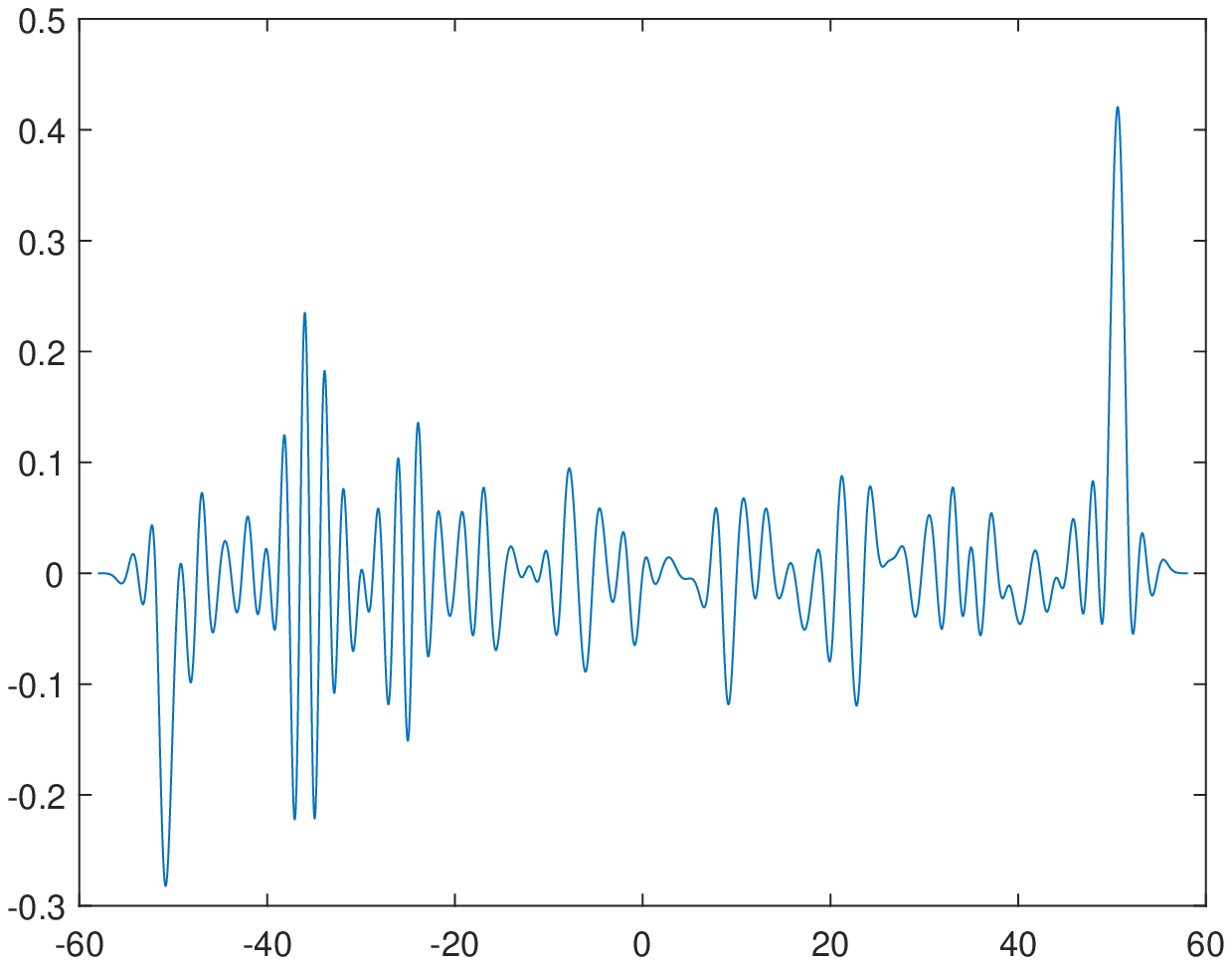}
 \hskip5mm
\includegraphics[width=58mm, height=42mm]{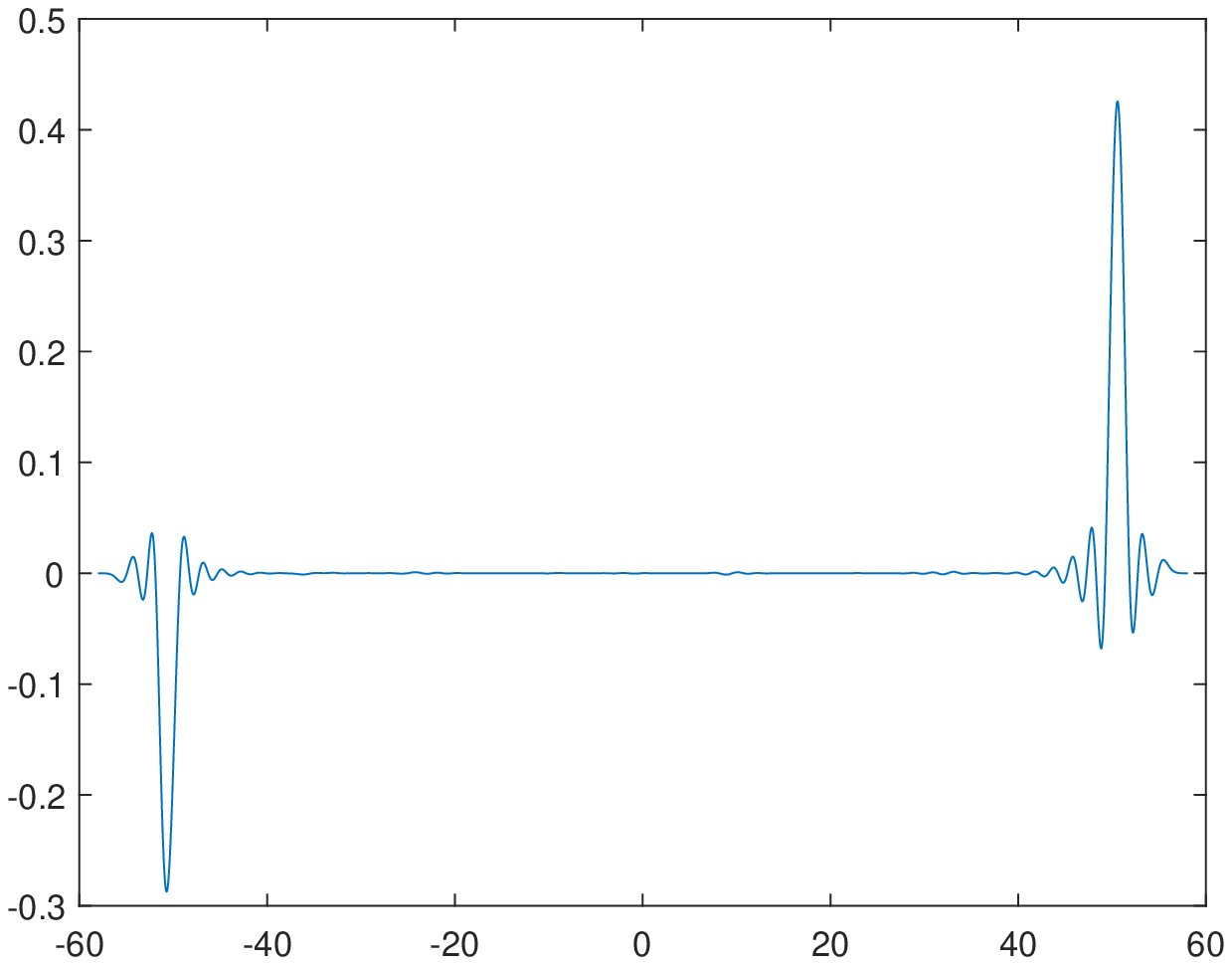}
\\
%\hspace{0.5mm}
\vspace{2mm}
\includegraphics[width=58mm, height=42mm]{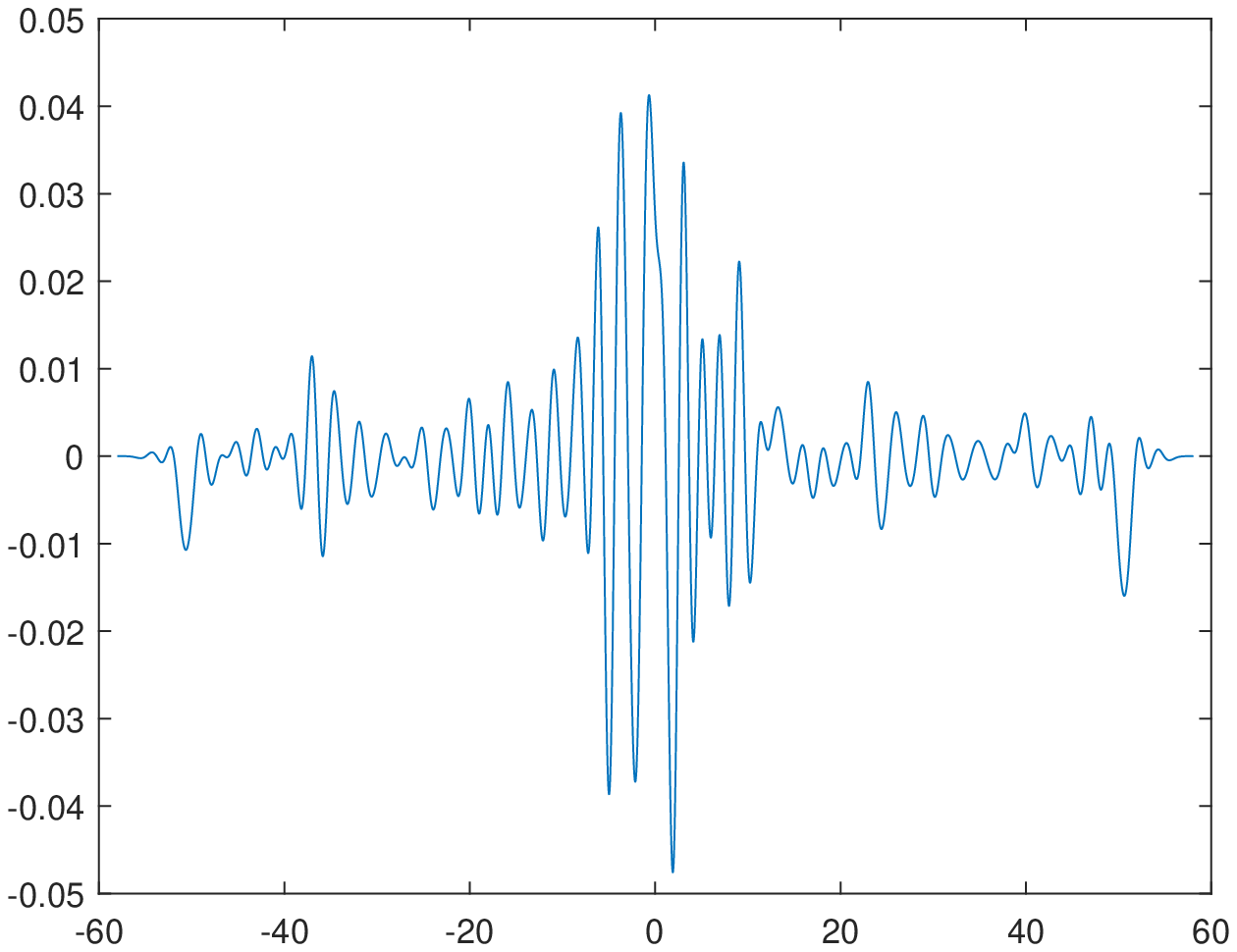}
 \hskip5mm
\includegraphics[width=58mm, height=42mm]{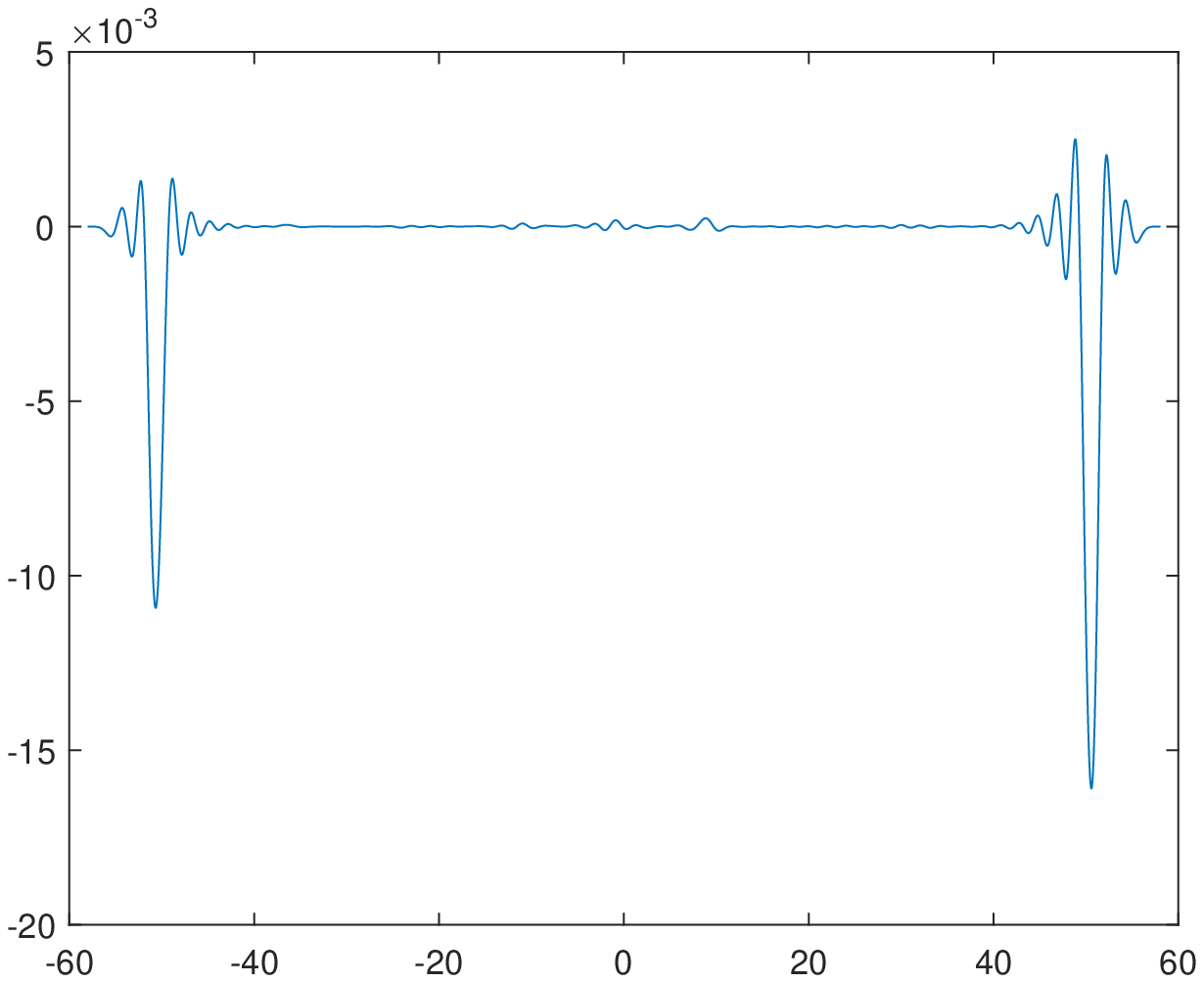}
\caption{Plotted on  the top left and bottom left  are
 the difference  $g_{0, L, \alpha}-f_{L, \alpha}$ between the preconstructed signal $g_{0, L, \alpha}$ and the original signal $f_{L, \alpha}$
 given in  Figure \ref{originalsignal} with  $\alpha=0$ (top) and $\alpha=0.8$ (bottom), while
   on the top right and bottom right are the difference  $g_{n, L, \alpha}-f_{L, \alpha}$ between the constructed signal $g_{n, L, \alpha}$ at the third iteration ($n=3$) and the original signal $f_{L, \alpha}$  with  $\alpha=0$ (top) and $\alpha=0.8$ (bottom). Here
 the number of samples in the reconstruction procedure is $4L+3=203$,
 and the relative preconstruction error
 $ \|g_{0, L, \alpha}-f_{L, \alpha}\|_2/\|f_{L, \alpha}\|_2$, the relative approximation error $\|g_{n, L, \alpha}-f_{L, \alpha}\|_2/\|f_{L, \alpha}\|_{2}$
 and the concentration ratio $\|f_{L, \alpha}\|_{2, \Omega_L^c}/\|f_{L, \alpha}\|_2$ of the original signal $f_{L, \alpha}$ are
$   %=0.7623/7.2628
 0.1050,
 %$\|g_{n, L, \alpha}-f_{L, \alpha}\|_2/\|f_{L, \alpha}\|_{2}
 %=0.5610/7.2628
 0.0772$ and
$ %\|f_{L, \alpha}\|_{2, [-L,L]^c}/\|f_{L, \alpha}\|_2=
0.0886$ respectively for the top figures,
 and
   $ %\|g_{0, L, \alpha}-f_{L, \alpha}\|_2/\|f_{L, \alpha}\|_{2}
   %=0.1008/1.6869
   0.0598$,
 $%\|g_{n, L, \alpha}-f_{L, \alpha}\|_2/\|f_{L, \alpha}\|_{2}
 %=0.0213/1.6869
  0.0126$ and
 $ %\|f_{L, \alpha}\|_{2, [-L,L]^c}/\|f_{L, \alpha}\|_2=
 0.0145$ respectively for the bottom figures.   }
 %{\color{blue}In Figure 3 there are two places ``are the difference'' or ``are the differences''. Also in Figures 4 and 5.}
\label{reconstructionerrordeterminisiticsampling}
\end{figure}

\begin{table}[t] %[!ht]
\begin{center}
\caption{Average of the relative approximation error  $E_{L, \alpha}(n)$ in \eqref{ELalpha.def00}  over  500  trails for $n=0$ and $3$. }
\label{tab:samplingone}
\begin{tabular}{c|ccccc}
\toprule
 \multicolumn{6}{c} {$n=0$} \\%& &$Gaussian$ & & &  & &$B-Spline$&  \\
\hline %\cmidrule(lr){2-6}
\backslashbox{L}{RAE } {$\alpha$}&0&0.2 & 0.4 & 0.6 & 0.8 \\
\hline %\midrule
50&0.1021&0.0891&0.0777&0.0674&0.0617\\
70&0.0926&0.0814&0.0722&0.0643&0.0608\\
90&0.0857&0.0766&0.0683&0.0615&0.0593\\
110&0.0811&0.0736&0.0665&0.0605&0.0582\\
%130&0.0788&0.0712&0.0654&0.0602&0.0586\\
%150&0.0762&0.0696&0.0642&0.0606&0.0578\\
170&0.0741&0.0682&0.0631&0.0589&0.0584\\
%190&0.0725&0.0672&0.0625&0.0595&0.0582\\
%210&0.0709&0.0664&0.0623&0.0592&0.0583\\
230&0.0701&0.0657&0.0623&0.0593&0.0580\\
%250&0.0693&0.0654&0.0614&0.0586&0.0578\\
%270&0.0685&0.0644&0.0608&0.0589&0.0585\\
290&0.0677&0.0639&0.0609&0.0586&0.0579\\
%310&0.0671&0.0635&0.0605&0.0589&0.0568\\
%330&0.0666&0.0634&0.0605&0.0590&0.0591\\
350&0.0665&0.0630&0.0604&0.0581&0.0580\\
\hline
\midrule
\multicolumn{6}{c} {$n=3$} \\%& &$Gaussian$ & & &  & &$B-Spline$&  \\
%\cmidrule(lr){2-6}
\hline \backslashbox{L}{RAE } {$\alpha$}&0&0.2 & 0.4 & 0.6 & 0.8 \\
\hline %\midrule
50&0.0846&0.0679&0.0515&0.0343&0.0213\\
70&0.0725&0.0574&0.0427&0.0279&0.0165\\
90&0.0634&0.0501&0.0368&0.0235&0.0133\\
110&0.0567&0.0453&0.0330&0.0207&0.0111\\
%130&0.0537&0.0416&0.0303&0.0182&0.0097\\
%150&0.0493&0.0388&0.0276&0.0168&0.0088\\
170&0.0461&0.0363&0.0256&0.0152&0.0077\\
%190&0.0438&0.0340&0.0242&0.0142&0.0073\\
%210&0.0409&0.0326&0.0228&0.0134&0.0065\\
230&0.0397&0.0315&0.0223&0.0128&0.0063\\
%250&0.0379&0.0299&0.0203&0.0120&0.0058\\
%270&0.0365&0.0288&0.0197&0.0115&0.0054\\
290&0.0352&0.0275&0.0190&0.0108&0.0053\\
%310&0.0342&0.0266&0.0184&0.0105&0.0048\\
%330&0.0332&0.0258&0.0178&0.0101&0.0047\\
350&0.0328&0.0248&0.0174&0.0097&0.0044\\
\hline\bottomrule
\end{tabular}
\end{center}
\end{table}
%%%%%%%%%%%%%%%%%%%%%%%%%%%%%%%%%%
Our  numerical simulations show that
\begin{equation*}
E_{L, \alpha}(n)\le \varepsilon_{L, \alpha}, \ n\ge 3,
\end{equation*}
for all $50\le L\le 350$ and $\alpha=i/5, 0\le i\le 4$,  see Table \ref{tab:samplingone} and Figure \ref{reconstructionerrordeterminisiticsampling}.
This demonstrates the conclusion in Theorem \ref{approximation.thm} on the approximation  property of $g_{n, L, \alpha}, n\ge 0$,
 %in the iterative algorithm  \eqref{frame.algorithm.eq1}
 to the original concentrated signal  $f_{L, \alpha}$ for large $n$.
 We observe from Figure \ref{reconstructionerrordeterminisiticsampling} that  $g_{n, L, \alpha}, n\ge 1$,
 %{\color{blue}Figure 3 is for $n=0,3$. Can we observe ``$n\ge 1$''?}
 in
 the iterative algorithm \eqref{framealgorithm} provide better approximations to the original signal $f_{L, \alpha}$
  than the preconstruction signal  $g_{0, L, \alpha}$ does, and that
 $g_{n, L, \alpha}, n\ge 3$,
 have almost perfect approximations % {\color{blue}``approximation'' to ``approximations''?}
  to the original signal $f_{L, \alpha}$ inside the domain far from the boundary.

\begin{figure}[t] %[htbp]
\centering
\includegraphics[width=58mm, height=42mm]{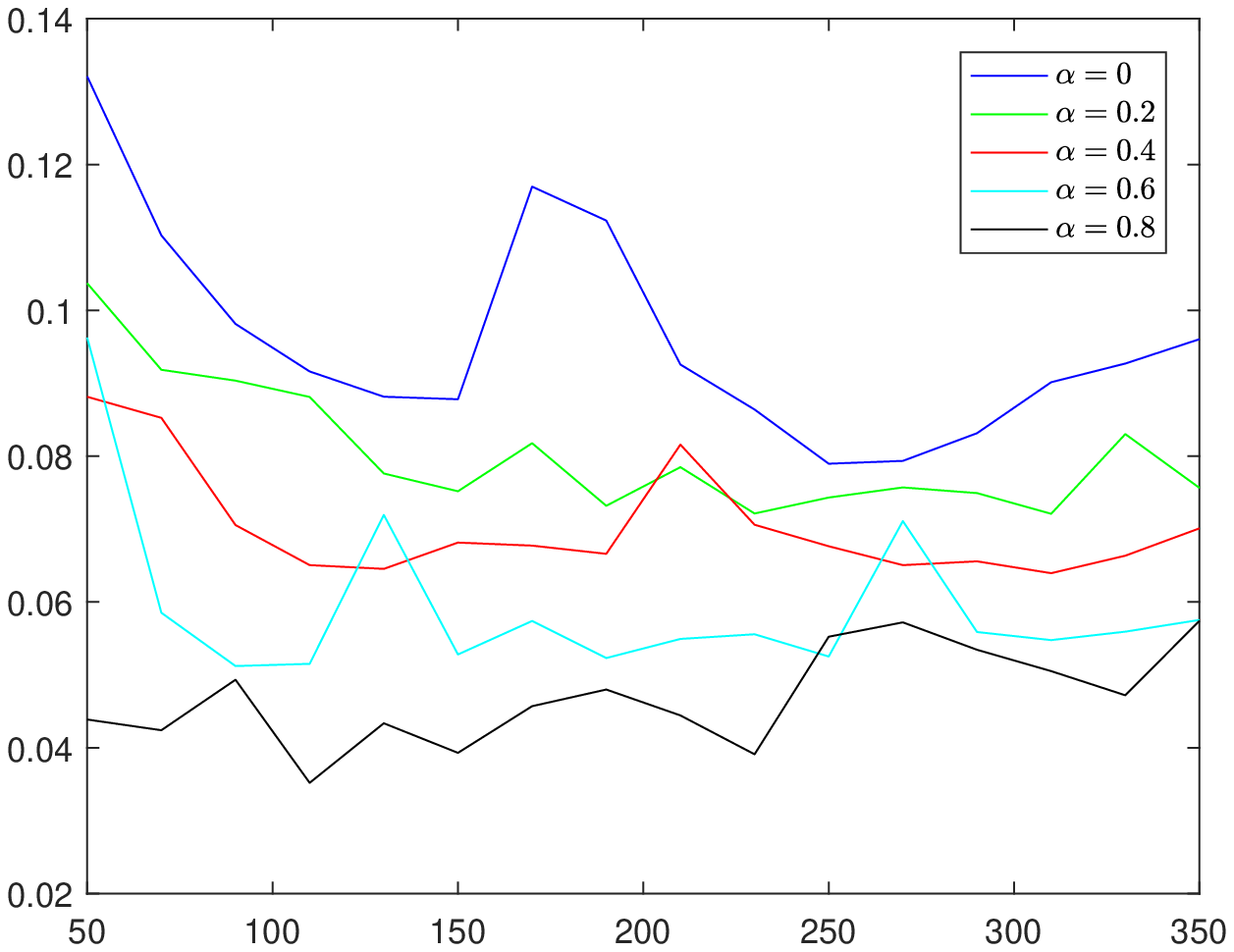}%meanL8sr22
\hskip5mm
\includegraphics[width=58mm, height=42mm]{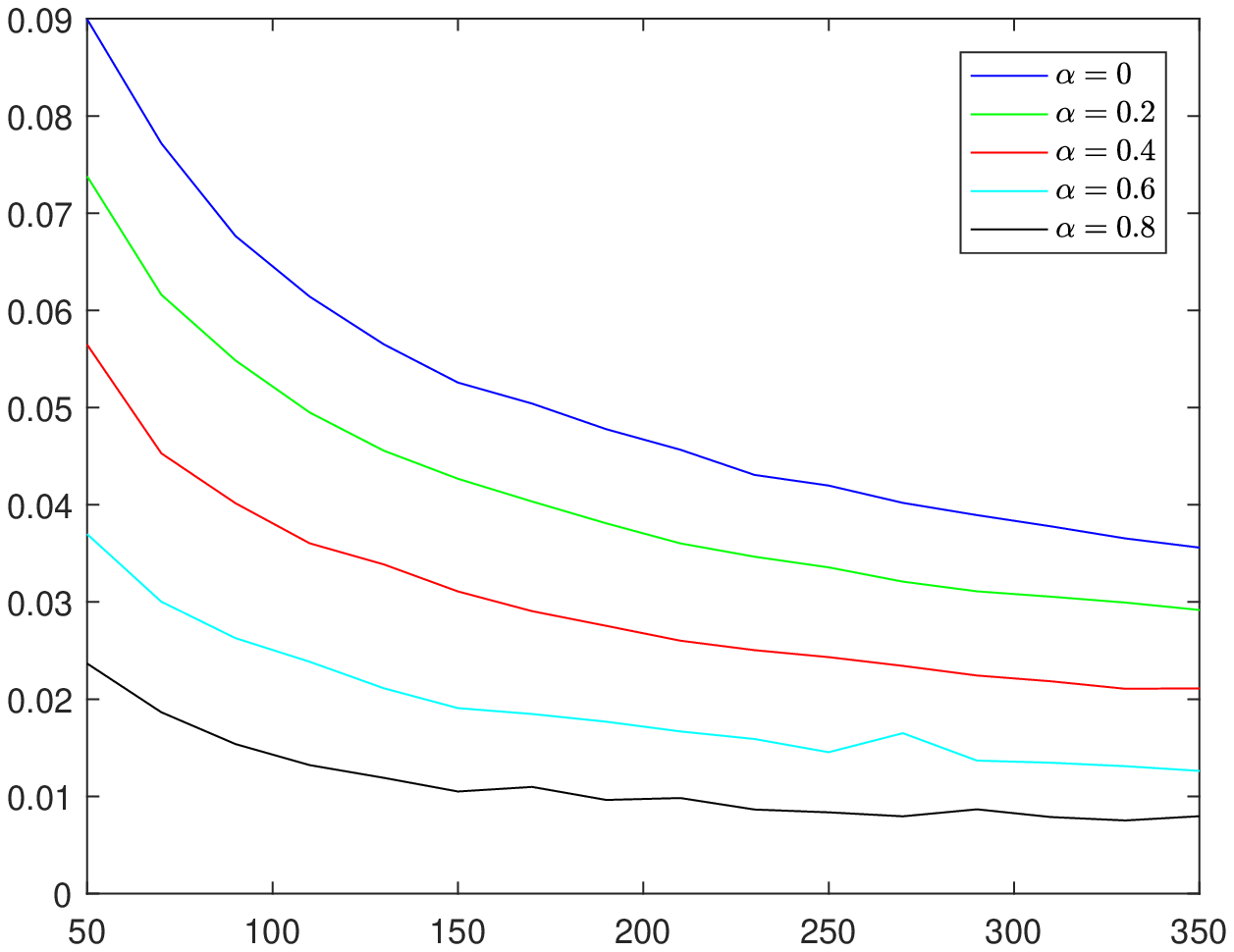}%meanL12sr22
\\
\vskip0.1in
\includegraphics[width=58mm, height=42mm]{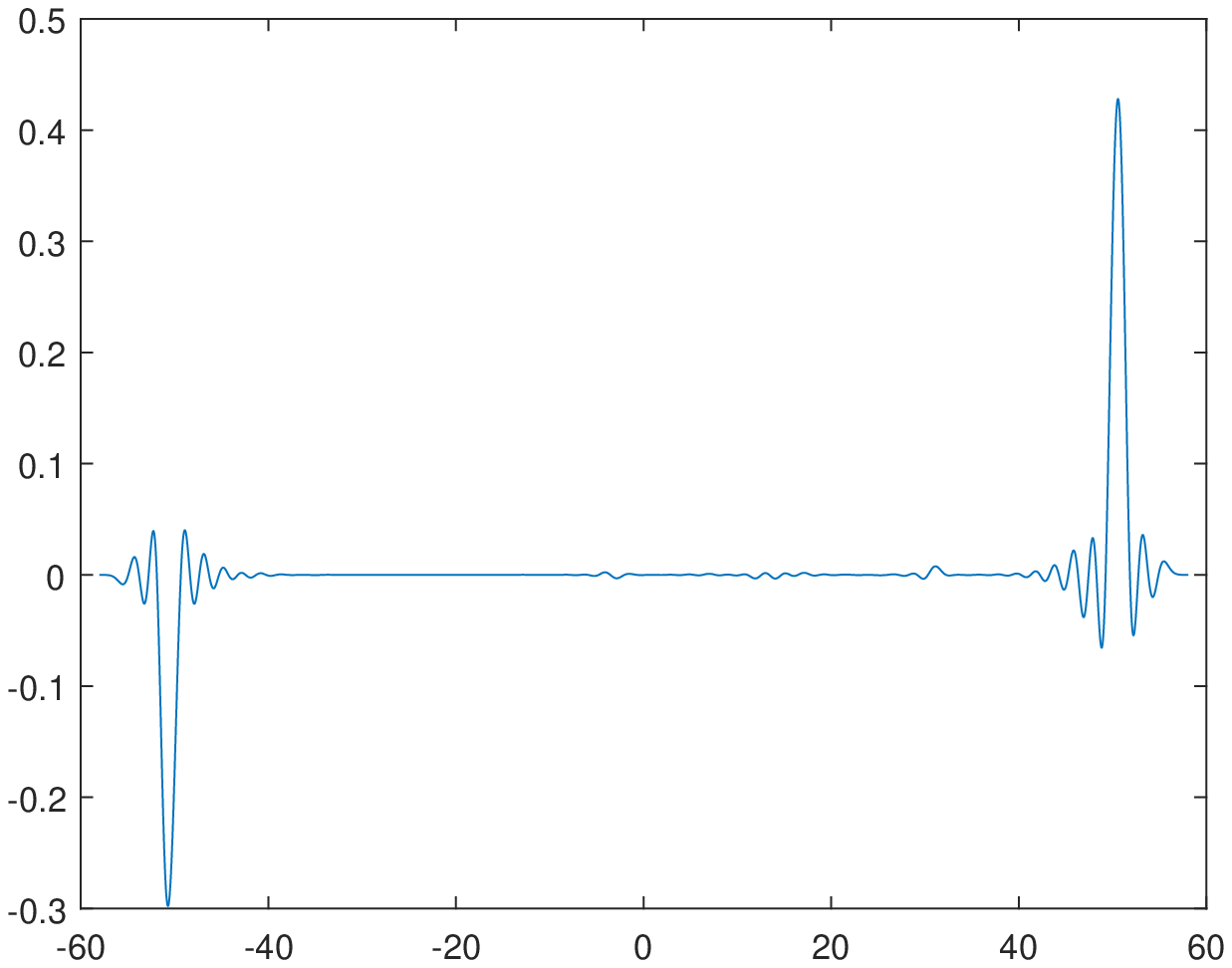}%meanL8sr22
\hskip5mm
\includegraphics[width=58mm, height=42mm]{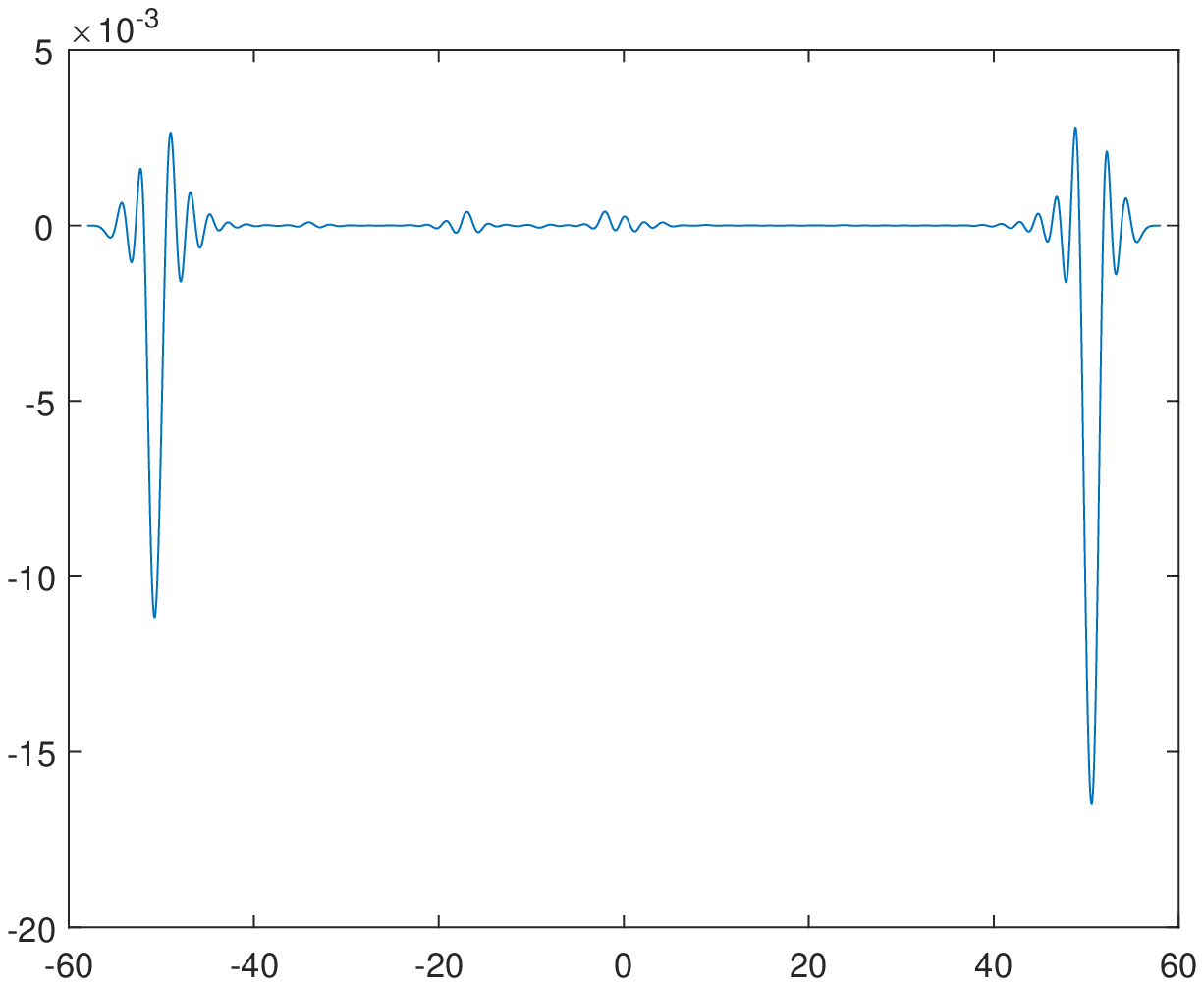}
%\includegraphics[width=38mm, height=33mm]{meanL16sr22.eps}%meanL16sr22
%%%%%\\
%%%%\includegraphics[width=48mm, height=38mm]{Paperg6minusf0L=50alpha=0Random.eps}
%\includegraphics[width=38mm, height=33mm]{Paperg2minusf0L=50alpha=08random.eps}
%%%%\includegraphics[width=48mm, height=38mm]{Paperg6minusf0L=50alpha=08random.eps}
\caption{%Shown on the top left  and right are  average of the relative approximation
% error $ E_{N, L, \alpha}(n)=\|g_{N, L, \alpha}^{(n)}-f_{L, \alpha}\|_2/\|f_{L, \alpha}\|_{2}$ over 100 trials
%with $n=6$ and $N=8L$ (left) and $N=12L$ (right).
%Plotted on the  bottom left and right
%are the difference between   $g_{N, L, \alpha}^{(n)}-f_{L, \alpha}$ between the reconstructed signal $g_{N, L, \alpha}^{(n)}$
% in the sixth iteration ($n=6$) and the original signal $f_{L, \alpha}$
% given in  Figure \ref{originalsignal} with $L=50,  \alpha=0$ (left) and $\alpha=0.8$ (right), where
% the number of random samples in the reconstruction procedure is  $N=8L=400$,
% $\|g_{N, L, \alpha}^{(n)}-f_{L, \alpha}\|_2/\|f_{L, \alpha}\|_{2}=0.5692/7.2628=0.0783$ for the bottom left figures,
% and
% $\|g_{N, L, \alpha}^{(n)}-f_{L, \alpha}\|_2/\|f_{L, \alpha}\|_{2}=0.0216/1.6869=0.0128$ for the bottom figures.
Shown on the top left  and right are the  average of the relative approximation
 error $ E_{N, L, \alpha}(n)$ over 500 trials
with $n=6$ and $N=8L$ (left) and $N=12L$ (right).
Plotted on the  bottom left and right
are the difference   $g_{N, L, \alpha}^{(n)}-f_{L, \alpha}$ between the reconstructed signal $g_{N, L, \alpha}^{(n)}$
 at the sixth iteration ($n=6$) and the original signal $f_{L, \alpha}$
 given in  Figure \ref{originalsignal}, where $L=50,  \alpha=0$ (left) and $\alpha=0.8$ (right), cf. Figure \ref{reconstructionerrordeterminisiticsampling}. Here
 the number of random samples in the reconstruction procedure is  $N=8L=400$,
and the relative approximation error $\|g_{N, L, \alpha}^{(n)}-f_{L, \alpha}\|_2/\|f_{L, \alpha}\|_{2}$
is $ %=0.5692/7.2628=
0.0783$ for the bottom left figure
 and
 %$\|g_{N, L, \alpha}^{(n)}-f_{L, \alpha}\|_2/\|f_{L, \alpha}\|_{2}=0.0216/1.6869=
 $0.0128$ for the bottom  right figure. }
\label{varepsilon.fig}
\end{figure}

 In the second part of our numerical simulations, we
 consider
  the sampling set
$\Gamma_{N,L}=\{\gamma_k, 1\le k\le N\}$ with
$\gamma_k, 1\le k\le N$, being independently selected on $[-L, L]$ with uniform distribution. We order these  random sampling positions in increasing order and denote by
$-L\le \mu_1\le \ldots\le \mu_N\le L$.  Similar to our first simulation,
 we take
uniformly  sampling set on the complement $[-L, L]^c$ with  gap
$\delta_{L, \alpha}:=C_\alpha L^{-\max(\alpha, 1/2)}/2$ and $C_\alpha$  given in \eqref{calpha.selection}.
%$\varepsilon_{L, \alpha}/2$.
Under the above setting,  the preconstruction operator in \eqref{S.Gamma.U} becomes
\begin{eqnarray*}
  \tilde S_{N, L, \alpha} f(x) & \hskip-0.08in = & \hskip-0.08in  \sum_{k=1}^N |I_{\mu_k}| f(\mu_k)K_\Phi(x,\mu_k)  +
\delta_{L, \alpha} \sum_{m=0}^{\infty} f(\gamma_m^+)  K_\Phi(x, \gamma_m^+)\\
& \hskip-0.08in & \hskip-0.08in  +
\delta_{L, \alpha}\sum_{m=0}^{\infty} f(\gamma_m^-)  K_\Phi(x, \gamma_m^-), \ f\in V_2(\Phi),
\end{eqnarray*}
where $\gamma_m^\pm= \pm (L+(m+1/2)\delta_{L, \alpha}), m\ge 0$, $|I_{\mu_1}|=L+\frac{\mu_2+\mu_1}{2}$,
$|I_{\mu_N}|=L-\frac{\mu_{N}+\mu_{N-1}}{2}$
and $|I_{\mu_k}|= \frac{\mu_{k+1}-\mu_{k-1}}{2}, 2\le k\le N-1$.
Let $g_{N, L, \alpha}^{(n)}, n \ge 0$, be  the $n$-th term  in the iterative algorithm \eqref{framealgorithm} with the original signal being $f_{L, \alpha}$ and the random sampling set being $\Gamma_{N,L}$ of size $N$.  Our simulations indicate that
most of signals $g_{N, L, \alpha}^{(n)}, n \ge 6$,  reconstructed from the iterative algorithm \eqref{framealgorithm} provide good approximations  to the original
signal $f_{L, \alpha}$ when $ N\ge 12L$, %see Figure \ref{varepsilon.fig}
%for the average of the relative approximation
% error $ E_{N, L, \alpha}(n)=\|g_{N, L, \alpha}^{(n)}-f_{L, \alpha}\|_2/\|f_{L, \alpha}\|_{2}$ for two concentrated signals
% in Figure \ref{originalsignal}.
 see Figure \ref{varepsilon.fig}
for the average of the relative approximation error
$$ E_{N, L, \alpha}(n)=\|g_{N, L, \alpha}^{(n)}-f_{L, \alpha}\|_2/\|f_{L, \alpha}\|_{2}$$
to two concentrated signals
 in Figure \ref{originalsignal}
 over 500 trials.
%%%%%%%%%%%%%%%%%%%%%%%%%%%%%%%%%%%%%%%%%%% %%% average relative error 2/2
%%%%%%%%%%%%%%%%%%%%%%%%%%%%%%%%%
 Shown in Table \ref{tab:samplingrandom} is the   success rate of  the iterative algorithm \eqref{framealgorithm}
 to approximate the original signal $f_{L, \alpha}$ over 500 trials, where a trial is considered as successful
 if  the relative approximation error satisfies $ E_{N, L, \alpha}(n)\le \varepsilon_{L, \alpha}$ for $n=6$.
 We observe that the success rate is higher as the random sampling size $N$ increases.
  %and it tends to stable  with  a certain high level when random sampling size $N$ larger than some number.
Recall from Figure \ref{varepsilon.fig.addyaxu} that $\varepsilon_{L, \alpha}=C_\alpha L^{-\max(\alpha, 1/2)}$ decreases as $L$ and $\alpha$ increase.
This together with Table  \ref{tab:samplingrandom}
 demonstrates the conclusion in Theorem \ref{maintheorem.tm} that with high probability,
 $g_{N, L, \alpha}^{(n)}, n \ge O(\ln L)$,  provide good approximations to the original
signal $f_{L, \alpha}$ when $ N\ge O(L \ln L)$.

%%%%%%%%%%%%%%%%%%%%%%%%%%%%%%%%%%%%%%%

\begin{table}[t] %[!h]
\begin{center}
\caption{  Success rate to reconstruct the concentrated signals  $f_{L, \alpha}$  from random samples
 of size  $N=8L, 12L$ over  500  trials.}
\label{tab:samplingrandom}
\begin{tabular}{c|ccccc|ccccc}
\toprule
%\hline
$N$ &  & & %\multicolumn{5}{c}
{$8L$} & &  &  \multicolumn{5}{c} {$12L$} \\%& &$Gaussian$ & & &  & &$B-Spline$&  \\
\hline
 %midrule
%\cmidrule(lr){2-6} \cmidrule(lr){7-11}
\backslashbox{L}{SR} {$\alpha$}&0&0.2 & 0.4 & 0.6 & 0.8& 0&0.2 & 0.4 & 0.6 & 0.8  \\
%\midrule
\hline
50&94.0&92.6&86.8&83.0&87.2&99.8&100.0&99.6&99.2&98.8\\
70&90.8&88.0&84.2&78.6&85.0&99.8&100.0&99.0&99.0&98.2\\
90&87.4&85.4&81.4&75.8&72.8&99.8&100.0&98.8&98.8&98.4\\
110&85.0&80.4&76.0&68.8&68.4&99.6&99.2&99.0&98.6&98.4\\
%130&81.4&80.6&73.6&61.6&62.4&99.8&99.4&98.2&98.4&98.6\\
%150&78.2&76.8&66.2&59.0&57.8&99.8&99.4&98.2&98.2&98.6\\
170&72.4&73.0&56.8&49.4&51.8&99.2&99.4&98.8&97.0&96.4\\
%190&68.2&68.4&60.4&47.6&45.4&99.8&99.4&98.6&97.0&96.0\\
%210&65.0&60.2&53.2&45.0&39.4&98.6&99.4&99.0&97.4&95.8\\
230&60.4&57.4&51.2&35.8&36.8&98.8&99.6&98.2&96.8&96.6\\
%250&57.0&54.0&46.4&34.6&31.2&99.6&98.8&98.0&97.8&97.2\\
%270&54.6&54.0&42.8&29.2&29.2&100.0&99.2&98.6&95.2&95.6\\
290&52.2&48.6&39.4&30.6&24.4&99.4&99.8&98.4&96.8&93.2\\
%310&49.6&43.6&36.6&23.6&24.6&99.0&99.2&98.0&96.4&94.0\\
%330&48.6&39.6&29.8&21.6&19.6&99.8&98.6&98.4&96.6&94.8\\
350&39.0&36.2&31.0&15.6&19.4&99.2&98.2&96.8&95.6&91.6\\
\bottomrule
\end{tabular}
\end{center}
\end{table}

\begin{figure}[htbp]
\includegraphics[width=58mm, height=42mm]{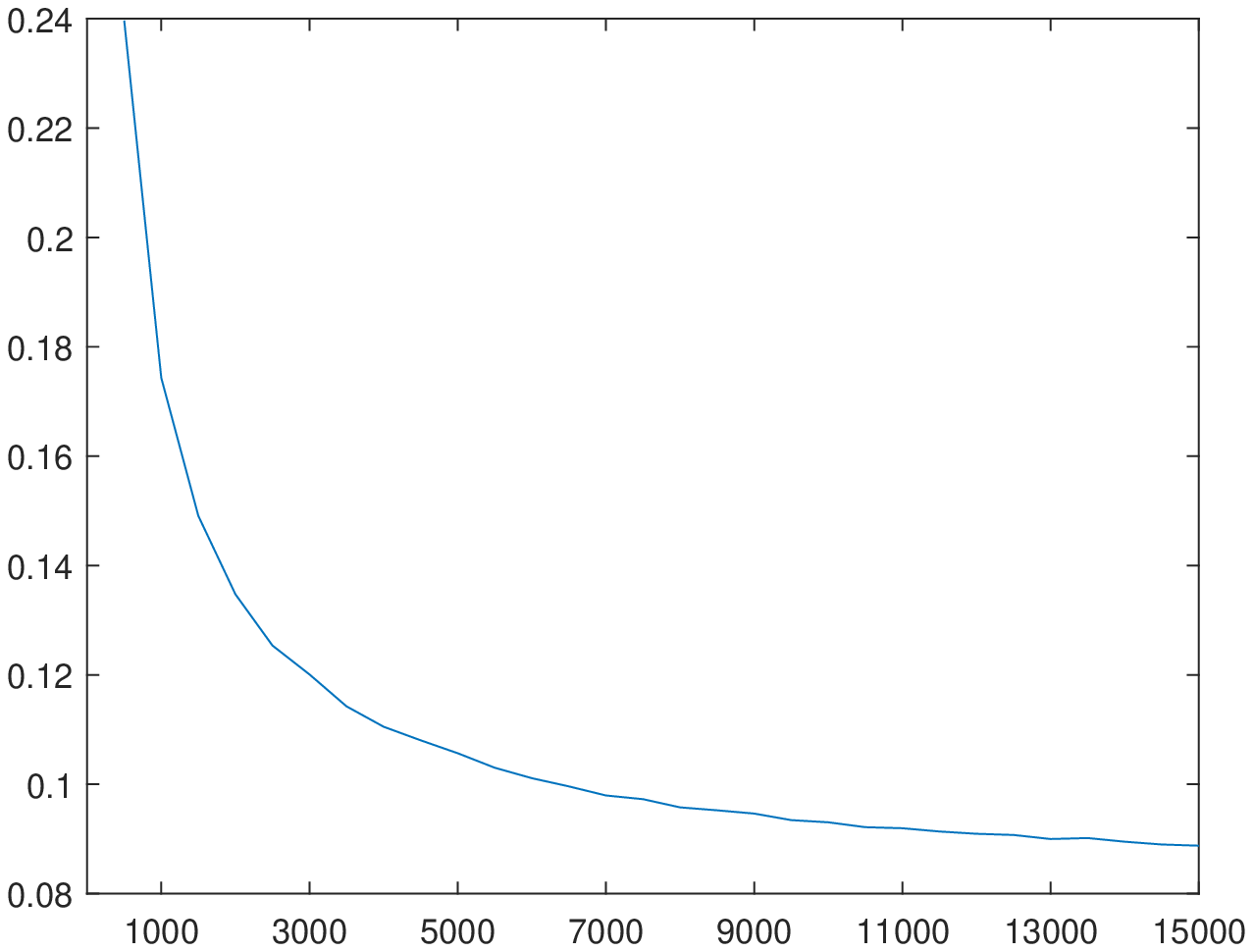}\hskip5mm
\includegraphics[width=58mm, height=42mm]{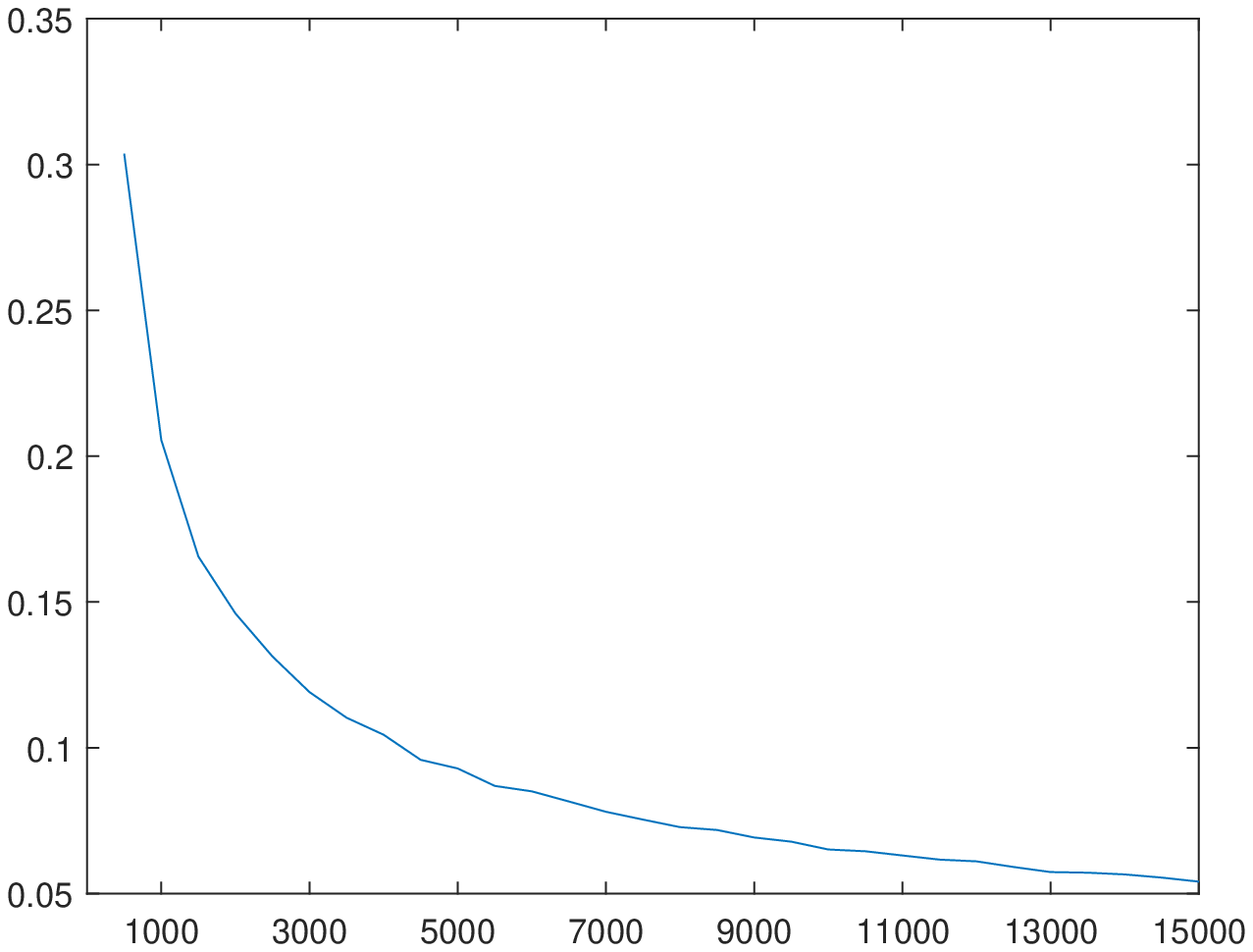}\\
\vskip0.1in
\includegraphics[width=58mm,height=42mm]{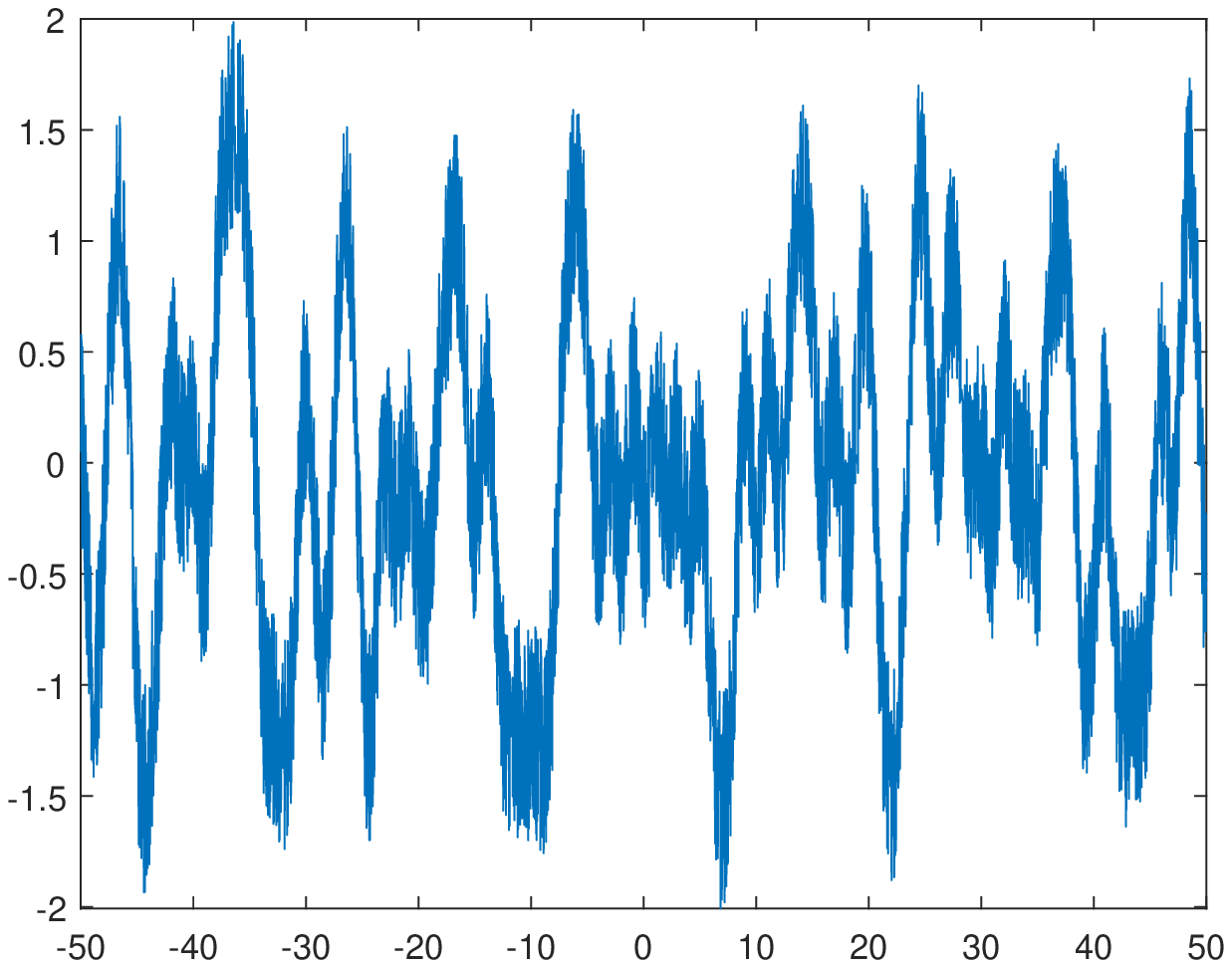} \hskip5mm %meanL8sr22
\includegraphics[width=58mm, height=42mm]{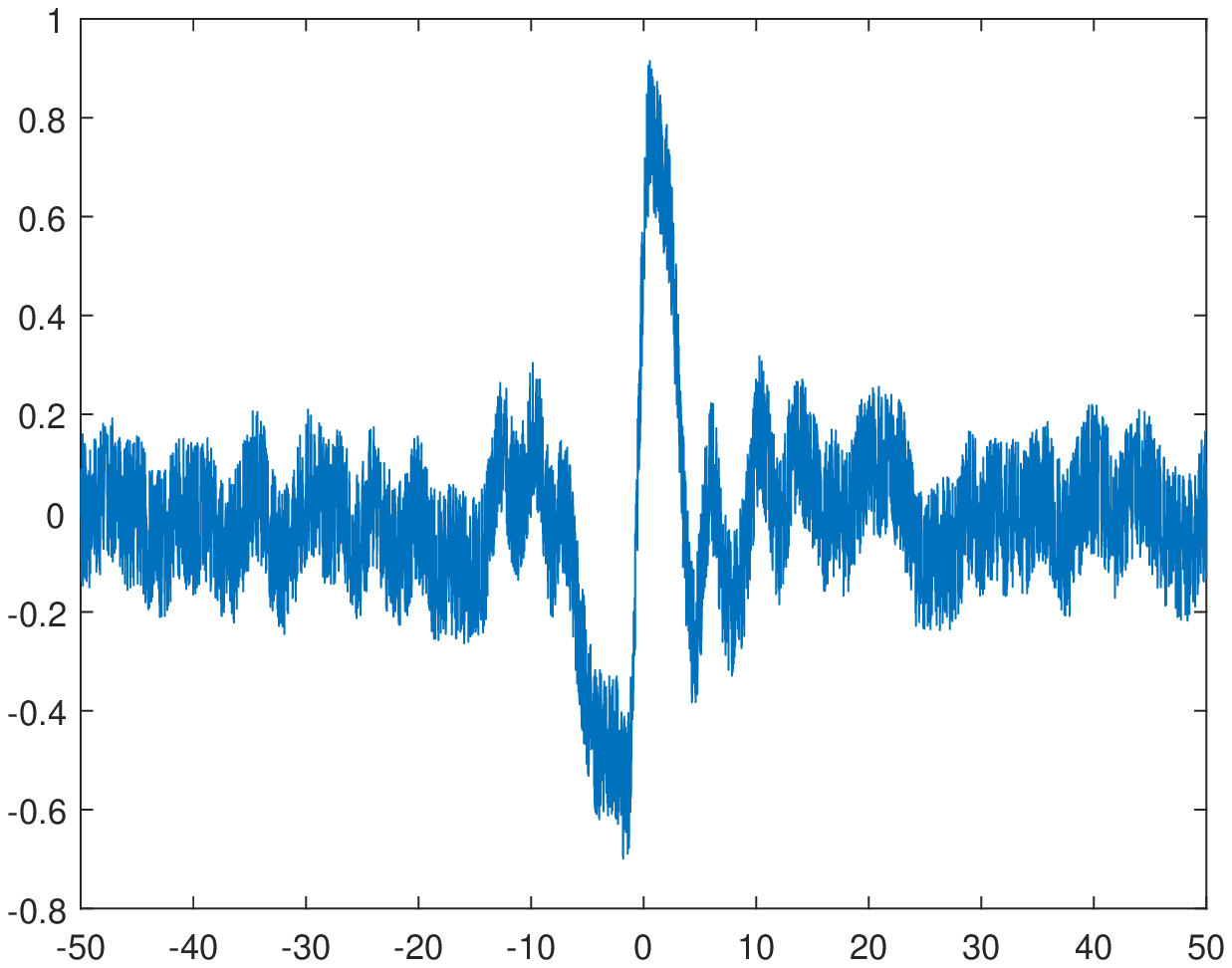}\\
%\includegraphics[width=48mm, height=38mm]{100Lalpha00gnnoise_example.eps}%meanL12sr22
%\\
%\includegraphics[width=48mm, height=38mm]{100Lalpha00gnf0noise_example.eps}%meanL8sr22
\vskip0.1in
\includegraphics[width=58mm, height=42mm]{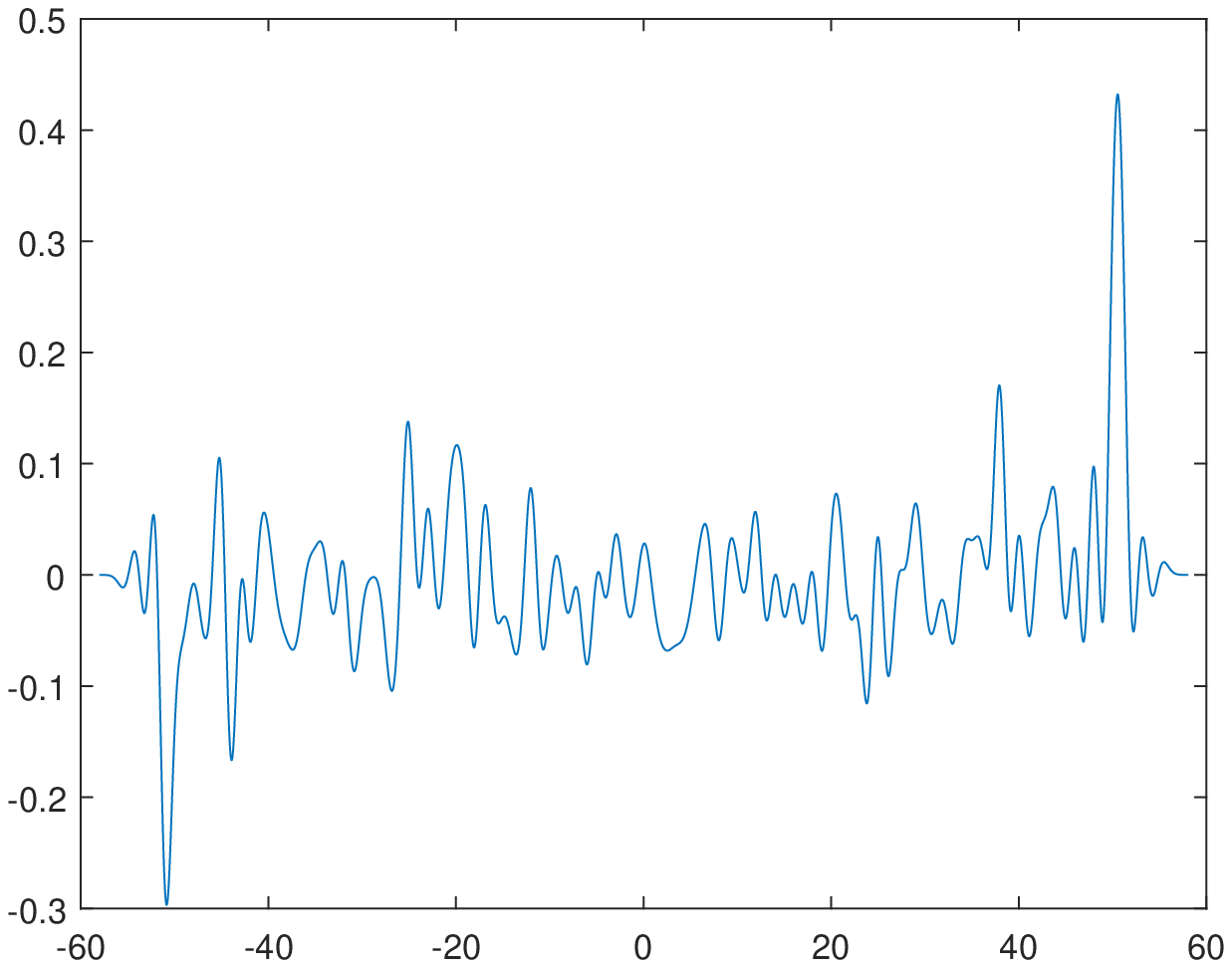}
\hskip5mm
%vskip0.1in
%\includegraphics[width=48mm, height=38mm]{100Lalpha08gnnoise_example.eps}
%\\
%\includegraphics[width=48mm, height=38mm]{100Lalpha08gnf0noise_example.eps}
\includegraphics[width=58mm, height=42mm]{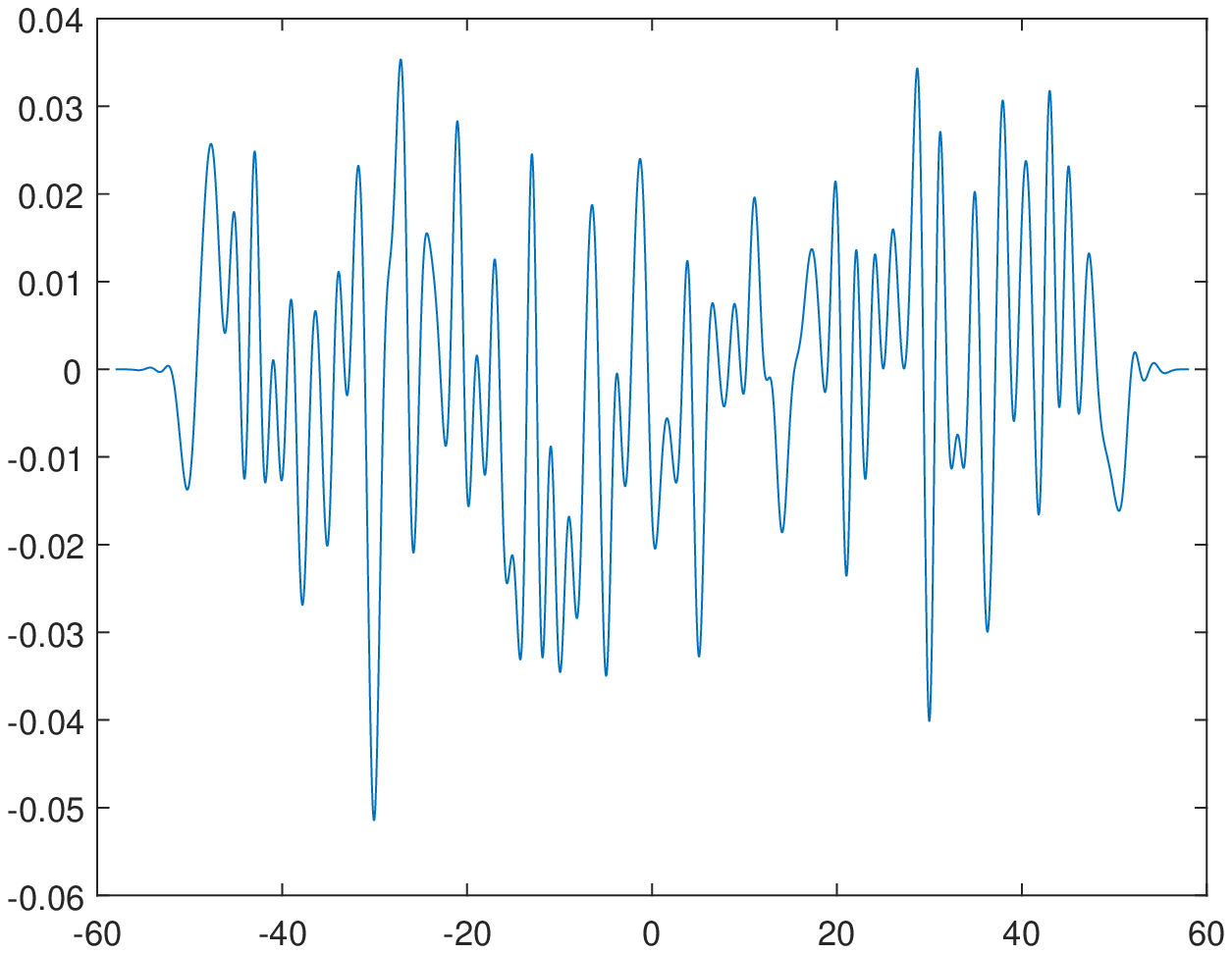}

\caption{Presented on the top left and right
 are the average of the relative approximation error $\tilde E_{N, L, 0}$ and  $\tilde E_{N, L, 0.8},
 L^2/5\le N\le 6 L^2$, over 100 trials respectively.
 Plotted on the middle left and right are
 noisy sampling data %$f_{L, \alpha}(\gamma)+\xi(\gamma)$
  in \eqref{noisedata.numerical}
with  $N=2 L^2=5000$ and $\alpha=0, 0.8$ respectively.
Shown on the bottom left and right are the difference   $\tilde g_{N, L, \alpha}^{(n)}-f_{L, \alpha}$
between the reconstructed signal $\tilde g_{N, L, \alpha}^{(n)}$
 at the sixth iteration ($n=6$) from noisy sampling data  in the middle figures
  and the original signal $f_{L, \alpha}$,
  where  $\alpha=0, 0.8$ for top/bottom figures respectively, cf. Figures \ref{reconstructionerrordeterminisiticsampling} and  \ref{varepsilon.fig}.
The relative approximation error  $\|g_{N, L, \alpha}^{(n)}-f_{L, \alpha}\|_2/\|f_{L, \alpha}\|_{2}$
and the concentration ratio $\|f_{L, \alpha}\|_{2, \Omega_L^c}/\|f_{L, \alpha}\|_{2}$
are 0.1036 and 0.0886 for the bottom left figure and  0.0948 %0.094794
and 0.0145 for the bottom right figure.
 }
\label{noisereconstructionaverage.fig}
\end{figure}

 In the  third part of our numerical simulations, we
 follow numerical simulations in the second part, except that
 the sampling data of a concentrated signal $f$ on the sampling set
$\Gamma_{N,L}=\{\gamma_k, 1\le k\le N\}$ being corrupted by i.i.d random noises
$\xi(\gamma)\in [-\delta, \delta], {\gamma \in \Gamma_{N, L}}$,     with uniform distribution, %  mean zero and  variance $\sigma^2$,
\begin{equation}\label{noisedata.numerical}
\tilde f_\gamma= f(\gamma)+ \xi(\gamma), \ \gamma\in \Gamma_{N, L}.
\end{equation}
Let $\tilde g_{N, L, \alpha}^{(n)}, n \ge 0$, be  the $n$-th term  in the algorithm  \eqref{noiseframealgorithm} with the original concentrated signal  $f_{L, \alpha}$  in  Figure \ref{originalsignal}  and the noisy data given in \eqref{noisedata.numerical}.
By Theorem \ref{random.thm}, the reconstructed signals $\tilde g_{N, L, \alpha}^{(n)}, n \ge O(\ln L)$,
provide good approximations to the original signal $f_{L, \alpha}$ when
 $$N\ge O(L \delta^2/\varepsilon_{L, \alpha}^2 \ln (L \delta^2/\varepsilon_{L, \alpha}^2))=
O(L^{\max(2\alpha+1, 2)}\delta^2 \ln (L^{\max(2\alpha+1, 2)}\delta^2)).$$
The above conclusion is  observed from Figure \ref{noisereconstructionaverage.fig}, where  $L=50, \delta=L^{\min(1/2-\alpha, 0)}/2$ and
  $\tilde E_{N, L, \alpha}=\|\tilde g_{N, L, \alpha}^{(n)}-f_{L, \alpha}\|/\|f_{L,\alpha}\|_2$
  is the relative approximation error between the reconstructed signal $\tilde g_{N, L, \alpha}^{(n)}$
 at the sixth iteration ($n=6$) from noisy data and the original signal $f_{L, \alpha}$
 given in  Figure \ref{originalsignal}.

\section{Proofs}
\label{proofs.section}

In this section, we collect the proofs of   Propositions \ref{Omegacovering.pr}, \ref{RksProperty.pr}, \ref{rkssampling.prop} and \ref{A.Gamma.Omega.Gap}, and Theorems
\ref{stability.thm}, \ref{approximation.thm}, \ref{deterministicnoise.thm} and \ref{random.thm}.

\subsection{Proof of Proposition \ref{Omegacovering.pr}}\label{Omegacovering.pr.pfsection}
Let $X_{\delta}$ be a discrete set of $X$ such that
\eqref{disjoint.def}
and \eqref{covering.eq1} hold, and
set
\begin{equation} \label {Omegacovering.pr.pfeq0}
\tilde X_{\delta}=\{x_i\in X_\delta\cap \Omega, \ \rho(x_i, \partial\Omega)>\delta\}.\end{equation}
Let $Y_\delta=\{y_k\}\subset \partial\Omega$ and $\tilde Y_\delta=\{z_k\}\subset \Omega$ be chosen so that
\begin{equation}\label{boundarycovering.eq}
\partial\Omega\subset \cup_{y_k\in Y_\delta} B(y_k, \delta) \end{equation}
and
\begin{equation}\label {Omegacovering.pr.pfeq1}
B(z_k, c\delta)\subset B(y_k, \delta)\cap \Omega,\   y_k\in Y_\delta,\ z_k\in \tilde Y_\delta.
\end{equation}
The existence of the set $\tilde Y_\delta$ follows from the Corkscrew condition \eqref{corkscrew.condition}
for the domain $\Omega$.

Let $ \Omega_\delta \subset  \tilde X_\delta\cup \tilde Y_{\delta}$ be a maximal set
such that
\begin{equation}\label {Omegacovering.pr.pfeq2}
\tilde X_\delta\subset \Omega_\delta,
\end{equation}
\begin{equation} \label {Omegacovering.pr.pfeq3}
B(x_i, \delta/2)\cap B(x_j, \delta/2)=\emptyset \ {\rm for \ all\ distinct} \ x_i, x_j\in \Omega_\delta,
\end{equation}
and
\begin{equation} \label {Omegacovering.pr.pfeq4}
B(y, \delta/2)\cap \big(\cup_{x_i\in \Omega_\delta} B(x_i, \delta/2)\big)\ne \emptyset \ \ {\rm for \ all} \ y\in \tilde X_\delta\cup \tilde Y_\delta.
\end{equation}

Now we show  that the  above maximal set $\Omega_\delta$ satisfies \eqref{Omegacovering.pr.eq}.
By
\eqref{Omegacovering.pr.pfeq0}, \eqref{Omegacovering.pr.pfeq1} and \eqref  {Omegacovering.pr.pfeq3}, we see that
the  maximal set $\Omega_\delta$
satisfies %\eqref{Omegacovering.pr.eq}.
\eqref{Omegacovering.pr.eq1} and \eqref{Omegacovering.pr.eq2}.

Next we establish the first inequality in \eqref{Omegacovering.pr.eq3}.
 Take  $x\in \Omega$. For the case that  $\rho(x, \partial \Omega)> 2\delta$,
 there exists $x_i\in X_\delta$ by \eqref{covering.eq1} such that
$\rho(x, x_i)\le \delta$, which implies that
$\rho(x_i, \partial \Omega)> \delta$ and hence $x_i\in \tilde X_\delta$. Then
for the case that $\rho(x, \partial \Omega)> 2\delta$,
\begin{equation} \label {Omegacovering.pr.pfeq5}
\rho(x,  \Omega_\delta)\le \rho(x,  \tilde X_\delta)\le \delta
\end{equation}
by \eqref{Omegacovering.pr.pfeq2}.
For  the case that $\rho(x, \partial \Omega)\le  2\delta$, there exists
$y_k\in  Y_\delta$ such that $\rho(y_k, x)\le 3\delta$ by the covering property \eqref{boundarycovering.eq}, which together with
  \eqref{Omegacovering.pr.pfeq1} implies
that
\begin{equation} \label {Omegacovering.pr.pfeq6}
\rho(x, \tilde Y_\delta)\le 4\delta.
\end{equation}
By the maximal property \eqref{Omegacovering.pr.pfeq4},  we have
\begin{equation}  \label {Omegacovering.pr.pfeq7}
\rho(y, \Omega_\delta)\le \delta \ {\rm for \ all} \ y\in \tilde X_\delta\cup \tilde Y_\delta.
\end{equation}
Combining \eqref{Omegacovering.pr.pfeq6} and \eqref {Omegacovering.pr.pfeq7}, we show that
\begin{equation} \label {Omegacovering.pr.pfeq8}
\rho(x, \Omega_\delta)\le  5 \delta
\end{equation}
for the case that $\rho(x, \partial \Omega)\le 2\delta$.
Therefore the first inequality in \eqref{Omegacovering.pr.eq3} follows from
\eqref{Omegacovering.pr.pfeq5} and \eqref{Omegacovering.pr.pfeq8}.

Finally we establish the second inequality in \eqref{Omegacovering.pr.eq3}.
Take $x\in \Omega$.  We obtain from  \eqref{regular.def} and \eqref{Omegacovering.pr.pfeq3} that
\begin{eqnarray*}\sum_{x_i\in \Omega_\delta} \chi_{B(x_i, 5\delta)}(x)
 & \hskip-0.08in \le &  \hskip-0.08in  \sum_{x_i\in  B(x, 5\delta)\cap \Omega_\delta}\frac{\mu(B(x_i, \delta/2))}{D_1(\mu) (\delta/2)^d}
 =    \frac{\mu\big(\cup_{x_i\in B(x, 5\delta)\cap \Omega_\delta} B(x_i, \delta/2)\big)}{D_1(\mu) (\delta/2)^d}\\
& \hskip-0.08in  \le & \hskip-0.08in    \frac{\mu(B(x, 11\delta/2))}{D_1(\mu) (\delta/2)^d} \le 11^d \frac{D_2(\mu)}{D_1(\mu)}.
\end{eqnarray*}
This completes the proof.

\subsection{Proof of Proposition \ref{RksProperty.pr}} \label{RksProperty.pr.pfsection}
 %We follow the argument used in \cite{nashedsun2010}, where the metric measure space is the standard Euclidean space
%and we include a sketch of the proof for  the completeness of this paper.
 Following the argument used in
\cite{nashedsun2010}, we have
\begin{equation}\label{boundedness.eq}
\|Tf\|_p\le \|K\|_{\mathcal S} \|f\|_p \ {\rm for \ all}\ f\in L^p.
\end{equation}
Then $V$ is a closed subspace of $L^p$.

By the interpolation property between $L^p$ and $L^\infty$ \cite{Berghbook76},
it suffices to prove
 \begin{equation}\label{fx.bound.new}
   \|f\|_\infty \le  (D_1(\mu))^{-1/p} \| K\|_{{\mathcal S}, \theta} \|f\|_{p},\   f\in V_p.
 \end{equation}
  Take $x\in X$. By \eqref{Vrks}, we obtain
\begin{equation}\label{fxbound}
  |f(x)|
   =   \Big |\int_X K(x,y) f(y) d\mu(y) \Big |
 %  &\le &  \Big(\int|K(x,y)|^{p/(p-1)}d\mu(y)\Big)^{1-1/p}\Big(\int \Big |\int K(y,z)g(z)d\mu(z)\Big |^p d\mu(z)\Big)^{1/p} \nonumber \\
   \le   \|K(x,\cdot)\|_{p'}\|f\|_p,
\end{equation}
where $1/p+1/p'=1$.

By the definition of the modulus of continuity, we have
\begin{equation*}
  |K(x,y^\prime)|\le \omega_{1}(K)(x,y)+|K(x,y)|,\ y\in  B(y', 1).
\end{equation*}
This together with \eqref{regular.def} and \eqref{KStheta.def} implies that
\begin{eqnarray}
\label{Kernelinfinity.eq}
 \| K(x, \cdot)\|_{\infty} &\hskip-0,08in \le & \hskip-0,08in \sup_{y'\in X}
 \frac{1}{\mu(B(y^\prime,1))} \int_{B(y^\prime,1)}\big(\omega_{1}(K)(x,y)+|K(x,y)|\big)d\mu(y)\nonumber\\
&  \hskip-0,08in \le  &  \hskip-0,08in (D_1(\mu))^{-1} \| K\|_{{\mathcal S}, \theta}.
  \end{eqnarray}
  Obviously,  \begin{equation*}
\| K(x, \cdot)\|_{1}\le \| K\|_{{\mathcal S}, \theta}.
  \end{equation*}
Interpolating the $L^1$ and $L^{\infty}$ norms of $K(x,\cdot)$ yields
\begin{equation}\label{kqestimate}
  \sup_{x\in X} \|K(x,\cdot)\|_{p'} \le (D_1(\mu))^{-1/p} \| K\|_{{\mathcal S}, \theta}.
\end{equation}
 Combining   \eqref{fxbound} and  \eqref{kqestimate} proves \eqref{fx.bound.new} and hence
\eqref{fx.bound}.

\subsection{Proof of Proposition \ref{rkssampling.prop}} \label{rkssampling.prop.pfsection}
 We follow the argument used in \cite{nashedsun2010}, where a similar result is established for a reproducing kernel space on the Euclidean space ${\mathbb R}^d$. Set $\delta=\delta(\Gamma), \Gamma_\Omega=\Gamma\cap \Omega$ and $\Gamma_{\Omega^c}=\Gamma\cap \Omega^c$.
 Then it follows from
\eqref{Igamma.assumptionX1}
that $\rho(y, \gamma)=  \rho(y, \Gamma_\Omega)\le  \delta$  for every $\gamma\in \Gamma_\Omega$ and $y\in I_\gamma$,
and  that  $\rho(y, \gamma)= \rho(y, \Gamma_{\Omega^c})\le  \delta$
 for every $\gamma\in \Gamma_{\Omega^c}$ and $y\in I_\gamma$.
Therefore
$\rho(y, \gamma)\le  \delta
$
for all $\gamma\in \Gamma$ and $y\in I_\gamma$.
Hence for all $x\in X$, we obtain
\begin{eqnarray}\label{SGamma.estimate}
 |S_\Gamma f(x)-f(x)|  & \hskip-0.08in  \le & \hskip-0.08in \sum_{\gamma\in \Gamma}
\int_{I_\gamma}| K(x, \gamma) f(\gamma)-  K(x, y) f(y)| d\mu(y) \nonumber\\
& \hskip-0.08in \le & \hskip-0.08in  \sum_{\gamma\in \Gamma}
\int_{I_\gamma}  |K(x, \gamma)- K(x, y)| |f(y)|  d\mu(y)\nonumber\\
%+\big(|K(x, y)|+|K(x, \gamma)-K(x,y)|\big)
& \hskip-0.08in   & \hskip-0.08in  +\sum_{\gamma\in \Gamma}
\int_{I_\gamma}   |K(x, \gamma)| \Big(\int_X |K(\gamma, z)-K(y, z)| \ |f(z)| d\mu(z)\Big)  d\mu(y) \nonumber\\
\nonumber\\
& \hskip-0.08in \le  & \hskip-0.08in \int_X \omega_{\delta}(K)(x, y) |f(y)| d\mu(y)\nonumber\\
& \hskip-0.08in  & \hskip-0.08in + \int_X \int_X \big(|K(x,y)|+\omega_{\delta}(K)(x, y)\big)\omega_{\delta}(K)(y, z) |f(z)| d\mu(z) d\mu(y)\nonumber \\
\qquad & \hskip-0.08in =: & \hskip-0.08in \int_X K_\Gamma(x, y) |f(y)| d\mu(y), \ f\in V_p.
\end{eqnarray}
%for all  $f\in V_p$ and
Observe that
\begin{equation}\label{Kestimate00}
\|K_\Gamma\|_{\mathcal S}\le \|\omega_{\delta}(K)\|_{{\mathcal S}} (1+ \|K\|_{{\mathcal S}}+ \|\omega_{\delta}(K)\|_{{\mathcal S}})
\le \|K\|_{{\mathcal S}, \theta}^2 \delta^\theta.
\end{equation}
By \eqref{SGamma.estimate} and \eqref{Kestimate00}, we obtain the following crucial estimate in the proof,
\begin{equation} \label{rkssampling.prop.pfeq2}\|S_\Gamma f-f\|_p\le \|\omega_{\delta}(K)\|_{{\mathcal S}} (1+ \|K\|_{{\mathcal S}}+ \|\omega_{\delta}(K)\|_{{\mathcal S}}) \|f\|_p
\le \|K\|_{{\mathcal S}, \theta}^2 \delta^\theta  \|f\|_p
\end{equation}
for $f\in V_p$.

By  \eqref{frameX.algorithm}, we can prove by induction on $n\ge 1$ that
\begin{equation} \label{rkssampling.prop.pfeq3}f_{n}-f_{n-1}= (I-S_\Gamma)^{n-1} (f_1-f_0)=(I-S_\Gamma)^{n} S_\Gamma f
\end{equation}
and
\begin{equation}\label{rkssampling.prop.pfeq3+}
  f_n=\sum_{k=0}^n (I-S_\Gamma)^k f_0=\Big(T+\sum_{k=1}^n (T-S_\Gamma)^k\Big) S_\Gamma f, \  n\ge 1.
\end{equation}
Define
\begin{equation}\label{rkssampling.prop.pfeq5}
  R:=T+\sum_{k=1}^{\infty}(T-S_{\Gamma})^k.
\end{equation}
Then one may verify that
$R$ is a bounded operator on $L^p$  by \eqref{rkssampling.prop.pfeq2},
\begin{equation} \label{rkssampling.prop.pfeq6}
\|Rf\|_p\le  \sum_{k=0}^\infty \|(I-S_\Gamma)^k Tf\|_p\le \big(1- \|K\|_{{\mathcal S}, \theta}^2 \delta^\theta\big)^{-1} \|Tf\|_p,  \ f\in L^p, \end{equation}
and $R$ is  a pseudo-inverse of the preconstruction  operator $S_{\Gamma}$,
\begin{equation} \label{rkssampling.prop.pfeq7}
RS_{\Gamma}f=S_{\Gamma} Rf=f,\  f\in V_p.
\end{equation}

By  \eqref{rkssampling.prop.pfeq2}, \eqref{rkssampling.prop.pfeq3+}, \eqref{rkssampling.prop.pfeq5} and \eqref{rkssampling.prop.pfeq7}, we have
\begin{eqnarray*}
\|f_n-f\|_p & \hskip-0.08in \le & \hskip-0.08in  \sum_{k=n+1}^\infty  \|(I-S_\Gamma)^k  S_\Gamma f\|_p\\
& \hskip-0.08in \le & \hskip-0.08in \frac{ 1+ \|K\|_{{\mathcal S}, \theta}^2 \delta^\theta}  {1- \|K\|_{{\mathcal S}, \theta}^2 \delta^\theta}
\big(\|K\|_{{\mathcal S}, \theta}^2 \delta^\theta\big)^{n+1} \|f\|_p, \  f\in V_p.
\end{eqnarray*}
This proves that $f_n, n\ge 1$, converge to $f$ exponentially.

\subsection{Proof of Theorem \ref{stability.thm}}\label{stability.thm.pfsection}
For $f\in V_p$,  let  $f_I$ be as in \eqref{f0.def0} except that replacing $h$ by $f$,
and  set $\delta=d_H(\Gamma_\Omega, \Omega)$.
 Following the argument after the statement of Theorem
\ref{stability.thm}, it suffices to prove
\begin{equation} \label{stability.thm.pf.eq00}
\|f_I-f\|_{p, \Omega}\le \|K\|_{{\mathcal S}, \theta} \delta^\theta \|f\|_p, \ \ f\in V_p.
\end{equation}
For any $\gamma\in \Gamma_\Omega$ and $x\in I_\gamma$
 it follows from
the Voronoi partition property \eqref{voronoipartition.def} that
%\begin{equation} \label{stability.thm.pf.eq3}
%d(x, \gamma)\le \min (\rho(x, \gamma),\ \rho(x, \Gamma_\Omega\backslash \{\gamma\}) )= \rho(x, \Gamma_\Omega)
%\le 2\delta.
%\end{equation}
%\begin{equation} \label{stability.thm.pf.eq3}
$\rho(x, \gamma)= \rho(x, \Gamma_\Omega)
\le \delta$.
%\end{equation}
This together with  \eqref{Vrks} implies that
\begin{eqnarray}\label{stability.thm.pf.eq4}
|f_I(x)-f(x)|  &\hskip-0.08in  = &\hskip-0.08in
\Big|\int_{X} \sum_{\gamma\in \Gamma_\Omega}  \big(K(\gamma, y)-K(x, y)\big) \chi_{I_\gamma}(x) f(y) d\mu(y)\Big|\nonumber\\
 & \hskip-0.08in  \le & \hskip-0.08in    \int_X \omega_{\delta}(K)(x,y) |f(y)|d\mu(y), \ x\in \Omega.
%\ \  {\rm for\  all} \ x\in I_\gamma \ {\rm and} \ \gamma\in \Gamma.
\end{eqnarray}
Combining \eqref{boundedness.eq}  %, \eqref{stability.thm.pf.eq3}
and \eqref{stability.thm.pf.eq4}
proves \eqref{stability.thm.pf.eq00},  and hence completes the proof.

\subsection{Proof of Theorem \ref{approximation.thm}}\label{approximation.thm.pfsection}
 For $f\in V_{p, \Omega, \varepsilon}$ and a sampling set $\Gamma_{\Omega^c}$ outside the domain $\Omega$, define
\begin{equation*}\label{approximation.thm.pfeq1}
f_0^c=\sum_{\gamma \in \Gamma_{\Omega^c} }\mu(I_\gamma) f(\gamma)  K(\cdot,\gamma)\in V_p
\ \ {\rm and} \ \ f_I^c=\sum_{\gamma \in \Gamma_{\Omega^c} }  f(\gamma)  \chi_{I_\gamma}. \end{equation*}
One may  verify easily  that
\begin{equation}  \label{approximation.thm.pfeq3}
\|f_0^c\|_p\le \|K\|_{{\mathcal S}, \theta} \|f_I^c\|_p=\|K\|_{{\mathcal S}, \theta} \|f_I^c\|_{p, \Omega^c}.
\end{equation}
  Set  $\tilde \delta= d_H(\Gamma_{\Omega^c}, \Omega^c)$.
  Applying similar argument used to prove \eqref{stability.thm.pf.eq2} and using \eqref{approximation.thm.eq4}, we obtain
\begin{equation*} \label{approximation.thm.pfeq4} \|f_I^c-f\|_{p, \Omega^c}\le  \|K\|_{{\mathcal S}, \theta} \tilde \delta^\theta  \|f\|_p\le \varepsilon \|f\|_p.
 \end{equation*}
 This together with \eqref{VOmega} and \eqref{approximation.thm.pfeq3}  implies that
 \begin{equation} \label{approximation.thm.pfeq5}
 \|f_0^c\|_p \le  \|K\|_{{\mathcal S}, \theta} \|f_I^c\|_{p, \Omega^c}\le  2
 \|K\|_{{\mathcal S}, \theta} \varepsilon \|f\|_p.
  \end{equation}

Set $\delta= d_H(\Gamma_\Omega, \Omega)$. Define
$f_n, n\ge 0$, as in \eqref{frameX.algorithm} with $\Gamma=\Gamma_{\Omega}\cup \Gamma_{\Omega^c}$.
Then it follows from  \eqref{approximation.thm.eq4}, \eqref{approximation.thm.eq1}  and Proposition  \ref{rkssampling.prop} that
$f_n, n\ge 0$, converge to $f$ exponentially,
\begin{equation}  \label{approximation.thm.pfeq6}
\|f_n-f\|_p\le  % \frac  { 1+ \|K\|_{{\mathcal S}, \theta}^2 \delta^\theta}
\frac{2}
 {1- \|K\|_{{\mathcal S}, \theta}^2 \delta^\theta}
\big(\|K\|_{{\mathcal S}, \theta}^2 \delta ^\theta\big)^{n+1} \|f\|_p, \ n\ge 0.
\end{equation}
Observe that
$f_0=g_0+f_0^c$ and
$$f_n=g_n+\Big(\sum_{k=0}^n (I-S_\Gamma)^k\Big) f_0^c, n\ge 1.$$
Therefore
\begin{equation}  \label{approximation.thm.pfeq7}
\|f_n-g_n\|_p  \le   \sum_{k=0}^n \big(\|K\|_{{\mathcal S}, \theta}^2 \delta ^\theta \big)^k \|f_0^c\|_p
\le
\frac{ \|f_0^c\|_p}
  {1- \|K\|_{{\mathcal S}, \theta}^2 \delta^\theta}
\le   2C_0 \varepsilon \|f\|_p
\end{equation}
by \eqref{approximation.thm.eq1}, \eqref{rkssampling.prop.pfeq2}  %\eqref{approximation.thm.eq4},
  and  \eqref{approximation.thm.pfeq5}.

Combining \eqref{approximation.thm.pfeq6}, \eqref{approximation.thm.pfeq7}
and then using \eqref{approximation.thm.eq2},
we have
\begin{equation*}
\|g_n-f\|_p\le \|f_n-f\|_p+\|g_n-f_n\|_p\le  4C_0 % \frac{  4
% \|K\|_{{\mathcal S}, \theta} }
%  {1- \|K\|_{{\mathcal S}, \theta}^2 \delta^\theta }
  \varepsilon \|f\|_p, \ n\ge 1.
\end{equation*}
This proves \eqref{approximation.thm.eq3}.

The conclusion \eqref{approximation.thm.eq3a} is trivial for $\varepsilon\ge 1/(9C_0)$. Now we consider the case that $\varepsilon< 1/(9C_0)$.  By
\eqref{approximation.thm.eq3} and the assumption $f\in V_{p, \Omega, \varepsilon}$, we have
\begin{equation}  \label{approximation.prop.pfeq4}
\|g_n\|_{p, \Omega^c} \le \|f-g_n\|_p+\|f\|_{p, \Omega^c}\le  5 C_0
%\frac{5 \|K\|_{{\mathcal S}, \theta} }  {1- \|K\|_{{\mathcal S}, \theta}^2 (d_{H}(\Gamma_\Omega, \Omega))^\theta}
\varepsilon \|f\|_p
\end{equation}
and
\begin{equation}  \label{approximation.prop.pfeq5}
\|g_n\|_p\ge \|f\|_p- \|g_n-f\|_p\ge
(1- 4C_0\varepsilon) %\frac{4 \|K\|_{{\mathcal S}, \theta}\varepsilon }  {1- \|K\|_{{\mathcal S}, \theta}^2 (d_{H}(\Gamma_\Omega, \Omega))^\theta}\Big)
\|f\|_p\ge \frac{5}{9} \|f\|_p.
\end{equation}
Combining \eqref{approximation.prop.pfeq4} and \eqref{approximation.prop.pfeq5}
proves \eqref {approximation.thm.eq3a}, and hence
the  reconstructed  signals  $g_n$ in \eqref{framealgorithm} are
$(9C_0\varepsilon)$-concentrated signals in $V_p$.

\subsection{Proof of Theorem  \ref{deterministicnoise.thm}}\label{deterministicnoise.thm.pfsection}
Let  $g_n, n\ge 0$ be as in \eqref{framealgorithm}. By Theorem \ref{approximation.thm}, it suffices to prove that
\begin{equation}\label{deterministicnoise.thm.pfeq1}
\|g_n-\tilde g_n\|_p\le C_0  \|\pmb\xi\|_{p, \mu(\Gamma_\Omega)}.
\end{equation}
Following similar argument used to establish  \eqref{approximation.thm.pfeq7}, we obtain
\begin{equation*}
\|g_n-\tilde g_n\|_p\le \big(1- \|K\|_{{\mathcal S}, \theta}^2 \delta^\theta\big)^{-1}\Big\|
\sum_{\gamma \in \Gamma_\Omega}\mu(I_\gamma)\xi(\gamma)  K(\cdot,\gamma)\Big\|_p.
\end{equation*}
This together with \eqref{approximation.thm.pfeq3} proves \eqref{deterministicnoise.thm.pfeq1} and hence completes the proof. %{``This together with \eqref{approximation.thm.pfeq3}'' I know what you mean}

\subsection{Proof of Proposition \ref{A.Gamma.Omega.Gap}}\label{A.Gamma.Omega.Gap.proofsection}
Let $\Omega_{\delta_1/10}$ be the  discrete set in  Proposition \ref{Omegacovering.pr} with $\delta$ replaced by $\delta_1/10$.
%For any  $x\not\in \cup_{\gamma\in \Gamma_\Omega} B(\gamma, \delta_1)$, we have that
 By \eqref{Omegacovering.pr.eq3}, we have
  \begin{eqnarray} \label{A.Gamma.Omega.Gap.pfeq1}
  \mathbb{P}\big \{ d_H(\Gamma_\Omega, \Omega)> \delta_1
        \big\} & \hskip-0.08in \le  & \hskip-0.08in   \mathbb{P}\big\{ B(x_i, \delta_1/2)\cap \Gamma_\Omega=\emptyset\ {\rm for \ some}\ x_i\in \Omega_{\delta_1/10}\big\}\nonumber\\
  & \hskip-0.08in \le  & \hskip-0.08in \sum_{x_i\in \Omega_{\delta_1/10}}    \mathbb{P}
\big\{ B(x_i, \delta_1/2)\cap \Gamma_\Omega=\emptyset\big\}.
   \end{eqnarray}
   We observe that
     \begin{equation} \label{A.Gamma.Omega.Gap.pfeq1+}
     \mathbb{P} \big\{ B(x_i, \delta_1/2)\cap \Gamma_\Omega=\emptyset\big\}\le \Big(1-\frac{\mu (B(x_i, \delta_1/2)\cap \Omega)}{{\mu(\Omega)}}\Big)^N
     \le \Big(1-\frac{ D_1(\mu) (c\delta_1/10)^d}{{\mu(\Omega)}}\Big)^N\end{equation}
     by  \eqref{Omegacovering.pr.eq1} and the assumption on the random sampling,
and also that
\begin{equation}  \label{A.Gamma.Omega.Gap.pfeq2}
\#\Omega_{\delta_1/10}
 \le  \sum_{x_i\in \Omega_{\delta_1/10}} \frac{\mu( B(x_i, c\delta_1/10))}{D_1(\mu) (c\delta_1/10 )^d}  =  \frac{\mu(\cup_{x_i\in \Omega_{\delta_1/10}} B(x_i, c\delta_1/10))} {D_1(\mu) (c\delta/10)^d}\le \frac{ 10^d {\mu(\Omega)}}{c^d D_1(\mu) \delta_1^d}
\end{equation}
by   Assumption \ref{metricspace.assump} and  Proposition \ref{Omegacovering.pr}.
Combining \eqref{A.Gamma.Omega.Gap.pfeq1}, \eqref{A.Gamma.Omega.Gap.pfeq1+} and
\eqref{A.Gamma.Omega.Gap.pfeq2} completes the proof.

\subsection{Proof of Theorem \ref{random.thm}}
By \eqref{random.thm.pfeq1} and Proposition \ref{A.Gamma.Omega.Gap}, we have
\begin{equation} \label{random.thm.pfeq1s}
{\mathbb P}\big \{d_H(\Gamma_\Omega, \Omega)> \tilde \delta_1\big \}
 \le  \frac{ 10^d {\mu(\Omega)}}{c^d D_1(\mu) {\tilde \delta}_1^d}
 \Big(1-\frac{  c^d D_1(\mu) {\tilde \delta}_1^d}{ 10^d \mu(\Omega)}\Big)^N\le \tau.
  \end{equation}
  Therefore it suffices to establish the conclusion under the hypothesis that
 \begin{equation}\label{random.thm.pfeq1s++} d_H(\Gamma_\Omega, \Omega)\le\tilde \delta_1.\end{equation}

Let
$g_n$ and $\tilde g_n, n\ge 0$, be defined  by  \eqref{framealgorithm} and \eqref{noiseframealgorithm} respecitvely. %\eqref{frame.algorithm.eq0} and \eqref{frame.algorithm.eq1}.
For a sampling set $\Gamma_\Omega$ with $d_H(\Gamma_\Omega, \Omega)\le \tilde \delta_1$, we obtain from Proposition
\ref{RksProperty.pr} and  Theorem  \ref{approximation.thm}
 that
\begin{equation} \label{random.thm.pfeq2}
|g_n(x)-f(x)| \le   8  (D_1(\mu))^{-1/p} \| K\|_{{\mathcal S}, \theta}^{2}  \varepsilon \|f\|_p
\end{equation}
for all integers $n$ satisfying \eqref{noiseinteger}.

Set $h_n=\tilde g_n-g_n, n\ge 0$. %{\color{blue} [[[The reason that I add this sentence is that when we define the operator $K_{n, \Gamma_\Omega}(x, y)$, we need $\delta$ and in (6.48) we also need to use $\delta$. ]]]}
Following the argument used in the proof of
Proposition \ref{rkssampling.prop}, we can show that
\begin{equation} \label{random.thm.pfeq3}
h_n(x)=\sum_{\gamma\in \Gamma_\Omega} \xi(\gamma) \mu(I_\gamma) \int_X K_{n, \Gamma_\Omega}(x, y)  K(y, \gamma) d\mu(y),
\end{equation}
and
\begin{equation}  \label{random.thm.pfeq4}
\|K_{n, \Gamma_\Omega}\|_{\mathcal S}\le \sum_{k=0}^n \Big(\|K\|_{{\mathcal S}, \theta}^2 \big(\max(d_H(\Gamma_\Omega, \Omega), d_H(\Gamma_{\Omega^c}, \Omega^c))\big)^\theta\Big)^k\le 2,
\end{equation}
where the last inequality follows from \eqref{approximation.thm.eq4}, \eqref{random.thm.pfeq0} and \eqref{random.thm.pfeq1s++}.

%{\color{blue}I add $\|K_{n, \Gamma_\Omega}\|_{\mathcal S}\le 2.$}
By  \eqref{random.thm.eq1} and \eqref{random.thm.pfeq3}, we have
\begin{equation} \label{random.thm.pfeq5}
 {\mathbb E}_{\pmb \xi} \big\{h_n(x) |d_H(\Gamma_\Omega,\Omega)\le \tilde\delta_1\big\} =
  \sum_{\gamma\in \Gamma_\Omega} {\mathbb E}_{\pmb \xi}(\xi(\gamma)) \mu(I_\gamma) \int_X K_{n, \Gamma_\Omega}(x, y)  K(y, \gamma) d\mu(y)=0,
  \end{equation}
  and
\begin{equation} \label{random.thm.pfeq6}
  {\rm Var}_{\pmb \xi} \big\{h_n(x)|d_H(\Gamma_\Omega,\Omega)\le \tilde\delta_1\big\} =
   \sigma^2 \sum_{\gamma\in \Gamma_\Omega} |\mu(I_\gamma)|^2
  \Big| \int_X K_{n, \Gamma_\Omega}(x, y)  K(y, \gamma) d\mu(y)\Big|^2.
  \end{equation}
  For a sampling set $\Gamma_\Omega$ satisfying \eqref{random.thm.pfeq1s++}, %$d_H(\Gamma_\Omega, \Omega)\le \tilde \delta_1$,
  we obtain from \eqref{regular.def}, \eqref{voronoipartition.def}, \eqref{Kernelinfinity.eq} and \eqref{random.thm.pfeq4} that
    \begin{equation} \label{random.thm.pfeq7}
  \mu(I_\gamma)\le D_2(\mu) \tilde \delta_1^d
   \end{equation}
  and
   \begin{equation} \label{random.thm.pfeq8}
  \Big| \int_X K_{n, \Gamma_\Omega}(x, y)  K(y, \gamma) d\mu(y)\Big|\le  \|K_{n, \Gamma_\Omega}\|_{\mathcal S} \|K(\cdot, \gamma)\|_\infty
  \le 2 (D_1(\mu))^{-1} \| K\|_{{\mathcal S}, \theta}
  \end{equation}
  for all $\gamma\in \Gamma_\Omega$.  Similarly, we have
  \begin{eqnarray}  \label{random.thm.pfeq9} & &
\sum_{\gamma\in \Gamma_\Omega} \mu(I_\gamma)
  \Big| \int_X K_{n, \Gamma_\Omega}(x, y)  K(y, \gamma) d\mu(y)\Big|\nonumber\\
&\hskip-0.08in  \le  & \hskip-0.08in
\int_\Omega \int_X |K_{n, \Gamma_\Omega}(x, y) | (|K(y, z)|+\omega_{\delta}(K)(y, z)|) d\mu(y) d\mu(z)\nonumber\\
&\hskip-0.08in  \le  & \hskip-0.08in  \|K_{n, \Gamma_\Omega}\|_{\mathcal S} \big(\|K\|_{\mathcal S}+
\|\omega_{\delta}(K)\|_{\mathcal S}\big)
\le 2 \|K\|_{{\mathcal S}, \theta},
\end{eqnarray}
 where $\delta=\max(d_H(\Gamma_\Omega, \Omega), d_H(\Gamma_{\Omega^c}, \Omega^c))$.
  Combining \eqref{random.thm.pfeq5}--\eqref{random.thm.pfeq9}, we get
\begin{equation*} \label{random.thm.pfeq10}
  {\rm Var}_{\pmb \xi} \big\{h_n(x)|d_H(\Gamma_\Omega,\Omega)\le \tilde\delta_1 \big\}
  \le  4 \sigma^2 (D_1(\mu))^{-1}  D_2(\mu)\|K\|_{{\mathcal S}, \theta}^2 \tilde \delta_1^d.
  \end{equation*}
 % if $d_H(\Gamma_\Omega, \Omega)\le \tilde \delta_1$.
Then applying  Chebyshev inequality yields
%  \begin{equation} \label{random.thm.pfeq11}
%{\mathbb P}_{\pmb \xi}\big \{|h_0(x)|\ge 2(D_1(\mu))^{-1/p}  \|K\|_{{\mathcal S}, \theta} \varepsilon \|f\|_p \ |d_H(\Gamma_\Omega,\Omega)\le \tilde\delta_1\big\}
%\le \frac{ \sigma^2 D_2(\mu) \tilde \delta_1^d
%          }
%          {(D_1(\mu))^{1-2/p}  \varepsilon^2 \|f\|_p^2}
%          \le \tau.
%\end{equation}
  \begin{equation}  \label{random.thm.pfeq11}
{\mathbb P}_{\pmb \xi}\big \{|h_n(x)|\ge 2(D_1(\mu))^{-1/p}  \|K\|_{{\mathcal S}, \theta} \varepsilon \|f\|_p \ |d_H(\Gamma_\Omega,\Omega)\le \tilde\delta_1\big\}\le
\frac{ \sigma^2 D_2(\mu) \tilde \delta_1^d
          }
          {(D_1(\mu))^{1-2/p}  \varepsilon^2 \|f\|_p^2}
          \le \tau,
\end{equation}
where
the second inequality holds by
\eqref{random.thm.pfeq0}.
Combining \eqref{random.thm.pfeq1s}, \eqref{random.thm.pfeq2} and \eqref{random.thm.pfeq11} completes the proof.

\end{document}